\documentclass[fleqn,usenatbib]{mnras}

\usepackage{graphicx}
\usepackage[percent]{overpic}
\usepackage{lipsum}
\usepackage{multirow}
\usepackage{color}
\usepackage{rotating}
\usepackage{mwe,tikz}
\usepackage[percent]{overpic}
\usepackage{mathtools}
\usepackage{physics}
\usepackage{amsmath}
\usepackage[utf8x]{inputenc}
\usepackage{hyperref}
\usepackage{longtable}
\usepackage{wasysym}
\usepackage[normalem]{ulem}
\usepackage{lmodern}
\hypersetup{
    colorlinks=true,
    linkcolor=blue,
    filecolor=magenta,      
    urlcolor=blue,
}
\usepackage{soul}
\usepackage{pdflscape}
\usepackage{longtable}

\newcommand{\msun}{${\rm M_{\sun}}$}

\def\ltsima{$\; \buildrel < \over \sim \;$}
\def\simlt{\lower.5ex\hbox{\ltsima}}
\def\gtsima{$\; \buildrel > \over \sim \;$}
\def\simgt{\lower.5ex\hbox{\gtsima}}
\def\msun{{\rm\,M_\odot}}


\usepackage{graphicx}
\usepackage{color}

\usepackage{amsmath} %??
\usepackage{graphicx}
   \usepackage{multirow}
\usepackage{epstopdf}
\newcommand{\be}{\begin{equation}}
\newcommand{\ee}{\end{equation}}
\newcommand{\bea}{\begin{eqnarray}}
\newcommand{\eea}{\end{eqnarray}}

\newcommand{\figref}[1]{Fig.~\ref{fig:#1}}

\newcommand{\equref}[1]{Eq.~(\ref{eq:#1})}

\newcommand{\gsim}{\lower.7ex\hbox{$\;\stackrel{\textstyle>}{\sim}\;$}}
\newcommand{\lsim}{\lower.7ex\hbox{$\;\stackrel{\textstyle<}{\sim}\;$}}

\def\insitu{{\it\,in situ\,\,}}
\definecolor{indiagreen}{rgb}{0.07, 0.53, 0.03}
\definecolor{darkspringgreen}{rgb}{0.09, 0.45, 0.27}

\newcommand{\Enlink}{{\it Enlink}}
\newcommand{\nmin}{{$N_{\mathrm{min}}$}}
\newcommand{\tinfall}{$T_{\mathrm{infall}}$}
\newcommand{\mtot}{$M_{\mathrm{tot}}$}

\usepackage{enumerate}
\usepackage{amssymb,amsmath}

\title[Action Space Clustering for Constraining Accretion History]{Using Action Space Clustering to Constrain the Accretion History of Milky Way like Galaxies}

\author[Wu et al.]{Youjia Wu$^{1}$\thanks{E-mail: youjiawu@umich.edu},
Monica Valluri$^{2}$,
Nondh Panithanpaisal$^{3}$,
Robyn E. Sanderson$^{3,4}$,
\newauthor{Katherine Freese$^{5,6,7}$,
Andrew Wetzel$^{8}$,
Sanjib Sharma$^{9,10}$.}
\\
$^{1}$Leinweber Center for Theoretical Physics, Department of Physics, University of Michigan, Ann Arbor, MI 48109, USA\\
$^{2}$Department of Astronomy, University of Michigan, Ann Arbor, MI, 48109, USA\\
$^{3}$Department of Physics and Astronomy, University of Pennsylvania, 209 S 33rd Street, Philadelphia, PA 19104, USA\\
$^{4}$Center for Computational Astrophysics, Flatiron Institute, 162 Fifth Avenue, New York, NY 10010, USA\\
$^{5}$Theory Group, Department of Physics, The University of Texas at Austin, 2515 Speedway, C1600, Austin, TX 78712-0264, USA\\
$^{6}$The Oskar Klein Centre, Department of Physics, Stockholm University, AlbaNova, SE-10691 Stockholm, Sweden\\
$^{7}$Nordita, KTH Royal Institute of Technology and Stockholm University, Roslagstullsbacken 23, 10691 Stockholm, Sweden\\
$^{8}$Department of Physics \& Astronomy, University of California, Davis, CA 95616, USA\\
$^{9}$Center of Excellence for Astrophysics in Three Dimensions (ASTRO-3D), Australia\\
$^{10}$Sydney Institute for Astronomy, School of Physics, A28, The University of Sydney, NSW 2006, Australia
}

\date{Accepted XXX. Received YYY; in original form ZZZ}

\begin{document}
\label{firstpage}
\pagerange{\pageref{firstpage}--\pageref{lastpage}}
\maketitle

%--- Abstract ----------------%
\begin{abstract}
In the currently favored cosmological  paradigm galaxies form hierarchically through the accretion of numerous satellite galaxies. Since the satellites are much less massive than the host halo, they only occupy a small fraction of the volume in action space defined by the potential of the host halo. Since actions are conserved when the potential of the host halo changes adiabatically, stars from an accreted satellite are expected to remain clustered in action space as the host halo evolves. In this paper, we identify accreted satellites in three Milky Way like disk galaxies from the cosmological baryonic FIRE-2 simulations by tracking satellite galaxies through multiple simulation snapshots. We then  try to recover these satellites by applying the cluster analysis algorithm \Enlink\, to the orbital actions of  accreted star particles in the present-day snapshot. We define several metrics to quantify the success of the clustering algorithm and use these metrics to identify well-recovered and poorly-recovered satellites. We study the distribution of these satellites in the infall time-progenitor mass space and infall time-stellar mass space, and statistically determine the boundaries between the well-recovered and poorly-recovered satellites in these two spaces with classification tree method.  The groups found by \Enlink\, are more likely to correspond to a real (accreted) satellite if they have high values of {\em significance}, a quantity that measures the density of the group relative to the background, that does not depend on knowledge of properties of the accreted satellites. Since cosmological simulations predict that most stellar halos have a population of \insitu\, stars, we test the ability of \Enlink\, to recover satellites when the sample is contaminated by between 10-50\% of  \insitu star particles, and show that most of the satellites well-recovered by \Enlink\, in the absence of \insitu stars, stay well-recovered even with 50\% contamination. We thus expect that, in the future, cluster analysis in action space will be useful in upcoming data sets (e.g. Gaia) for identifying accreted satellites in the Milky Way.
\end{abstract}

\begin{keywords}
 galaxies: halos - stars: kinematics and dynamics - galaxies: formation - methods: statistical - methods: data analysis
\end{keywords}
%--- Abstract ----------------%

\section{Introduction}
\label{sec:Introduction}

In the currently favored cosmological paradigm galaxies form hierarchically through the accretion and merger of numerous satellite subhalos. N-body simulations of galaxy formation make strong predictions about the number of dark matter subhalos and their mass functions and the mass assembly rates of halos. Using cosmological simulations it has been shown that mass functions of subalos (at different redshifts) is a strong discriminator between cold dark matter (CDM) and alternative forms of dark matter (e.g. Warm Dark Matter (WDM), Self-interacting Dark Matter (SIDM)).   In particular the number of subhalos in present day Milky Way (MW) mass halos in the mass range $10^{6}-10^{10}\msun$ is sensitive to the nature of dark matter \citep[for  recent reviews see,][]{ bullock_boylan-kolchin_17,zavala_frenk_19}. In addition, in $\Lambda$CDM the merger rate of dark matter halos per unit mass ratio (relative to the host halo at the time of accretion), per redshift interval has a nearly universal functional form \citep{fakhouri_08}. 

Several novel methods are being used to detect  dark (or nearly dark) subhalos in the Milky Way's halo: 
modeling gaps in thin stellar streams that may have resulted from impact with a dark subhalo \citep[e.g.][]{Erkal2016_gaps,WhelanBonacaGD12018,Bonaca_spur_2018}, perturbations to the Galactic disk by dark matter subhalos, which might produce bending modes or wiggles or corrugations in the density of the disk \citep{feldmann_spolyar_15} which may already have been detected in the solar vicinity \citep{widrow_etal_12} and on slightly larger scales \citep{antoja_etal_18}. Dark matter subhalos of masses $\gsim 3\times10^9\msun$ tend to be massive enough to retain their baryons and form stars \citep[e.g.][]{lazar_etal_20}. Many of these subhalos are currently detectable as satellites (dwarf spheroidal galaxies and ultra faint dwarf galaxies) in the Local Group. In addition, satellites that were accreted throughout the Milky Way's history have been tidally disrupted by the Galactic potential and now form much of the stellar halo of the Milky Way. If it is possible to observationally dissect the stellar halo of the Milky Way to reliably identify the stellar debris of such satellites, one could use the kinematics and chemical abundance signatures in the debris to determine properties (e.g. masses) of the progenitors. We would also be able to use this information to determine the accretion history of our Galaxy and possibly quantify the mass function of accreted satellites.

Early work \citep[e.g.][]{1996ApJ...465..278J,1999MNRAS.307..877T,helmi_dezeeuw_00,harding_etal_01} showed that  merger remnants  remain coherent in phase space (or integrals-of-motion space) long after they have become so phase mixed that they are impossible to detect via their spatial distributions. Most early works have focused on finding coherent structures in energy, angular-momentum, or velocity space. Since stars in a galaxy are collisionless, the space of orbital actions, integrals-of-motion that are conserved even under adiabatic changes to the underlying gravitational potential, are particularly promising \citep[for a detailed introduction see,][]{BT08}. Since accreted substructures are much less massive than the host halo,  they occupy a much smaller volume in the action space defined by the gravitational potential of the host halo, implying that stars from the same progenitor could still be clustered in the action space at $z=0$, making it possible to detect these mergers events. 

The advent of Gaia \citep{2001A&A...369..339P,2016A&A...595A...3F,2016A&A...595A...4L,2018A&A...616A...2L,2018A&A...616A...4E, 2010MNRAS.408..935G} is making it possible to obtain 6-dimensional phase space information for hundreds of thousands to tens of millions of halos stars. Numerous efforts are underway to automatically identify substructures in the phase space. With the SDSS-Gaia DR1 catalogue of $\sim 80000$ main sequence turn-off halo stars in 7D phase space (3D position+3D velocity+metallicity), \citet{2018MNRAS.478..611B} showed that  metal-rich halo stars were on significantly more radial orbits than metal poor stars. They inferred that this highly anisotropic velocity distribution was consistent with a single, head-on accretion event by a satellite with mass above $10^{10} \msun$. This satellite was named the ``Gaia Sausage'' at that time. It was soon after discovered in Gaia-DR2 data, and was named ``Gaia Enceladus'' \citep{2018Natur.563...85H}. Now this satellite is referred to as ``Gaia Enceladus Sausage" (GES).

\citet{2018ApJ...856L..26M} analyzed the same SDSS-Gaia catalog in action space ($\{J_r,J_z,J_{\phi}\}$) to show that the metal-rich stars were more extended toward high radial action $J_r$ and more concentrated around the $J_{\phi}=0$, showing different patterns from metal-poor stars, in agreement with \citet{2018MNRAS.478..611B}. \citet{10.1093/mnras/sty1403} developed an algorithm to find over-densities in action space, and used the metallicity as a secondary check.  They identified 21 substructures in the SDSS-Gaia catalogue, and argued that 5 of them are associated with the accretion of the progenitor of $\omega$ Centauri, inferring the minimum mass of the $\omega$ Centauri progenitor to be $5 \times 10^8 \msun$. \citet{2020MNRAS.492.1370B} applied the clustering algorithm DBSCAN to search for streams in integrals of motion space. \citet{2018AJ....156..179R}  applied several clustering algorithms to  energy+actions space coordinates for 35 nearby r-process-enhanced field halo stars (obtained using Gaia proper motions, radial velocities and parallaxes) and were able to identify 8 separate clusters with statistically distinct iron abundances, supporting the view that the 35 stars were accreted in at least 8 distinct satellites (clusters) in the action space.  \citet{yuan_etal_20} searched for dynamical substructures in the LAMOST DR3 catalog of very metal-poor stars cross-matched with Gaia DR2 by applying the self-organizing map algorithm StarGO. They identified 57 dynamically tagged groups, many of which belonged to previously identified accretion events. \citet{2021ApJ...907...10L} applied the clustering algorithm HBSCAN to 4D energy-action data of \~ 1500 very metal poor stars based on spectroscopic data from the HK and Hamburg/ESO surveys, and found 38 dynamically tagged groups, with many of them corresponding to previously known substructures and 10 of them being new. \citet{2021ApJ...908...79G} used HBSCAN on 4D energy-action data of 446 r-process-enhanced stars in the halo and the disk of the Milky Way, and found 30 chemo-dynamically tagged groups (CDTGs),  with stars from the same CDTG showing statistically significant similarities in their metallicities, indicating that stars from the same CDTG have experienced common chemical evolution histories   in their parent substructures prior to entering the Milky Way halo. \citet{necib_etal_19} and \citet{necib_etal_20_nyx} used clustering algorithms to identify clusters in phase space (primarily position-velocity space) and have found debris of several previously known and newly discovered satellites, including evidence for the accretion of a prograde satellite close to the disk plane that they named ``Nyx''. 
 
Hamiltonian dynamics  tells us that accretion events should remain coherent in the action space for a very long time, as long as the potential changes slowly enough. However, it is as yet uncertain how long one can expect structures  to remain coherent in a Milky Way like galaxy that grows hierarchically.  The discovery of fairly massive past mergers like the ``Gaia Sausage''  and the ``Sequoia'' \citep{2019MNRAS.488.1235M} galaxy, the on-going mergers like the Sagittarius stream \citep{LyndenBell:1995zz} and future mergers like the LMC \citep{2007ApJ...668..949B, 2010ApJ...721L..97B} raises questions about how ``adiabatic'' the evolution of the Milky Way has been and therefore how well one might distinguish other individual merger events. Is there a boundary in infall time such that satellites that fell in before this time have experienced so much phase mixing, that we can no longer find them through cluster analysis in the action space?  The average density of satellite within its tidal radius relative to the mean density of the host within the satellite's orbit roughly determines its rate of tidal disruption. However, the mass of the satellite also determines the initial dispersion of its stars in action space and hence its expected degree of clustering. Therefore is there an optimal range of mass for satellites that can be detected via cluster analysis in phase space? The reliability of cluster analysis algorithms in finding substructures is also poorly understood: while numerous clusters are often identified by such algorithms, it is unclear how many of them correspond to discrete building blocks (individual satellites), how many are comprised of multiple satellites, how many are subcomponents of individual satellites and how many are spurious. Therefore we need to find a metric (or metrics)  computed from cluster analysis  that quantify the correspondence between groups identified by cluster analysis and real galactic building blocks in the hierarchical galaxy formation framework. This paper is  motivated by these questions.

%In this paper we test the ability of a cluster finding algorithm, \Enlink\ \citep{Sharma_2009}, to find the population of accreted satellites by searching for groups in phase space for halo stars in three MW-like galaxies in a suite of cosmological hydrodynamics simulations  \citep{10.1093/mnras/sty1690}. We then assess how the ability of \Enlink\, to detect a merger remnant depends on the mass of the satellite at accretion and the time at which the accretion event occurred. We further identify a metric provided by the \Enlink\, algorithm that is correlated with well-recovered satellites. 

The cosmological hydrodynamic zoom-in simulations of Milky Way (MW)-mass galaxies from the Feedback In Realistic Environments (FIRE) project \footnote{website: \url{https://fire.northwestern.edu/}} \citep{10.1093/mnras/sty1690} provide a great testbed for addressing these questions. We focus on three MW-mass galaxies from the Latte suite of FIRE-2 simulations which have different merger histories, ranging from one with a very quiescent recent history to one with a very active recent history. In Section \ref{sec:data} we describe how the simulation snapshots were analyzed to identify the accreted stars in the three halos at $z=0$. In Section \ref{ss:action} we provide a brief introduction to the calculation of actions $\{J_r, J_z, J_{\phi}\}$  in cylindrical coordinates using publicly available AGAMA code \citep{10.1093/mnras/sty2672}.  In Section \ref{ss:cluster} we describe the density-based hierarchical cluster analysis algorithm \Enlink\ \citep{Sharma_2009}  which we use to perform the cluster analysis in 3D action space. In Section \ref{subsec:acc} we study the overlap between the real stellar building blocks (satellites) of the three MW-mass galaxies and the groups found by \Enlink. We define several metrics that  we use to evaluate the accuracy of recovery of substructure with \Enlink.   We also use a binary classification tree to statistically determine boundaries in  infall time (\tinfall), total progenitor mass (\mtot) and stellar mass at $z=0$ ($M_{\rm stellar}$) for well-recovered and poorly-recovered satellites. The stellar halo of the host galaxy contains both accreted stars and \insitu stars. The \insitu stars are formed in the host galaxy and are in the stellar halo of the host galaxy at $z=0$. In Section \ref{subsec:insitu} we include varying fractions of \insitu star particles into our analysis and study the robustness of our results against contamination from these stars. Finally, in Section \ref{sec:con} we discuss how the boundaries (in \tinfall\, \mtot\, and $M_{\rm stellar}$) separating well-recovered and poorly recovered satellites depend on the dynamical history of the host MW-like galaxies. We then summarize our results and conclude.

\section{Simulation Data}
\label{sec:data}
Our analysis uses cosmological baryonic zoom-in simulations of MW-mass galaxies from the FIRE project \citep{10.1093/mnras/sty1690}. These simulations are run with \texttt{GIZMO} \citep{Hopkins2015} which uses an optimized TREE+PM gravity solver and a Lagrangian mesh-free, finite-mass method for accurate hydrodynamics. Star formation and stellar feedback are also implemented. All halos were simulated in $\Lambda$CDM cosmology at particle mass resolution of $\sim7100 \msun$ and spatial resolution of 1--4 pc for star/gas particles, and a particle mass resolution of $\sim35,000 \msun$ and spatial resolution of 40 pc for dark matter particles. The complete sample currently consists of 8 MW-mass galaxies and 3 Local-Group-like pairs. \citet{Wetzel_2016} and \citet{10.1093/mnras/stz1317} show that when baryonic physics is included, the properties of dwarf galaxies in the FIRE-2 simulations agree well with observations of Local Group (LG) satellites down to the resolution limit (just below classical dwarf mass). Of particular importance for this work, the simulations produce satellites with mass--size and mass--velocity dispersion relations consistent with observations of the MW and M31 \citep{10.1093/mnras/stz1317} distributed similarly with respect to their massive hosts \citep{2020MNRAS.491.1471S,2020arXiv201008571S}. This implies that we expect the sizes and relative positions of the accreted structures in action space to resemble those in the Milky Way.

To assign the accreted star particles to particular progenitor galaxies, the dark matter particles in each snapshot of the simulations are first processed with \texttt{Rockstar} to produce halo catalogs, which then are connected in time to form a merger tree \citep{2013ApJ...762..109B, 2013ApJ...763...18B}. \texttt{Rockstar} computes the maximum circular velocity $v_\mathrm{max}$ and the virial radius for each halo and subhalo identified in the dark matter distribution. A star particle is considered part of a halo or subhalo if it is within the virial radius \emph{and} its velocity with respect to the center of that halo or subhalo is less than $2 v_\mathrm{max}$\citep{2020ascl.soft02015W, 2020ascl.soft02014W}. Within the host halo, this selection does a good job in picking out star particles gravitationally bound to a subhalo rather than the host halo.

One of the challenges with identifying substructure in the stellar halo is that there are indications from cosmological hydrodynamical simulations that some fraction of the stars in the halo of a Milky Way like galaxy were not accreted from satellites but were born \insitu in the host galaxy, both at very early times before the disk was well established  \citep[e.g.][]{2020MNRAS.497..747S} and in smaller proportions at later times, in gas propelled into the halo by star-forming winds \citep[e.g.][]{yu_etal_20}. These stars then remain in the stellar halos of these MW-like galaxies at $z=0$, a generic prediction of multiple simulations using independent codes and differing star-formation and feedback prescriptions \citep{Zolotov2009,cooper_etal_10, font_etal_11, Zolotov2012,Tissera2013,Pillepich15,monachesi_etal_19}. However, these various cosmological simulations predict a wide range of values for the \emph{fraction} of halo stars that were formed \insitu, which probably varies with assembly history and likely also depends on the feedback prescriptions adopted by different codes. While some simulations predict that up to 80\% of the stellar halo was formed \insitu, observations of  the MW stellar halo \citep{bell_etal_08,2020arXiv200608625N} find that almost all of it shows significant substructure, implying that a significant portion of the Milky Way stellar halo was accreted (although \insitu stars formed in outflows may also be clustered; \citealt{yu_etal_20}). 

Recently \citet{Ostdiek:2019gnb}  showed that they were able to train and validate a deep learning neural network  algorithm on 5D mock Gaia kinematical data \citep{sanderson_etal_20} from the same simulations we use here \citep{Wetzel_2016,10.1093/mnras/sty1690}   to separate \insitu halo stars from accreted stars. They then applied this method to 72 million stars in the Gaia DR2 catalog with parallax measurement errors of less than 10\% and were able to identify over 650,000 stars as accreted. They then used a cluster finding algorithm to identify clusters in phase space and their comparisons with other datasets allowed them to validate both robustness of their neural network algorithm for separating \insitu stars, and identify several new structures \citep{necib_etal_19,necib_etal_20_nyx}. Based on the success of such algorithms to separate accreted and \insitu\, stars, we assume in this paper that such separation is possible, and we use information derived from the analysis of simulation snapshots to identify accreted satellites in the simulations.

We focus on  three MW-mass galaxies: m12i, m12f and m12m. For each galaxy there are 600 snapshots from $z=99$ to $z=0$, with a time difference between snapshots of approximately 25 Myr at late times. This relatively high ``framerate'' allows us to track the time-evolution of accreted structures in much of the host galaxy with $\sim$10--100 snapshots per dynamical time. Since it takes several dynamical times for a bound satellite to be tidally disrupted and turn into a stream, we select luminous substructures that are bound between 2.7 to 6.5 Gyr ago and that are within the virial radius of the host galaxy at present day. These time scales correspond to redshift $z\approx 0.25-0.75$. We follow each substructure throughout its evolution to recover most of the star particles that once belonged to the satellite. These star particles are then tracked forward to $z=0$. The stellar mass of each substructure at $z=0$ is summed up to get $M_{\rm stellar}$. Star particles belonging to a particular substructure are also traced back in time until the substructure is no longer bound to the host halo. This time is defined as the infall time \tinfall\, of the substructure, and all the mass belonging to this substructure at \tinfall\, is summed up to get the total mass of the satellite (stars, gas and dark matter) \mtot. Such an accreted substructure is referred to as a ``satellite''. The number of star particles in a satellite is defined as the ``size" of this satellite. Extending the sampling time window to a time beyond 6.5 Gyr ago might help detect satellites that fall in very early. However, at earlier times ($z\lesssim$ 2--4), there is no clear ``host'' galaxy in terms of mass \citep{2020MNRAS.497..747S}, and the disk of what eventually becomes the most massive galaxy is not usually formed yet at that point \citep{2018MNRAS.481.4133G}, so the question of what is accreted onto what is not well-defined.

%mvc{I don't understand this - how can you require that the separation between any two star particles be {\em greater} than 120kpc - one would not get any stars in the galaxy}.
%\mvc{need more details to justify the equation below and mention that this is an empirical relation that has been found to work for separating streams from phase mixed satellites}.
Starting with the star particles assigned to each accreted structure, we make three selections on both real space and velocity space to identify coherent streams and phase-mixed satellites. In the first selection, we choose satellites with stellar mass between $M_\star \approx 10^6\msun$ and $10^9 \msun$ at $z=0$. In the second selection, we use the fact that the ``length" of disrupted substructures in physical space is larger than that of satellite galaxies that are still self-gravitating at $z=0$. In practice, the ``length" of a substructure can be measured by the maximum value of the pairwise distance between two star particles in a substructure. To take advantage of this fact and rule out the self-gravitating satellite galaxies in the $z=0$ snapshots, we require the ``length" of each substructure we tracked to be greater than 120 kpc, as 120 kpc is much larger than the characteristic size of self-gravitating dwarf galaxies. 

The remaining satellites are classified by eye into two categories, coherent streams and phase-mixed satellites, by viewing their configuration in position and velocity space. A forthcoming study     \citep{2021arXiv210409660P} shows that this by-eye classification corresponds to a selection in the space of stellar mass and local velocity dispersion, where  ``local'' velocity dispersion is defined for each particle using near neighbors in phase space rather than position (at fixed stellar mass, phase-mixed satellites have higher local velocity dispersion compared to coherent streams). Although this classification is mildly resolution-dependent, the overall results when this classification is applied to the full sample of streams from all 13 analyzed systems in \citep{2021arXiv210409660P} demonstrate the mass- and time-dependence expected from theory for the relative abundance of phase-mixed versus coherent debris, implying that resolution effects do not dominate our classification of phase-mixed satellites.

\section{Methods}
\label{sec:met}
The orbits in a galactic potential are largely quasi-periodic and regular. %motion of satellite halos orbiting around the center of the host halo is quasi-periodic. 
Therefore the orbits of most satellites and individual stars in the halo of a galaxy can be described by an elegant set of  variables -- the   action-angle variables. Orbital actions are particularly powerful for understanding the evolution of  a galaxy since they are conserved under adiabatic evolution of the potential. Section \ref{ss:action} gives a brief introduction to action-angle variables. We then use a publicly available  dynamical modeling toolbox AGAMA \citep{10.1093/mnras/sty2672}, to numerically compute actions (under the assumption of axisymmetry) for the accreted star particles in the three MW like galaxies in FIRE-2 described in Section~\ref{sec:data}. 

The initial actions of all the stars in a satellite as it orbits a larger galaxy have a small spread compared to the range of possible actions in the Galactic potential. Because actions are adiabatically invariant by construction, star particles from the same progenitor are expected to remain clustered in the action space at $z=0$ if the gravitational potential of the host halo has changed only adiabatically (i.e. slowly enough) following their infall. Therefore, the accreted satellites found in Section \ref{sec:data} should be recoverable through cluster analysis in the action space. In Section \ref{ss:cluster}, we feed the cluster analysis algorithm \Enlink\, with the three-dimensional orbital actions of accreted star particles, and find several groups. To measure how well the groups found by \Enlink\, recover the satellites tracked in Section \ref{sec:data}, we define various metrics in Section \ref{ss:cluster_metrics}. %Since it is unclear if the evolution of the potential of a MW like galaxy has really been adiabatic over the past 6.5 Gyr, we aim to test this with simulations. In Section \ref{ss:cluster} we introduce the basic ideas of  the cluster analysis algorithm \Enlink\, \citep{Sharma_2009} and then apply it to the actions calculated in Section \ref{ss:action} to try to recover the accreted satellites in the action space for each of the 3 MW-like galaxies from the  FIRE-2 simulations.

\subsection{Action Evaluation with AGAMA}
\label{ss:action}

In order to compute actions from the positions and velocities of star particles at $z=0$ one needs an estimate of the gravitational potential. In the case of the real MW, this is derived using a multitude of observational tracers which provide the masses of the various stellar components (the bulge, thin and thick disk and stellar halo) and kinematics of disk stars and halo objects to derive the mass and density profile of the dark matter halo. 

In this paper we study 3 simulated disk galaxies from the FIRE-2 simulations with different accretion histories. We use the masses and positions of all the particles (dark matter, gas, stars) within 600 kpc from the center of galaxy to compute the gravitational potential of each galaxy at $z=0$. This is done using AGAMA  \citep{10.1093/mnras/sty2672} which builds the gravitational potential via multipole expansion on a hybrid cylindrical-polar grid (to compute the potential of the flattened stellar disk) and spherical polar grid (to compute the potential of the dark matter halo). The potential generated by dark matter and hot gas in the dark matter halo is represented by an expansion in spherical harmonics with $l_{max}=4$; the potential from stars and cold gas in the disk is expanded in azimuthal-harmonics up to $m_{max}=4$. The star, gas and dark matter particles within 600 kpc of the galactic center are used to calculate the potential. \citep[for a detailed introduction to how these expansions work see,][]{10.1093/mnras/sty2672}. 

With the gravitational potential in hand, we use the positions and velocities of all accreted halo star particles (at $z=0$) to compute the three actions $J_r$, $J_\phi$ and $J_z$ defined by,
\be
J_{q} = \frac{1}{2\pi} \oint \frac{p_q}{m} \, dq
\ee
where $m$ is the mass of a star particle, $q = r, \phi$ or $z$, and $p_q$ is the canonical momentum corresponding to $q$. $\{J_r,J_\phi,J_z\}$ is the most useful set of actions in cylindrical or spherical coordinate systems. %Actions are a particularly useful set of integrals of motion because any distribution function for a collisionless system can be written as a function of the actions by Jeans Theorem. 
For context, $J_{\phi}$ is the same as angular momentum $L_z$. $J_r$ and $J_z$ describes the extent of oscillations in spherical radius and $z$ dimension  respectively.

In this work we have assumed that the gravitational potential is known almost perfectly, i.e. all space coordinates for dark matter particles, gas particles and star particles are assumed known and without error when computing the potential. The main deviation from the  potential used for computing the actions and the true gravitational potential  comes from the spherical harmonic expansion. When computing the actions we further assume that the phase space coordinates for all accreted star particles are known perfectly (i.e. no errors are added). What's more, the current version of AGAMA only computes actions in perfectly oblate potentials and so oblate axisymmetry is imposed, despite the fact that all three potentials are slightly triaxial. 

%\mvc{This whole paragraph is a repetition of material that has been said before. Consider deleting or add more detail: Star particles from the same accreted object in a galaxy should share similar orbital properties and thus would share similar actions and would be clustered in the $\{J_r, J_{\phi}, J_z\}$ action space at the time of accretion.  Since actions are adiabatic invariants, if the gravitational field changes slowly enough during its evolution, we could expect the actions of star particles from the same accreted object to be conserved and stay clustered in action space. it is then viable to find accreted objects through cluster analysis in action space at redshift $z=0$.}

\begin{figure*}
\begin{center}
\includegraphics[width=0.3\textwidth]{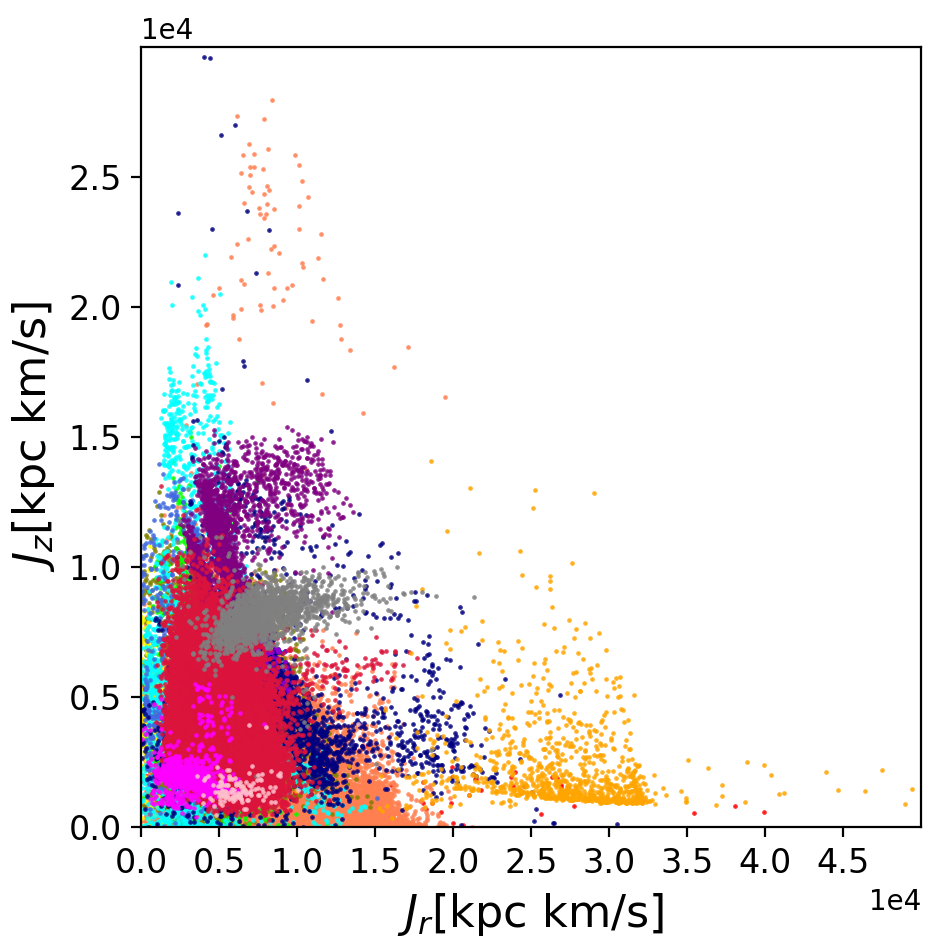}
\includegraphics[width=0.3\textwidth]{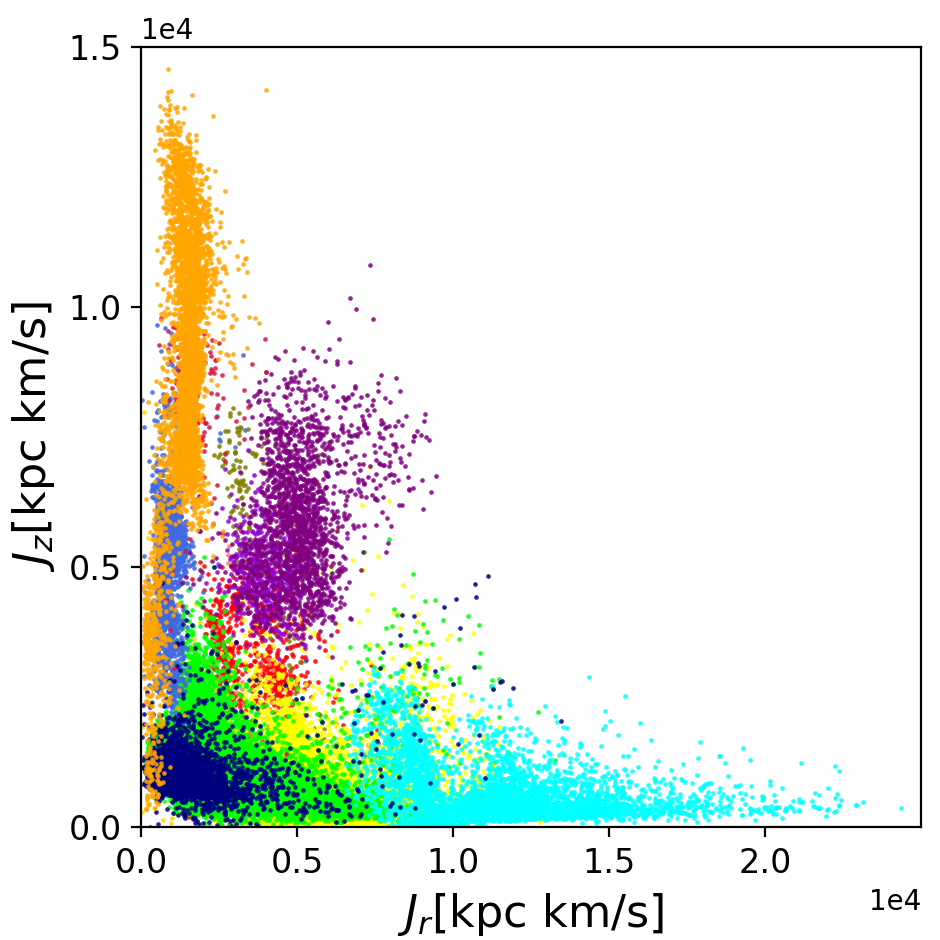}
\includegraphics[width=0.3\textwidth]{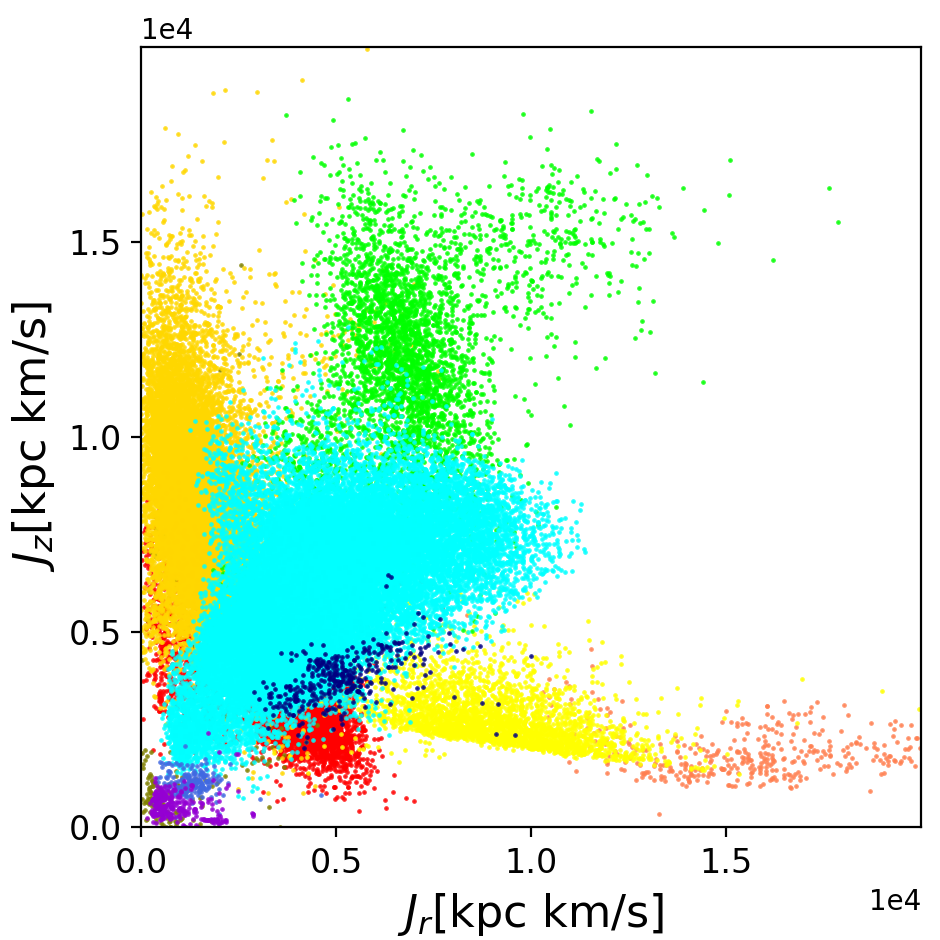}
\includegraphics[width=0.3\textwidth]{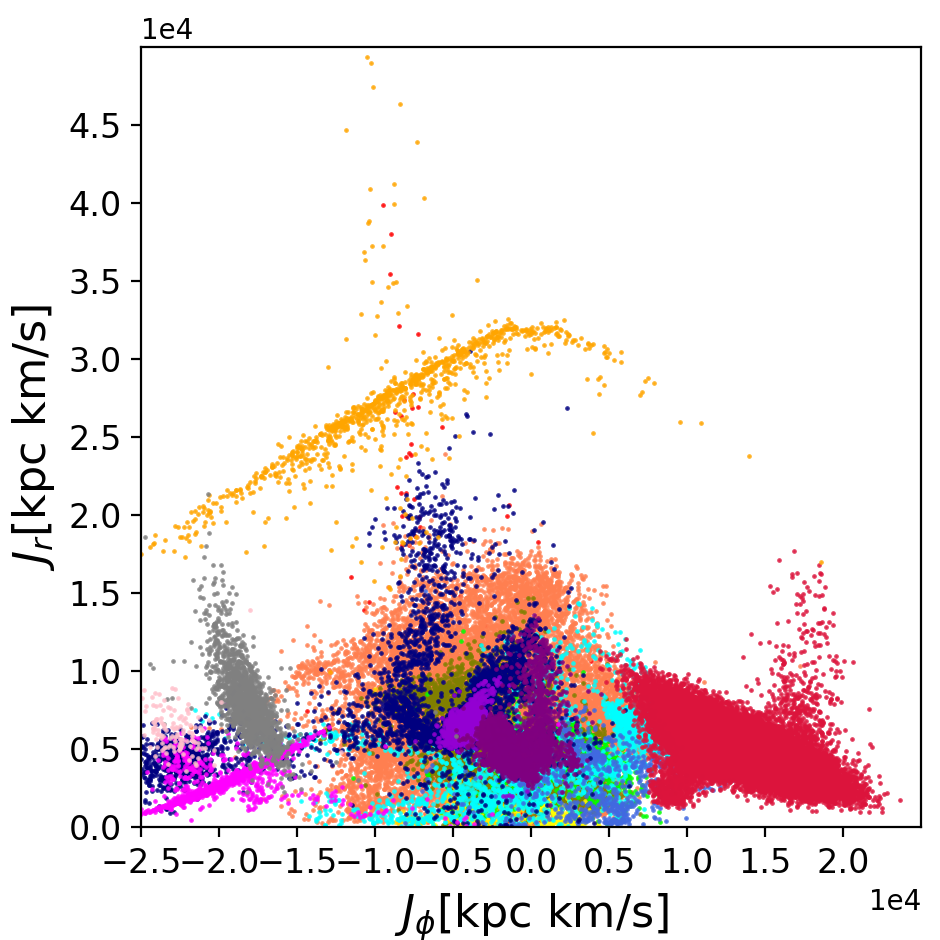}
\includegraphics[width=0.3\textwidth]{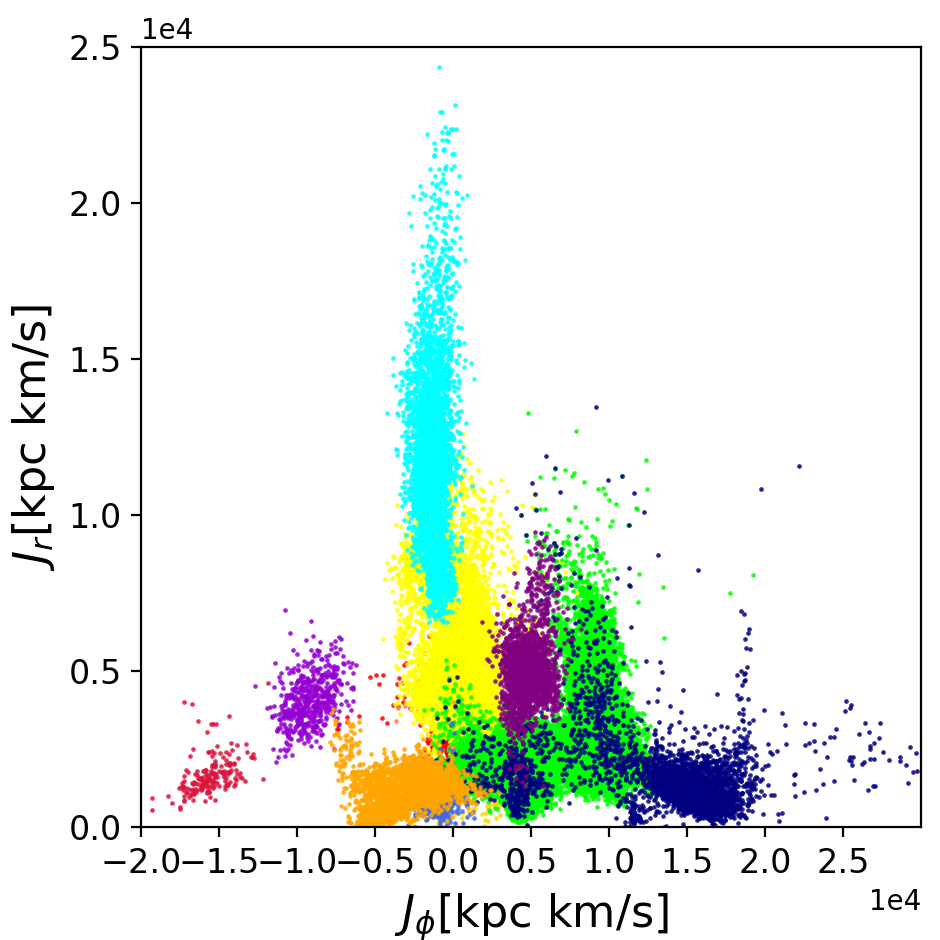}
\includegraphics[width=0.3\textwidth]{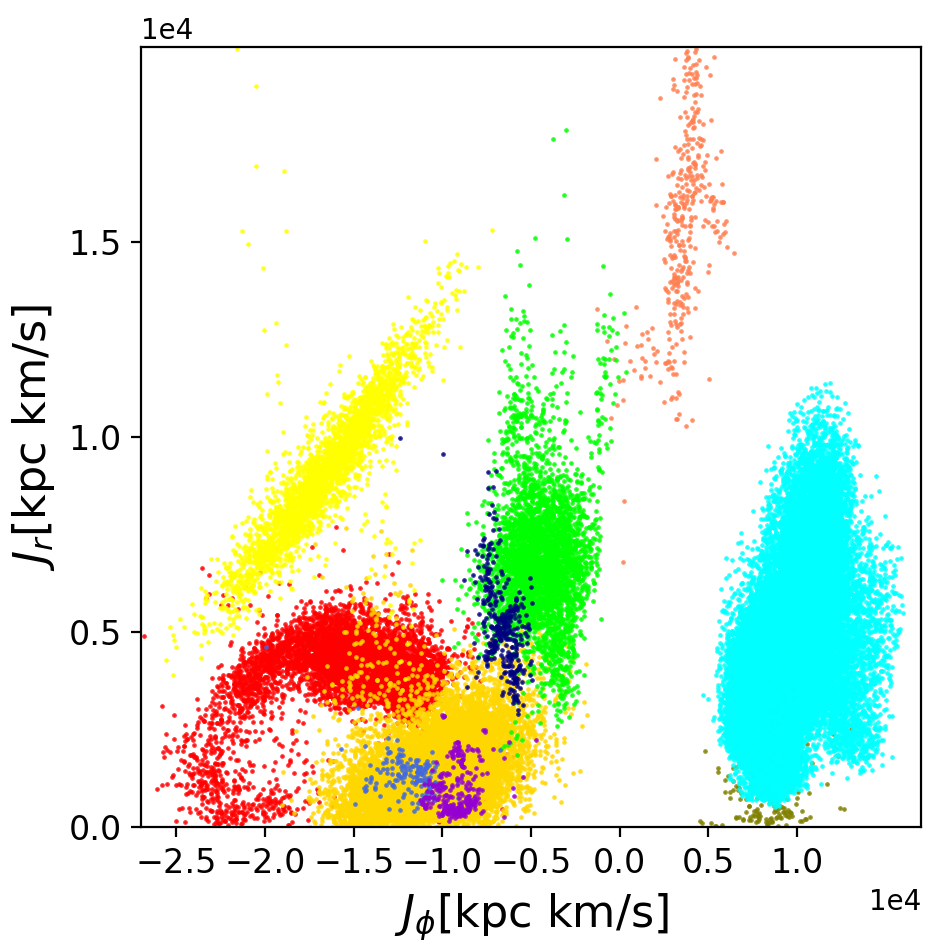}
\includegraphics[width=0.3\textwidth]{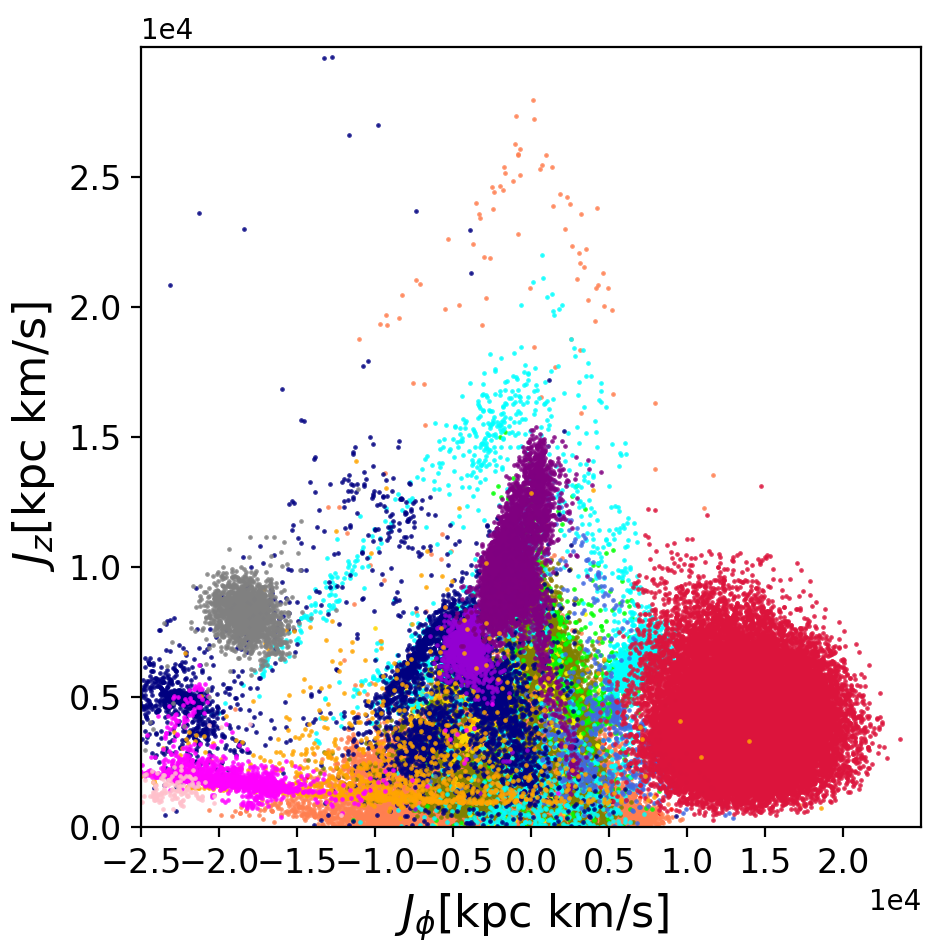}
\includegraphics[width=0.3\textwidth]{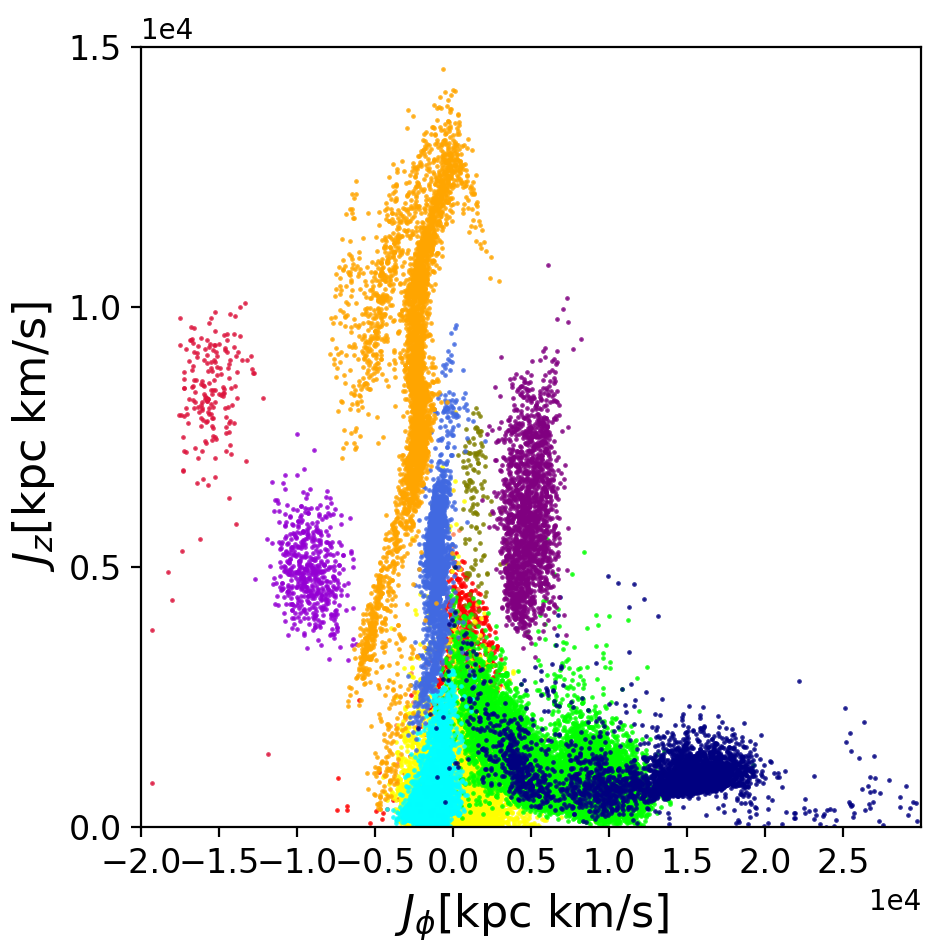}
\includegraphics[width=0.3\textwidth]{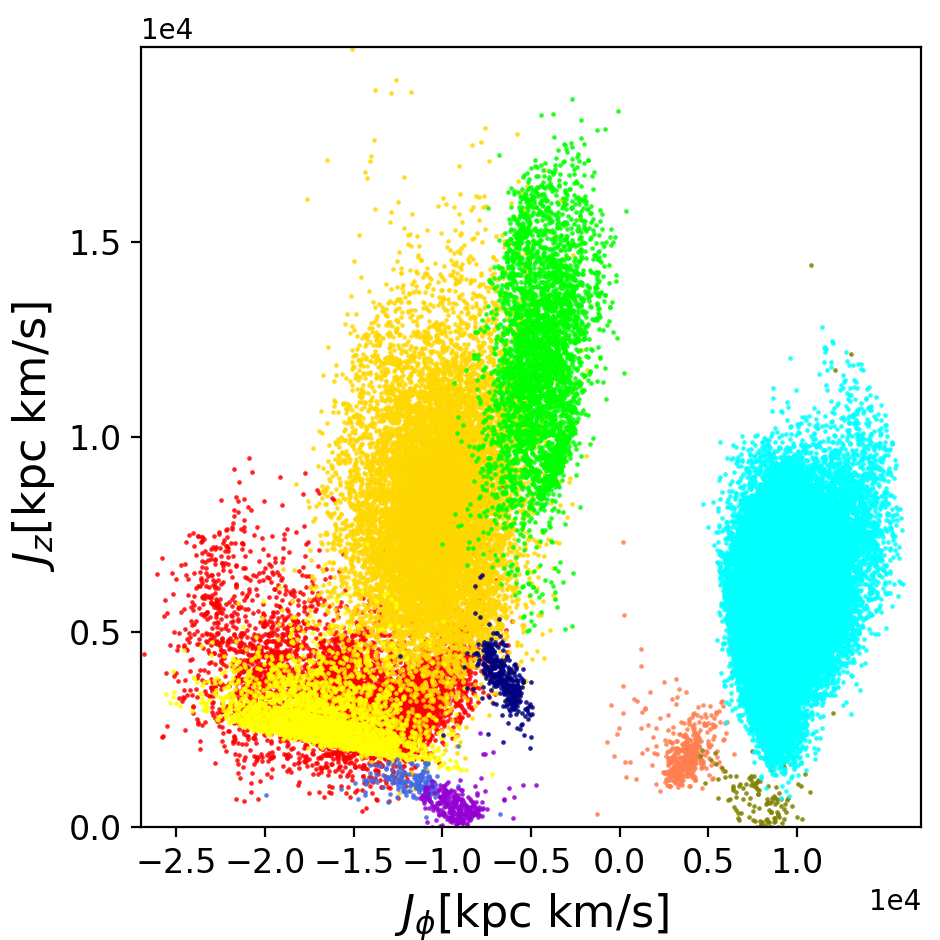}

\includegraphics[width=0.27\textwidth]{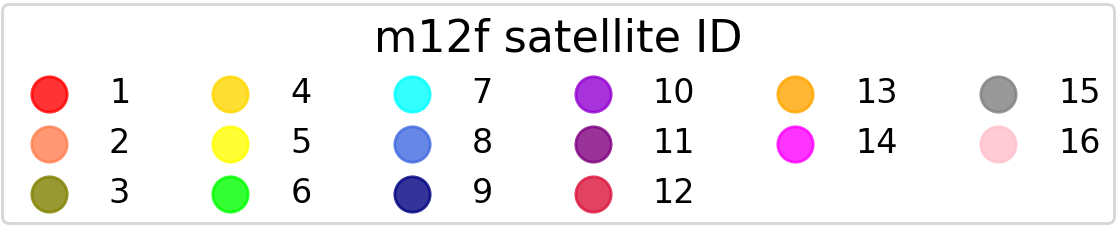}
\hspace{1.0cm}
\includegraphics[width=0.225\textwidth]{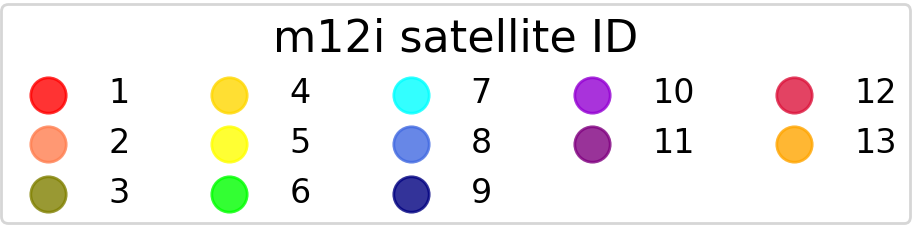}
\hspace{1.6cm}
\includegraphics[width=0.18\textwidth]{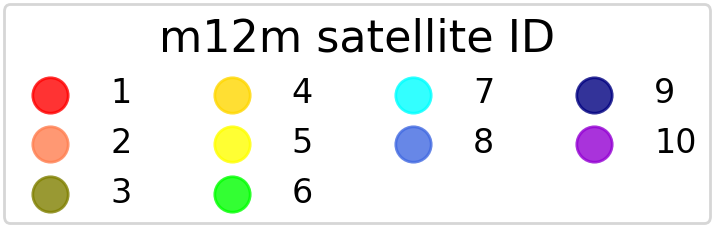}
\end{center}
\caption{Two dimensional projections of actions $\{J_r, J_z, J_{\phi}\}$ for accreted star particles from 3 MW-like galaxies in FIRE-2 simulations. The three columns correspond to the three galaxies in FIRE-2 simulations: m12f, m12i and m12m, from left to right. Each accreted satellite identified from analyzing snapshots from the simulation is indicated by a different color.}.
\label{fig:obj}
\end{figure*}

 \figref{obj} shows three projections of the actions $\{J_r,J_\phi,J_z\}$ for accreted star particles  for each of the three MW-like galaxies m12f, m12i and m12m  in FIRE-2 simulation. The color coding in \figref{obj} corresponds to the individual satellites that are identified by analyzing the snapshots from simulation data as described in Section \ref{sec:data} and is shown by the legend in the bottom row.

%Objects 1-8 are phase-mixed objects, and objects 9-16 are coherent streams. In galaxy m12f, different phase-mixed objects intersect with each other.  Coherent streams, on the other hand, tend to form distinct and easily identified structures, like objects 12, 14 and 15. In m12i, objects 1, 2, 4 and 5 are phase-mixed and all the other objects are coherent streams. In m12m, only objects 3 and 10 are phase-mixed. In m12i and m12m, no significant difference in distributions in phase space can be observed between phase-mixed objects and coherent streams.

%%%%%% FIGURE %%%%%%
%%%%%% FIGURE %%%%%%

\subsection{Finding Clusters with \Enlink}
\label{ss:cluster}

We apply algorithm \Enlink\, \citep{Sharma_2009}, a density-based hierarchical group finding algorithm capable of identifying structures of any shape and density in multi-dimensional datasets, to the action space variables  $\{J_r, J_z, J_{\phi}\}$ for the 3 galaxies from FIRE-2. \Enlink\, is especially useful for astrophysical data as it can effectively detect groups which are not globular. Here we summarize the basic principles behind \Enlink. For more details see \citet{Sharma_2009}.

\begin{enumerate}
\item{\textbf{Build a locally adaptive Mahalanobis (LAM) metric $\Sigma^{-1}(\mathbf{x})$.} The distance squared between two data points $\mathbf{x_i}$ and $\mathbf{x_j}$ is defined as: 
\be
s^2(\mathbf{x_i},\mathbf{x_j})=|\Sigma(\mathbf{x_i},\mathbf{x_j})|^{\frac{1}{d}} (\mathbf{x_i}-\mathbf{x_j})^{\intercal} \Sigma^{-1}(\mathbf{x_i},\mathbf{x_j}) (\mathbf{x_i}-\mathbf{x_j})
\label{eq:dis}
\ee
where $d$ is the dimension of data, $\Sigma^{-1}(\mathbf{x_i},\mathbf{x_j}) = 0.5(\Sigma^{-1}(\mathbf{x_i})+\Sigma^{-1}(\mathbf{x_j}))$. To build this LAM metric $\Sigma^{-1}(\mathbf{x})$, first divide the whole data set into regions with each region containing $(d+1)$ data points, and the particles in each region are distributed as uniformly as possible. Then calculate the local covariance matrix $\Sigma(\mathbf{x})$ of the data points in each region and smooth it so that this local matrix changes continuously and smoothly from region to region.}
\item{\textbf{Calculate a local density $\rho(x)$ for each data point.} For each data point, find $k$ nearest neighbours to it based on distance squared defined in \equref{dis}. The size of neighborhood $k$ has a pre-defined default value in \Enlink\, : $k=10$. We have tried $k=5-20$, and the results are robust against the variation of $k$ in this range. Calculate density as:
\be
\rho(\mathbf{x_j}) = \sum^k_{i=1}\frac{m_i}{h^d |\Sigma(\mathbf{x_j})|^{\frac{1}{2}}} W(\sqrt{(\mathbf{x_i}-\mathbf{x_j})^{\intercal} \Sigma^{-1}(\mathbf{x_j}) (\mathbf{x_i}-\mathbf{x_j})}/h)
\label{eq:dens}
\ee

%\be
%\rho(\mathbf{x_j}) = \sum^k_{i=1}\frac{m_i}{h^d |\Sigma(\mathbf{x_j})|^{\frac{1}{2}}} W(\sqrt{(\mathbf{x_i}-\mathbf{x_j})^{\intercal} \Sigma^{-1}(\mathbf{x_i}) (\mathbf{x_i}-\mathbf{x_j})}/h)
%\ee
%\mvc{the above two equations are almost identical - differing only in one term - it is not clear one needs both} 

where $h$ is the smoothing length corresponding to a given $k$, $d$ is the dimensionality of data space, $W$ is a kernel function to normalize the integral of density and $m_i$ is the mass of each particle.}
\item{ \textbf{Identify clusters based on density and preserve significant groups.} We start building groups at local density peaks, and these groups grow by absorbing nearby points. Once two groups try to absorb a common particle, a saddle point is reached and the smaller group is absorbed into the bigger one as a subgroup (the subgroup is not canceled and still treated as a separate group). After all the (sub)groups are identified, the  {\it significance} $S$ of each group is calculated:
\be
S = \frac{\ln(\rho_{\mathrm{max}})-\ln(\rho_{\mathrm{min}})}{\sigma_{\ln(\rho)}}
\label{eq:sign}
\ee
where $\rho_{\mathrm{max}}$ and $\rho_{\mathrm{min}}$ are maximum and minimum densities in that group. $\sigma_{\ln(\rho)}$ is the standard deviation of log of densities in one group. We define the size of a satellite as the number of star particles it contains. We denote the minimum size of satellites we tracked by $N_S$. The minimum number of particles in a group, \nmin\,, is chosen to be of the same order of magnitude as  $N_S$. Then any group with number of particles lower than \nmin\, would have its {\it significance} set to 0. For two intersecting groups, we compare their values of {\it significance} to a pre-determined  threshold $S_{Th}$. If both groups have $S>S_{Th}$, then keep both of them. Otherwise, the one with lower $S$ is eliminated and its particles are absorbed into the group with higher $S$.}
\end{enumerate}

As can be seen from the description above, four parameters  determine the result of cluster finding: size of neighborhood $k$, threshold {\it significance} $S_{Th}$, mass of each particle $m_{i}$ and minimum size of group \nmin. In this work, we neglect the mass of star particles, so all the $m_i$ is set to 1.  Three other parameters have their default values pre-defined in the \Enlink\, program. During our runs, $k$ and $S_{Th}$ are varied slightly around their default values and \nmin\, is chosen to be in the same order of magnitude as the minimum size of satellites, so that the number of groups that the algorithm yields is close to the number of accreted satellites in a galaxy. 
%\mvc{how is ``reasonable'' determined? This is very important for application to observations (which is the ultimate goal). One needs to  find a statistically  objective way to do this. The  number of stars will be much  larger and we will need to somehow decide if the chosen value of \nmin is reasonable.}

\begin{figure*}
\begin{center}
\includegraphics[width=0.3\textwidth]{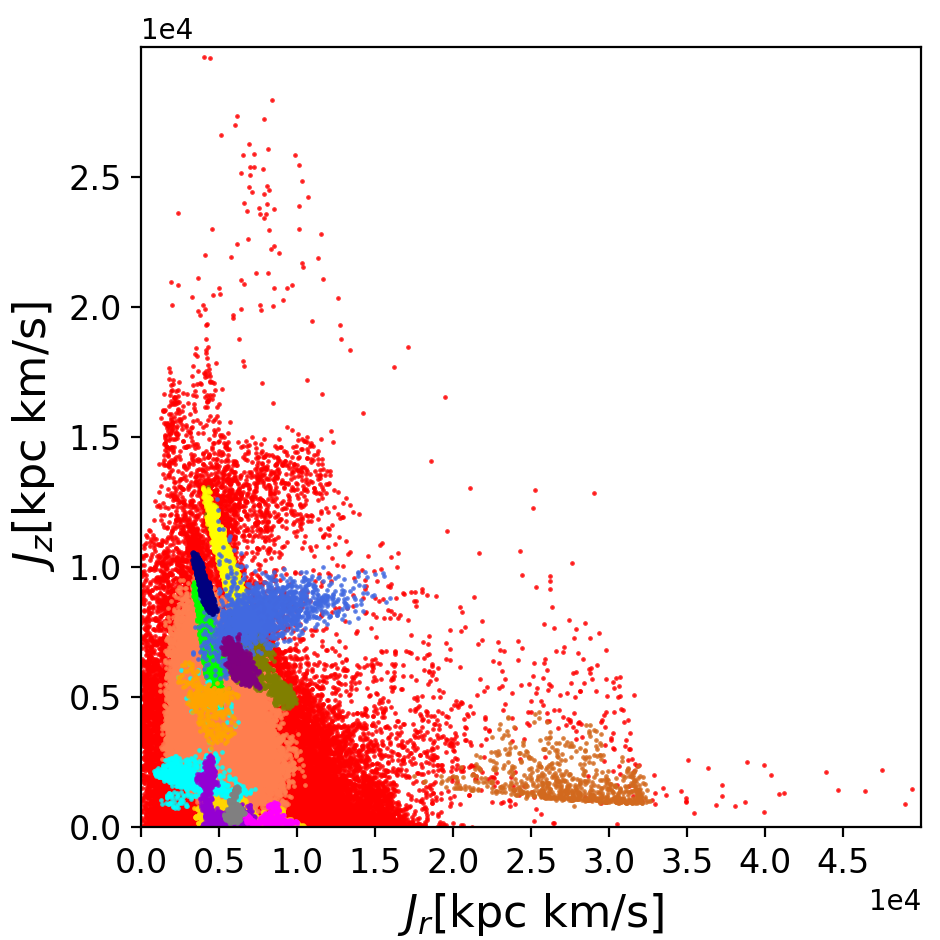}
\includegraphics[width=0.3\textwidth]{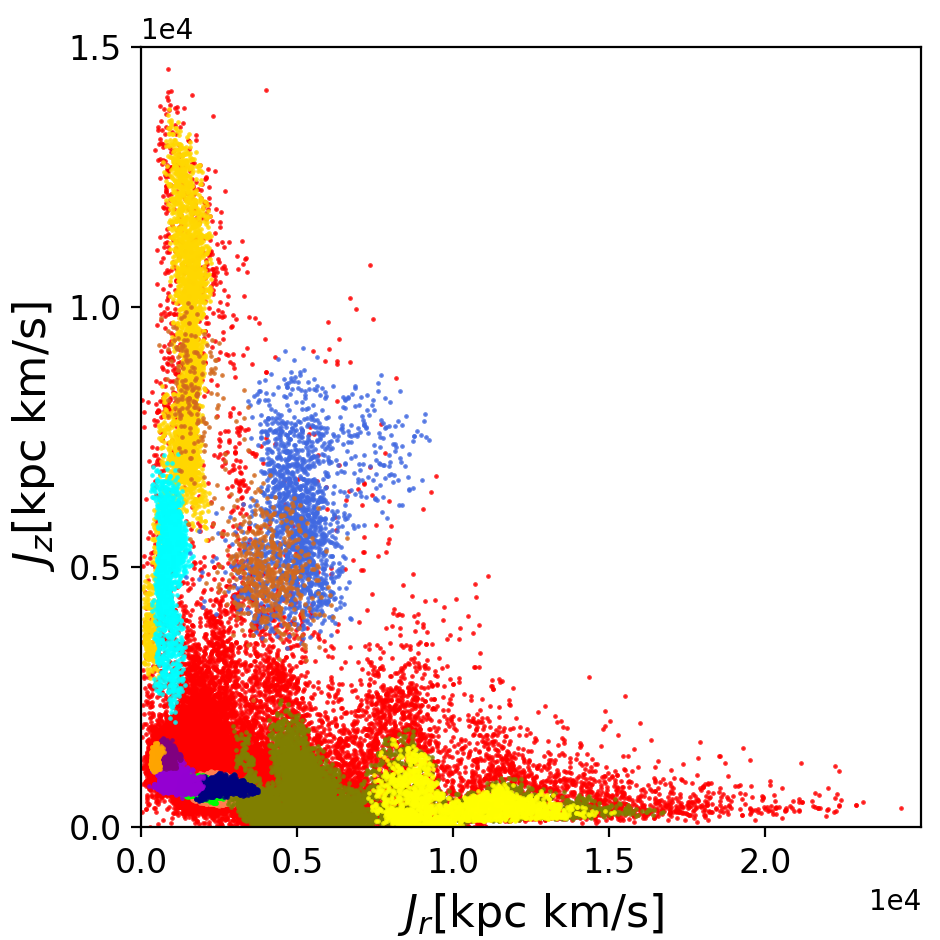}
\includegraphics[width=0.3\textwidth]{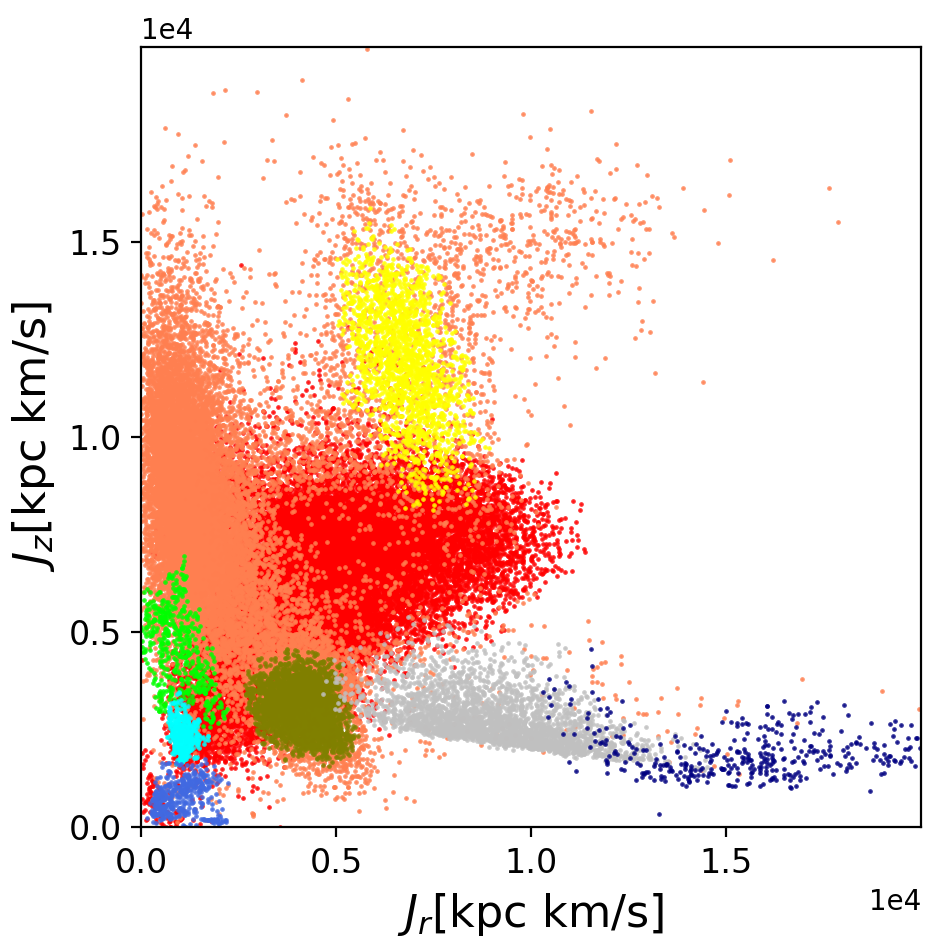}
\includegraphics[width=0.3\textwidth]{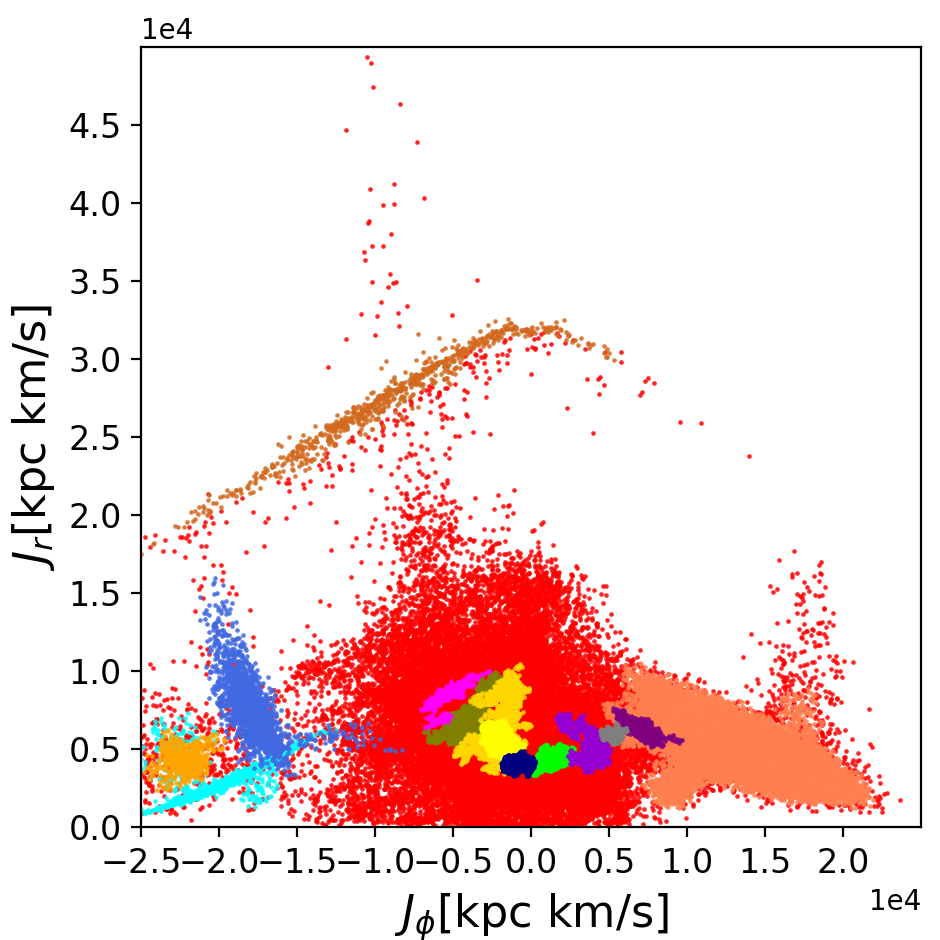}
\includegraphics[width=0.3\textwidth]{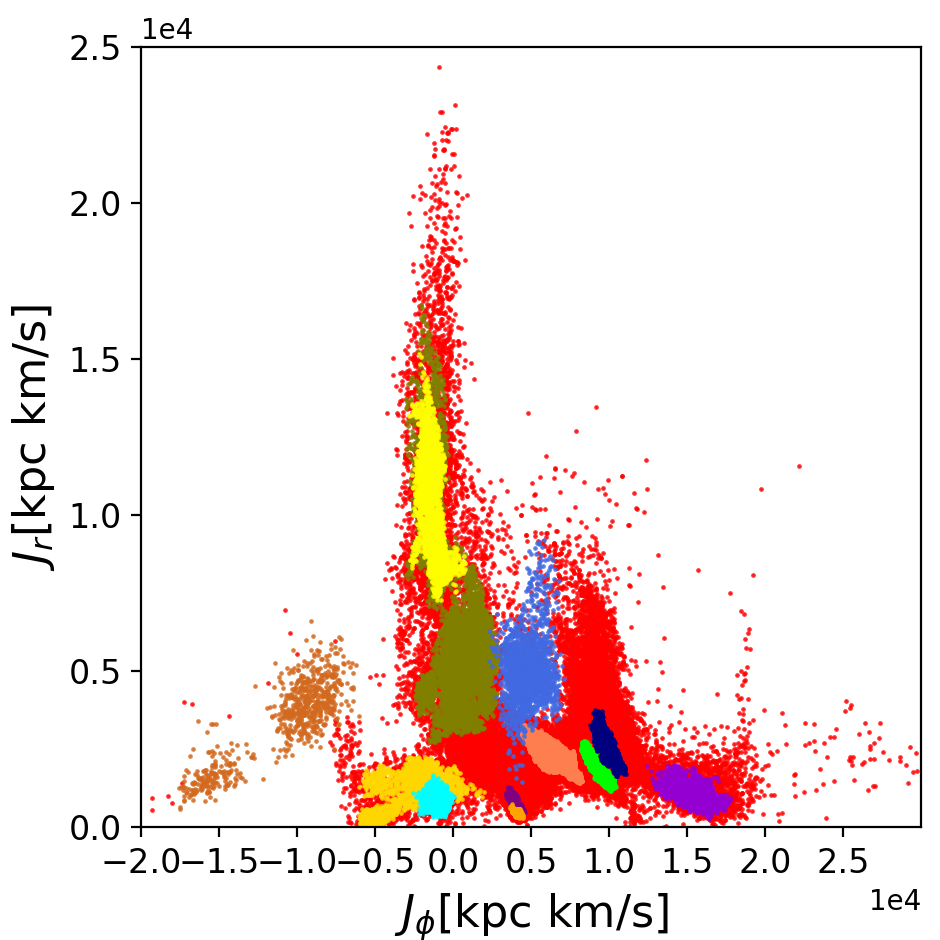}
\includegraphics[width=0.3\textwidth]{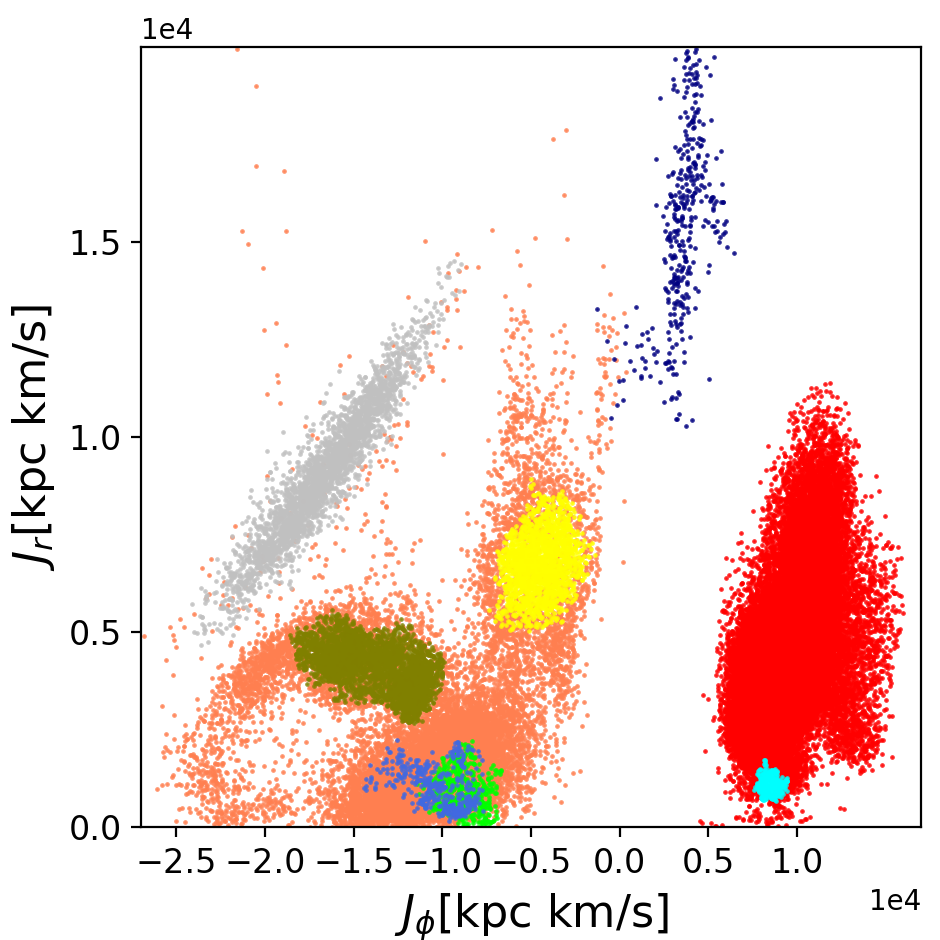}
\includegraphics[width=0.3\textwidth]{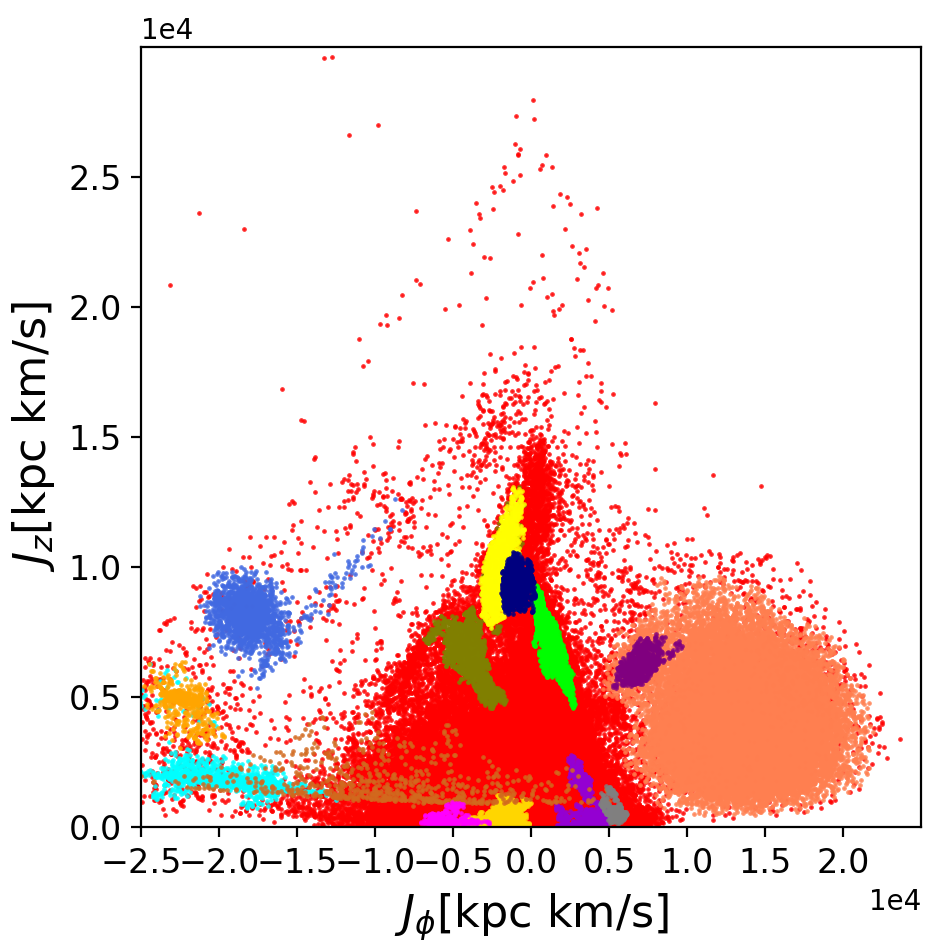}
\includegraphics[width=0.3\textwidth]{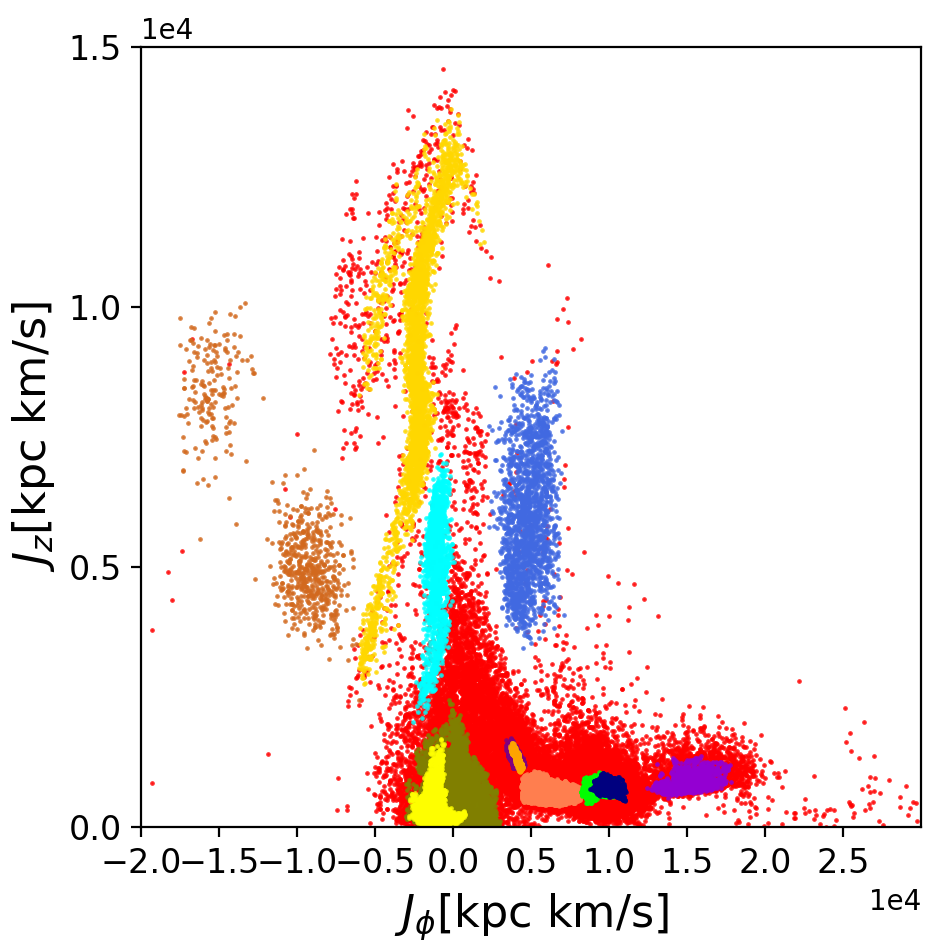}
\includegraphics[width=0.3\textwidth]{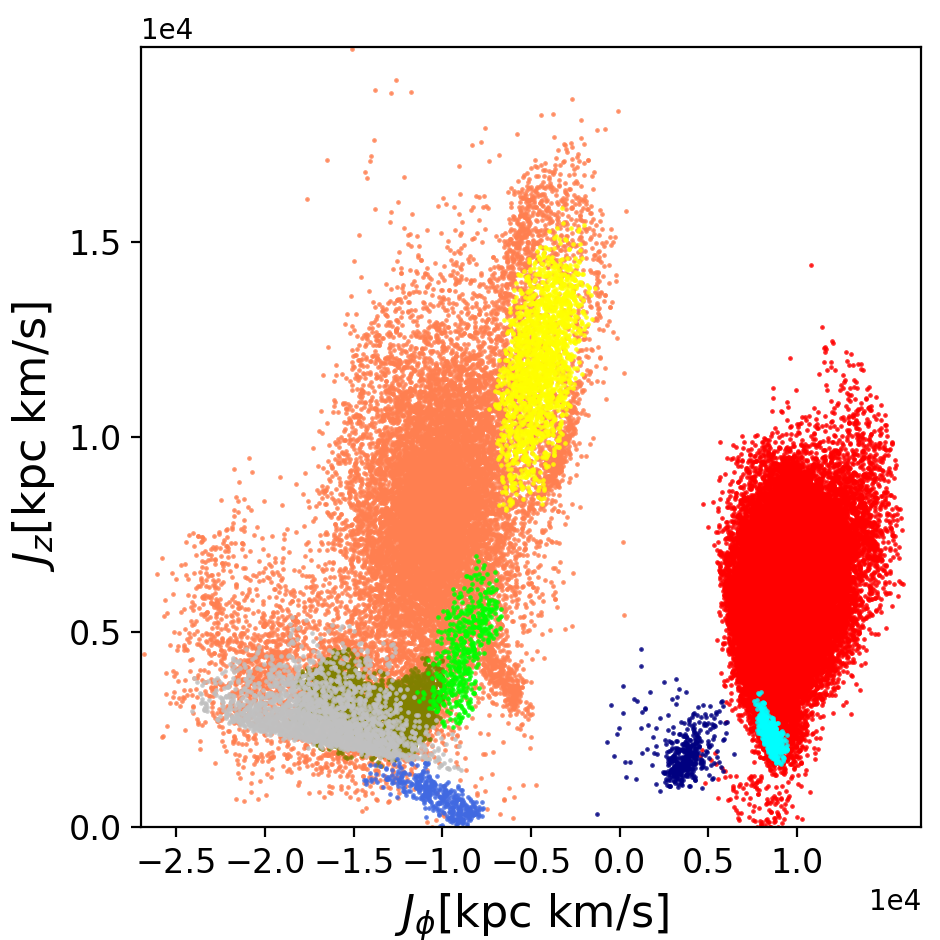}
\includegraphics[width=0.27\textwidth]{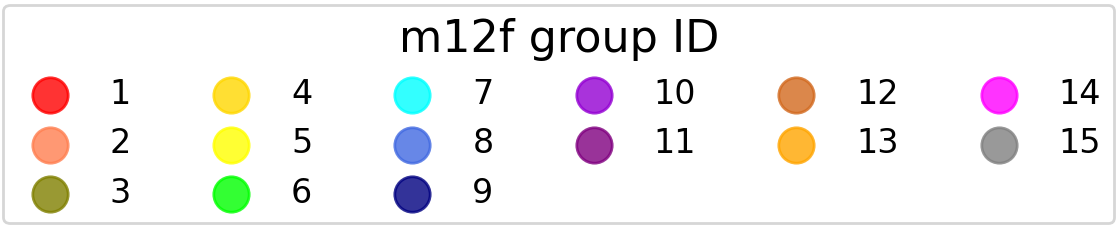}
\hspace{1.0cm}
\includegraphics[width=0.225\textwidth]{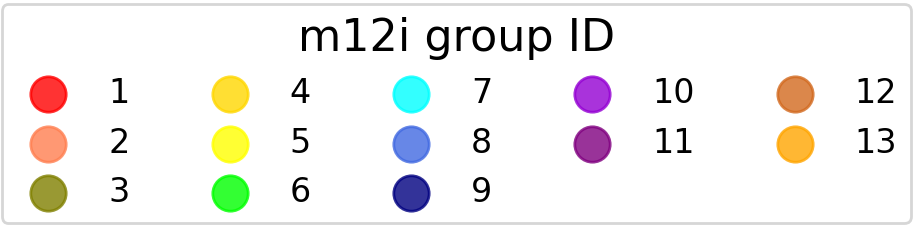}
\hspace{1.6cm}
 \includegraphics[trim=0. 0. 0. 0., clip,width=0.18\textwidth]{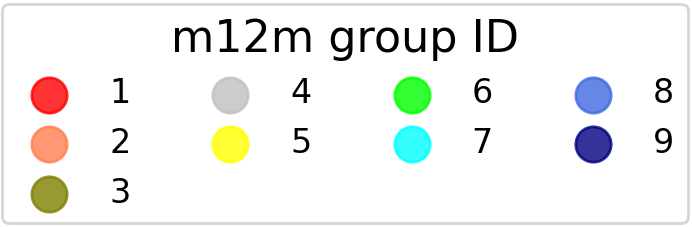}  
 \end{center}          
\caption{Same as \figref{obj}, but with substructures identified by \Enlink\, to see how well \Enlink\, can reproduce the actual satellites in the simulations as shown in \figref{obj}. Particles from different groups identified by \Enlink\, are indicated by different colors. Results plotted here are obtained from \Enlink\, runs with size of neighbourhood $k=10$, minimum group size $N_{\mathrm{min}}=300$ and {\it significance} threshold $S_{Th}$ at its default value.  }.
\label{fig:grp}
\end{figure*}

%\MV{Youjia - please check that the next paragraph is correct - I have rewritten it since it was confusing} 
Groups that were identified by \Enlink\, with $k=10$, \nmin$=300$ and $S_{Th}$ at its default value for the three galaxies from the FIRE-2 simulations are shown in \figref{grp}. The minimum number of star particles in each  of the satellites in the three galaxies is around 150 (corresponding to a minimum stellar mass of around $6.0\times10^5 \msun$), so a reasonable value for \nmin\, would seem to be 150.  With \nmin$=150$, many  small  random groups emerge in all three galaxies. With \nmin$=300$  the identification of large and clear satellites is not affected in m12f and m12m, but  in m12i, a satellite with size 184 (located in action space at around $J_{\phi}=-12000\mathrm{kpc\times km/s}, J_z = 8000 \mathrm{kpc\times km/s}, J_r=2000 \mathrm{kpc\times km/s}$) cannot be identified. However, since many small  spurious groups that emerge at \nmin$=150$ make it harder to  identify the large and clear satellites, we keep \nmin$=300$, corresponding to a minimum stellar mass of around $1.2\times10^6 \msun$ for groups found by \Enlink\,.
%\mvc{I think that there is another way to look at this. For instance with StarGO, Yuan et al found a lot (57) groups in Gaia+LAMOST data. They don't claim that all 57 correspond to individual satellites. They expect that several (sometimes up to 12) may belong to a single satellite. Maybe what we need to do is consider a lower \nmin\, that will yield many possible sub components and use an additional test to assess if multiple groups actually belong to a single larger satellite. One could do this iteratively with \Enlink\, - e.g. run it for  $N_{min}=150$ and find all the groups. Then run it for $N_{min}=300$ and find out how many of the groups for the smaller Nmin merged to form a new group. If one of the previous groups was merged with group 1 (background), then extract it and consider it a real group, but if two groups merge together with the larger Nmin then perhaps one can examine the significance parameter or max and min density that \Enlink\, gives to decide the significance of the larger group. I would be interested in knowing how many groups that appear with smaller Nmin are really ``spurious''. } 

In applications to observational data, since the sizes of datasets are expected to be much larger and the number of accreted satellites is completely unknown, these 4 parameters should be determined carefully. It is also likely that one satellite  could be split into multiple groups by cluster analysis \citep{yuan_etal_20}. Adding additional information, such as metallicity  or abundances of specific  elements \citep[e.g. Ca, N, Fe, Eu,][]{sanderson_etal_17,2018AJ....156..179R} can help to confirm the identification of a satellite or identify subcomponents of a single satellite, but this is beyond the scope of this paper. In this paper, we are testing how well cluster analysis in the action space alone can perform under optimum conditions: perfect knowledge of all 6 phase space coordinates, accurate representation of the galactic potential, and prior insights into the optimum values of parameters like \nmin. These simplifying assumptions enable us to assess how well cluster analysis can perform in the best-case scenario. %\mvc{This is a very true and important point - and I think we should investigate this further in this paper to provide a way to make an objective assessment, when the "truth'' is not known.}\YW{One might use the so called "silhouette plot" to estimate the quality of the cluster analysis. One may also mention the cross-validation with metallicity. Or, even more ambitiously, using the silhouette plot in the metallicity space to cross validate the choice of parameters in action space.}

\subsection{Metrics to Assess Identification of Satellites by Clusters}
\label{ss:cluster_metrics}

In order to quantify the ability of \Enlink\, to identify accreted satellites in action space, we define three matrices: $R_{ij}$, $P_{ij}$ and $M_{ij}$; and four quantities based on these matrices: {\it recovery, purity, merit} and {\it contrast}.  We refer to the clusters identified by \Enlink\, as ``groups''. The number of star particles contained in a group is called the size of the group. In order to quantify the degree to which a group identified by \Enlink\, matches one of the original accreted satellites we also define the  ``{\it best recovery group}'' (similarly for {\it purity, merit}) and best fit group. Based on {\it recovery, purity, merit} and best fit group, we select out the ``well-recovered satellites".

\begin{enumerate}
\item{\textbf{Three matrices $R_{ij}, P_{ij}, M_{ij}$:} The $ij$ component of these three %\mvc{general rule of thumb in journals, small numbers  up to 10 are to be spelled out except in e.g. 3-D}
matrices describes the similarity between satellite $j$ and group $i$:
\begin{eqnarray*}
R_{ij} &=&\frac{\mathrm{number~of~particles~shared~by~satellite}~\mathit{j}~\mathrm{and~group}~\mathit{i}}{\mathrm{number~of~particles~in~satellite}~\mathit{j}}, \\
P_{ij} &=&\frac{\mathrm{number~of~particles~shared~by~satellite}~\mathit{j}~\mathrm{and~group}~\mathit{i}}{\mathrm{number~of~particles~in~group}~\mathit{i}}, \\
 M_{ij} &=& R_{ij}\times P_{ij}
 \end{eqnarray*}
 }
 \item{\textbf{\textit{Recovery, purity, merit:}} For satellite $k$, the maximum values of $R_{ik}, P_{ik}$ and $M_{ik}$ are called the {\it  recovery} ($r_{k}$), {\it purity} ($p_{k}$) and {\it merit} ($m_{k}$) of this satellite:
 \begin{eqnarray*}
 r_{k} &=& \mathrm{max}_i \{R_{ik}\}, {\rm for\,i\, in\, range\, of\, group\, indices}\\
 p_{k} &=& \mathrm{max}_i\{P_{ik}\}, {\rm for\,i\, in\, range\, of\, group\, indices}\\
 m_{k} &=& \mathrm{max}_i\{M_{ik}\}, {\rm for\,i\, in\, range\, of\, group\, indices}
 \end{eqnarray*}
 }
 \item{{\bf Best {\it recovery, purity, merit} group and best fit group:} 
For satellite $k$, the group which yields $r_k$ is called the ``best {\it recovery} group'',  and likewise $p_k$ and $m_k$  define the ``best {\it purity} group'' and ``best {\it merit} group'' of satellite $k$, respectively. If a group is both the ``best {\it recovery} group'' and the ``best {\it purity} group" of satellite $k$, then it is defined as the ``best fit group'' of satellite $k$. }
\item{\textbf{\textit{Contrast}}: To compare the size of satellite $k$ and its best fit group, we define {\it contrast} ($c_k$) as:
\be
c_{k} = \frac{r_k-p_k}{\sqrt{r_k p_k}}
\ee
where $r_k$ and $p_k$ are {\it recovery} and {\it purity} of satellite $k$. Note {\it contrast} is only defined for a satellite with a best fit group. When $c_k$ is positive, satellite $k$ is smaller than its best fit group (the group contains contaminants - i.e. stars that were not originally part of the satellite), and when $c_k$ is negative, satellite $k$ is larger than its best fit group (not all of the members of the satellite have been identified as group members).}
\item{\textbf{Well-recovered satellite: }If the {\it recovery, purity} and {\it merit} of a satellite are all greater than 0.5 and this satellite has a best fit group, then this satellite is called {\it well-recovered} or is said to have a {\it high identifiability score}. A satellite that is not well-recovered is called poorly-recovered.}
\end{enumerate}
%\mvc{question: how do you go about deciding which group corresponds to which satellite? Do you just compute recovery, {\it purity}, etc for every group and every satellite and then use the "best fit" one as the identifier. I think this is what you do - so we should say this quite clearly.}

%%%%%% FIGURE %%%%%%

%%%%%% FIGURE %%%%%%

\section{Results}
\label{sec:res}
We do a cluster analysis with \Enlink\, on actions $\{J_r, J_z, J_{\phi}\}$ of accreted star particles from three MW-like galaxies in FIRE-2 simulations. In Section \ref{subsec:acc}, we calculate the values of {\it recovery, purity, merit} and {\it contrast} for each satellite in three galaxies to evaluate how well these satellites are recovered by \Enlink. We study the distribution of well-recovered satellites on the \mtot-\tinfall\,  plane and $M_{\rm stellar}$-\tinfall\, plane, and find the boundaries in \mtot\, $M_{\rm stellar}$ and \tinfall\ values that separate well-recovered and poorly-recovered satellites by the classification tree method. We also investigate the relation between {\it significance} of a group and the identification power of this group (that is, whether this group  corresponds to a well-recovered satellite or not). In Section \ref{subsec:insitu}, we include certain percentages of \insitu stars into the input data set for cluster analysis. We pick out the well-recovered satellites which are identified when \insitu\ star particles are absent. Calculate the values of {\it merit} of these satellites under different \insitu star contamination ratios and demonstrate the robustness of the identification of these satellites under the contamination of \insitu star particles.
\subsection{Cluster Analysis on Accreted Star Particles}
\label{subsec:acc}

We apply \Enlink\,  to the  actions $\{J_r, J_z, J_{\phi}\}$ for accreted star particles from  the MW-like galaxies m12f, m12i and m12m in FIRE-2 simulations. In a manner similar to \figref{obj} three projections of the actions for accreted star particles $\{J_r,J_\phi,J_z\}$ are shown in Figure~\ref{fig:grp}, but now the color coding corresponds to the individual {\it groups} that are identified by \Enlink. The legend in the bottom row now indicates the color coding for the individual groups identified. A comparison with  \figref{obj} shows visually that while in m12i and m12m the number of groups identified is one larger than the number of satellites., in m12f only 15 groups (of 16 satellites) are identified. Overall however, there is an excellent correspondence between the groups identified by \Enlink\, and the original satellites seen in \figref{obj}.

\begin{figure*}
\includegraphics[width=0.45\textwidth]{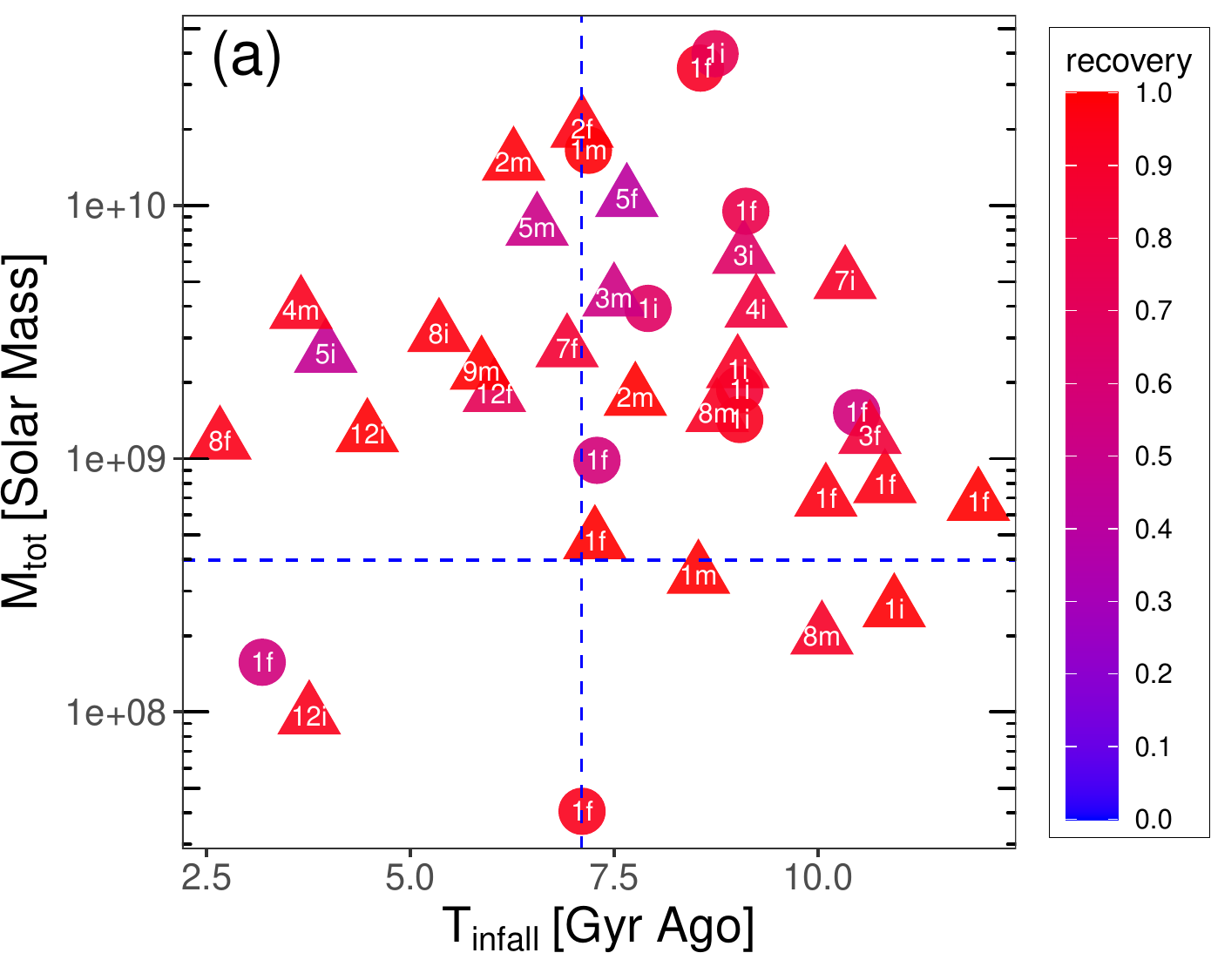}
\includegraphics[width=0.45\textwidth]{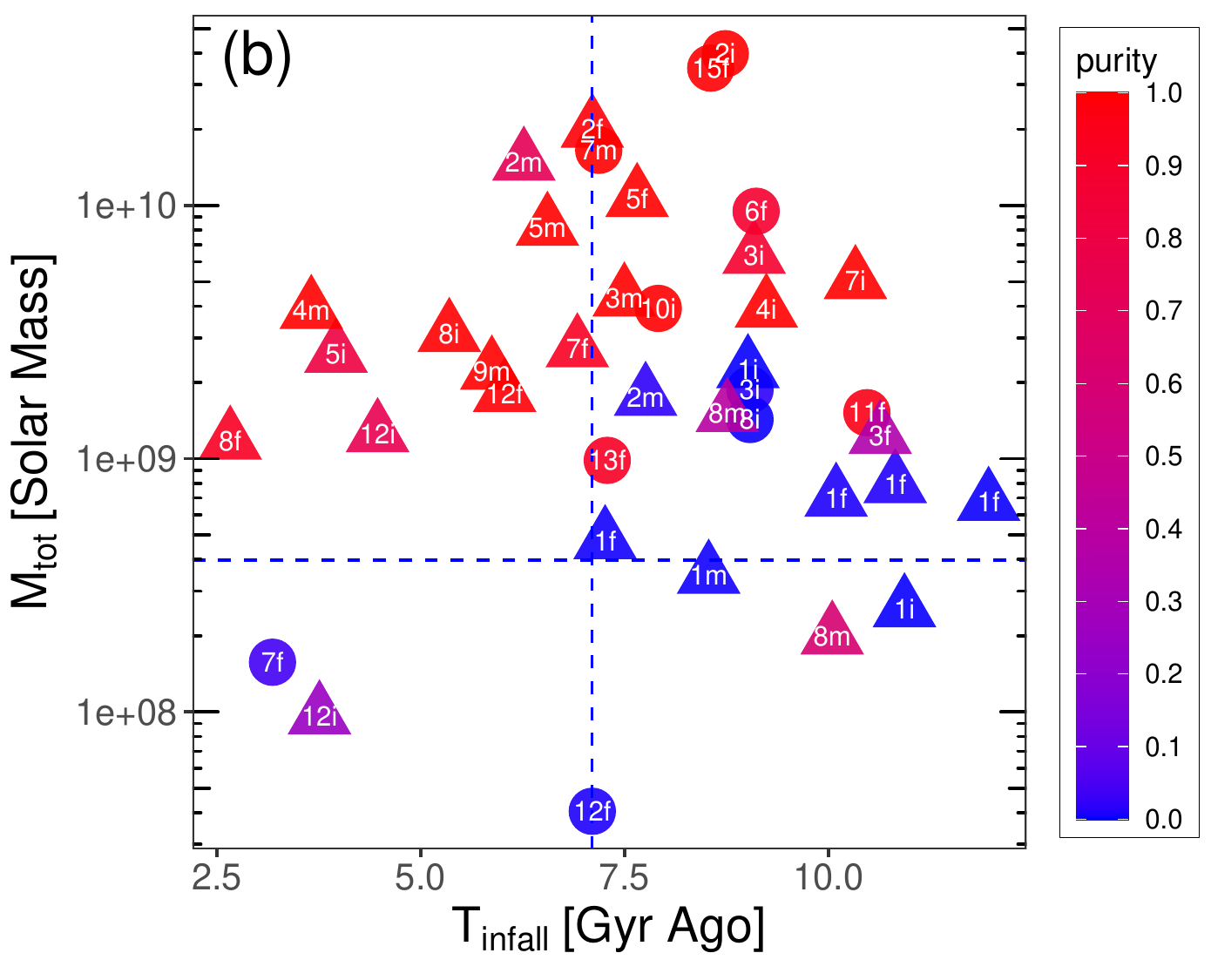}
\includegraphics[width=0.45\textwidth]{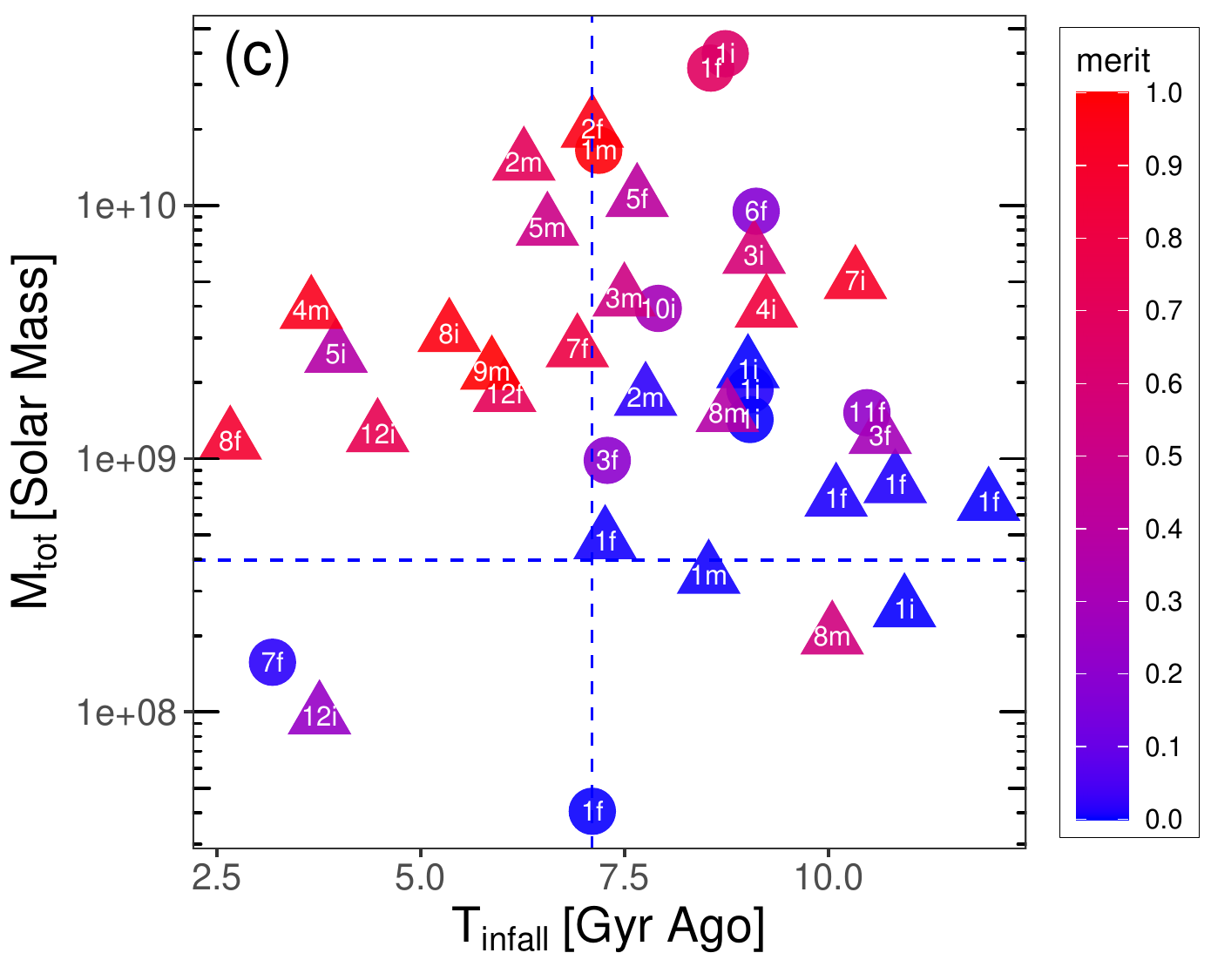}
\includegraphics[width=0.45\textwidth]{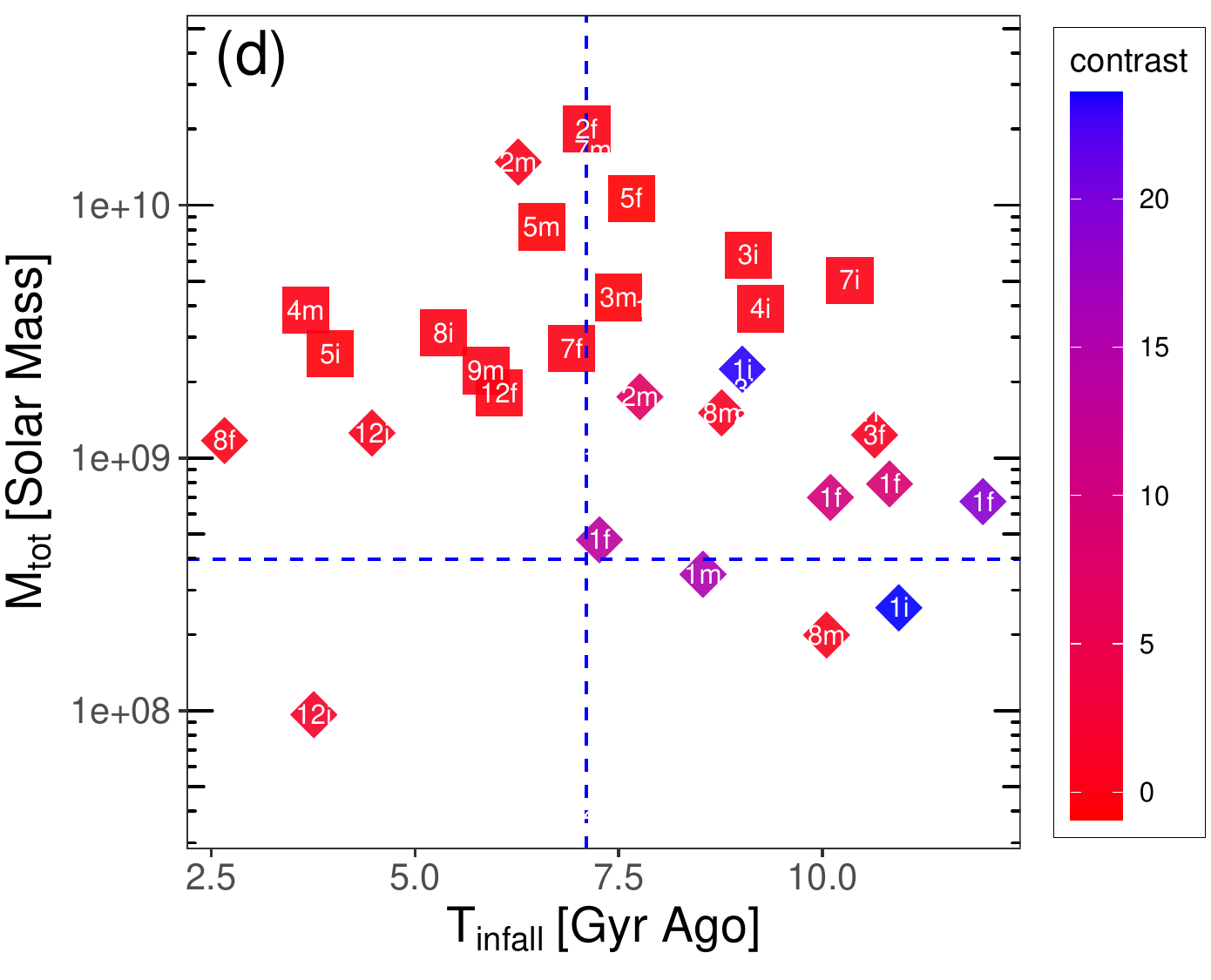}       
\caption{39 satellites in 3 MW-like galaxies from the FIRE-2 simulations plotted on the \mtot\, vs. \tinfall\,  plane. Data points are color coded by the {\it recovery} (panel a), {\it purity} (panel b), {\it merit} (panel c) or {\it contrast} (panel d), with red points indicating better results in all panels.  The white number on each symbol indicates the group ID (as shown in \figref{grp}) of the ``best {\it recovery}'' (panel a), ``best {\it purity}'' (panel b), ``best {\it merit}'' (panel c) and ``best fit'' (panel d) group corresponding to each satellite. The white letters associated with each data point indicates which galaxy each satellite is from: ``f'' for ``m12f'', ``i'' for ``m12i'' and ``m'' for ``m12m''. Group 1 in all three galaxies is the ``best'' group of many satellites because it is the largest group, as shown in \figref{grp}, and is generally considered as the ``background''. A satellite is marked as a triangle if it has a ``best fit group'' and as a circle otherwise. Satellites are from the analysis of simulation data in Section \ref{sec:data}, and groups are identified by \Enlink . In panel d, only satellites with best fit groups are plotted, and shapes indicate whether the {\it contrast} is positive (diamond) or negative (square). Among the 39 satellites, 14 of them are well-recovered and 25 of them are poorly recovered. The vertical and horizontal dashed lines mark \mtot$=4.0\times 10^8 \msun$ and \tinfall = 7.1 Gyr ago, which are the boundaries identified by the classification tree method. 91\% (10/11) of the satellites with \mtot\, greater than $4.0\times 10^8 \msun$ and fell into the halo less than 7.1 Gyr ago are well-recovered by \Enlink.} %, with the satellite marked by ``5i'' as an exception}.
\label{fig:acc4}
\end{figure*}

\figref{acc4} shows the values of {\it recovery}, {\it purity}, {\it merit} and {\it contrast} of 39 satellites in the three simulated galaxies m12f, m12i and m12m, on a  \mtot\, vs. \tinfall\ plot. Each symbol corresponds to one accreted satellite and symbols are color coded by the value of their {\it recovery} (panel a), {\it purity} (panel b), {\it merit} (panel c) and {\it contrast} (panel d), with red colored points indicating better results. In panels a, b, and c, satellites are also shape coded by whether their best {\it recovery} group match their best {\it purity} group (triangle) or not (circle) (in other words, whether the satellite has a best fit group (triangle) or not (circle)). White numbers indicate the group ID (as shown in \figref{grp}) signifying best {\it recovery} (panel a), best {\it purity} (panel b), best {\it merit} (panel c) or best fit (panel d) for each satellite. A white letter associated with each data point indicates which galaxy this satellite is from: `f' for `m12f', `i' for `m12i' and `m' for `m12m'. Among the 39 satellites, 14 of them are ``well-recovered" (their values of {\it recovery}, {\it purity} and {\it merit} are larger than 0.5 and they have best fit groups), while 25 of them are poorly-recovered.  With \tinfall\, and $\log_{10}(M_{\rm tot}/\msun)$ as inputs, we grow a classification tree to predict whether a satellite is well-recovered or not. (For a brief introduction to classification tree method, see Appendix \ref{sec:dtm}) The boundaries that the classification tree finds out are: \tinfall = 7.1 Gyr ago and \mtot = $10^{8.6} \msun$, which are shown as vertical and horizontal blue dashed lines in \figref{acc4}. 91\% (10/11) of the satellites in the region bounded by \mtot $>4.0\times 10^8 \msun$ and \tinfall $< 7.1$ Gyr ago are  ``well-recovered" satellites. Of the satellites outside this region, 86\% of the sample (24/28) are poorly-recovered. Three satellites in galaxy m12i (marked by 3i, 4i and 7i on the plot) fell into the host galaxy far more than 7.1 Gyr ago (\tinfall = 9.08-10.33 Gyr ago), but are still well-recovered. This can be attributed to the unique dynamic history of galaxy m12i. See Section \ref{sec:con} for more discussion. Group 1 in each galaxy is the ``best" group of many satellites, because Group 1 is the largest group and contains star particles from many ``poorly-recovered" satellites. Group 1 in each of the galaxies can be considered as the ``background" group. Panel d plots the {\it contrast} of satellites, shape coded by whether the {\it contrast} is positive (diamond) or negative (square). Only the satellites with a best fit group are plotted in panel d. Note that a positive {\it contrast} means that a satellite is smaller than its best fit group, and a negative {\it contrast} means the opposite. Most of the values of contrast are close to 0, with a few extremely large values (above 20). These extremely large values belong to satellites with Group 1 as best fit group, indicating that Group 1 is much larger than these satellites, consistent with the fact that Group 1 contains star particles from many satellites and is the background group. The values of contrast of ``well-recovered" satellites are close to 0, meaning the number of particles in these satellites is similar to that in their best groups.

\begin{figure*}
\includegraphics[width=0.45\textwidth]{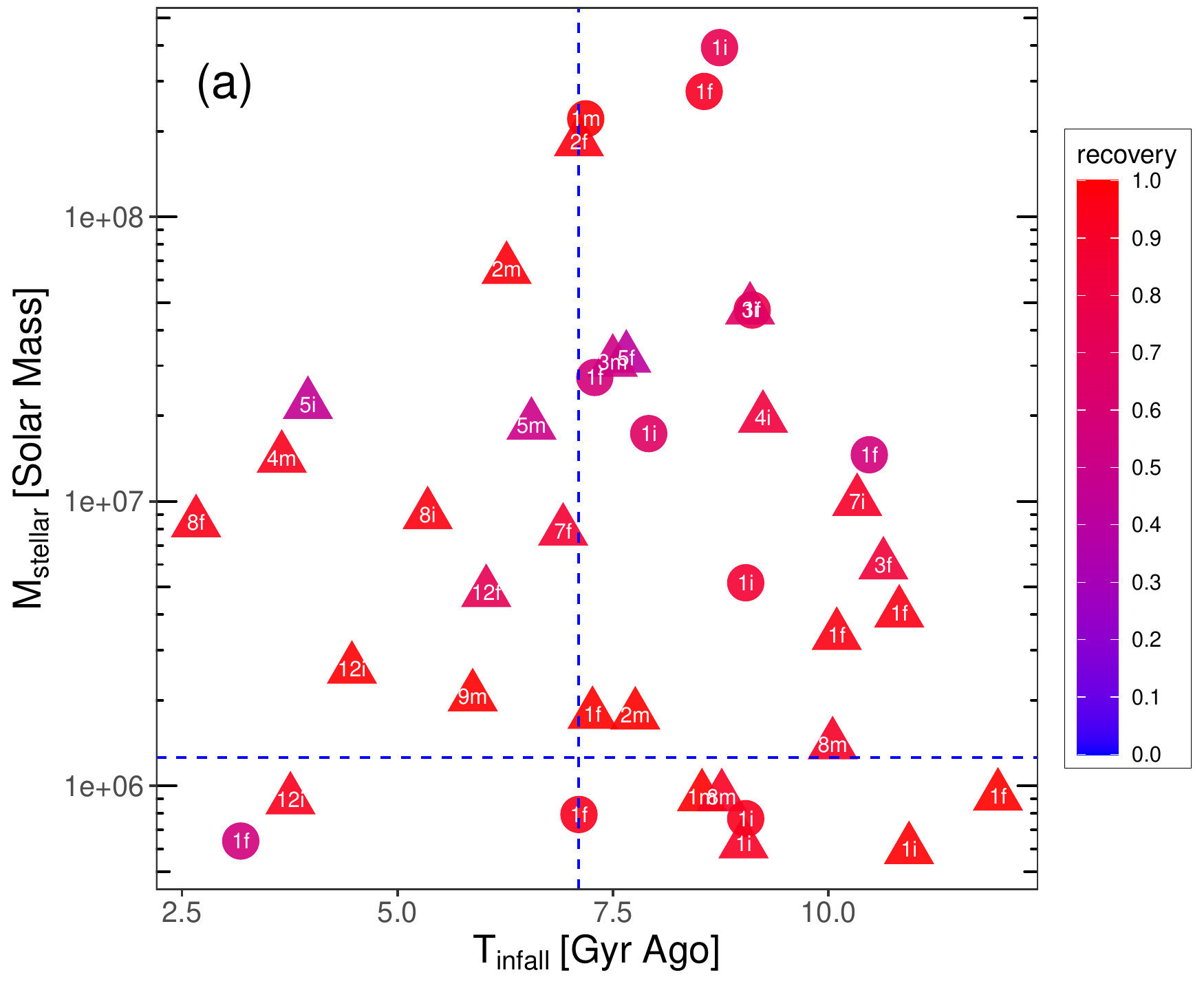}
\includegraphics[width=0.45\textwidth]{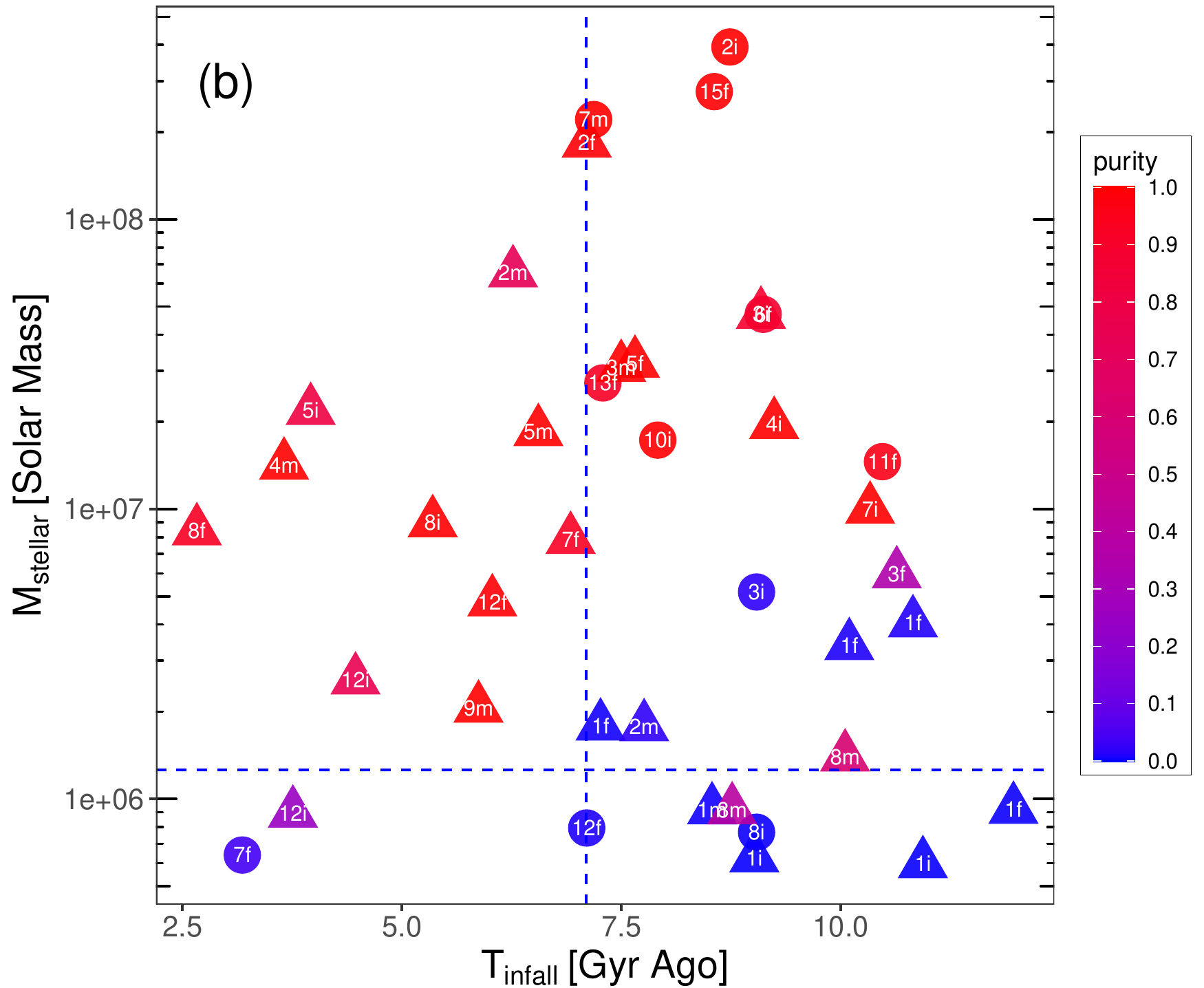}
\includegraphics[width=0.45\textwidth]{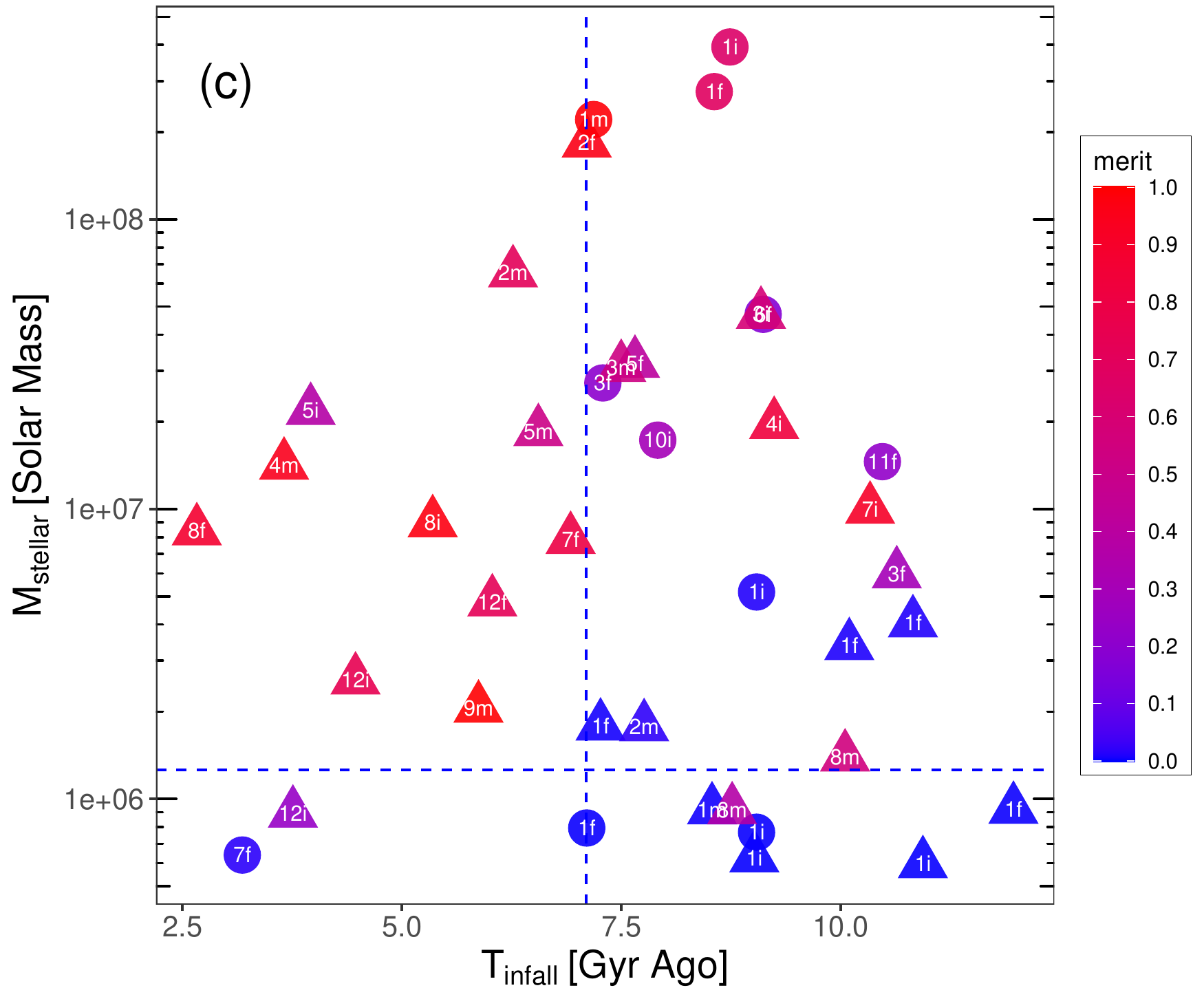}
\includegraphics[width=0.45\textwidth]{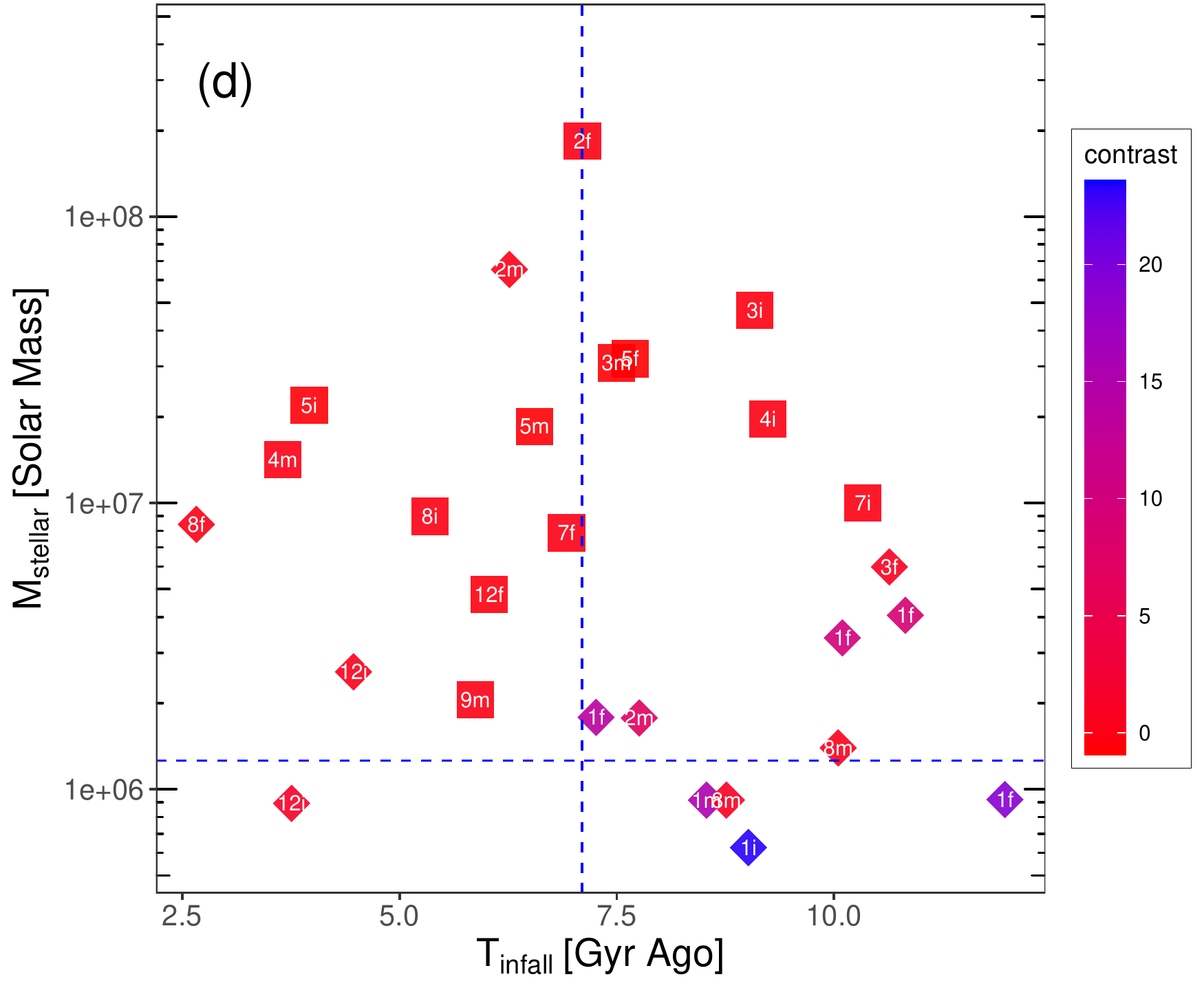}       
\caption{Same as \figref{acc4}, but the y axes of these plots are labeling $M_{\rm stellar}$. The vertical and horizontal dashed lines mark $M_{\rm stellar}=1.2\times 10^6 \msun$ and \tinfall = 7.1 Gyr ago, which are the boundaries separating well-recovered and poorly-recovered satellites identified by the classification tree method. 91\% (10/11) of the satellites with $M_{\rm stellar}$ greater than $1.2\times 10^6 \msun$ and fell into the halo less than 7.1 Gyr ago are well-recovered by \Enlink.} %, with the satellite marked by ``5i'' as an exception}.
\label{fig:acc5}
\end{figure*}

Similar to \figref{acc4}, in \figref{acc5} we plot the values of {\it recovery}, {\it purity}, {\it merit} and {\it contrast} of 39 satellites in the three simulated galaxies m12f, m12i and m12m, on the $M_{\rm stellar}$  vs. \tinfall\ plane. The color code, shape code and the meaning of the white numbers/letters are the same as those in \figref{acc4}. The boudaries separating well-recovered and poorly-recovered satellites found by the classification tree method in the $M_{\rm stellar}$  vs. \tinfall\ plane are: $M_{\rm stellar}=1.2\times 10^6 \msun$ and \tinfall = 7.1 Gyr ago, marked by the horizontal and vertical blue dashed lines in the four panels. The \tinfall\, boundaries found in \mtot\, vs. \tinfall\ and $M_{\rm stellar}$  vs. \tinfall\ planes agree with each other. 91\% (10/11) of the satellites with $M_{\rm stellar}$ greater than $1.2\times 10^6 \msun$ and fell into the halo less than 7.1 Gyr ago are well-recovered by \Enlink.

In Section \ref{ss:cluster}, we find that 150 star particles are corresponding to $M_{\rm stellar}\approx6\times10^5\msun$, so setting $N_{\rm min}=300$ is equivalent to setting the lower bound of $M_{\rm stellar}$ to be around $1.2\times10^6\msun$ for groups found by \Enlink\ . Any group with less stellar mass than $1.2\times10^6\msun$ (number of particles smaller than $N_{\rm min}=300$) will not be found by \Enlink\, under our choice of $N_{\rm min}$. This lower bound in $M_{\rm stellar}$ agrees with the $M_{\rm stellar}=1.2\times 10^6 \msun$ boundary for well-recovered satellites. This agreement raises a caveat that the boundary on $M_{\rm stellar}$ (probably also on \mtot) might be an artifact due to a particular choice of $N_{\rm min}$. 
%%%%%% FIGURE %%%%%%
%%%%%% FIGURE %%%%%%

%%%%%% FIGURE %%%%%%
%%%%%% FIGURE %%%%%%

%%%%%% FIGURE %%%%%%
\begin{figure*}
\includegraphics[width=0.3\textwidth]{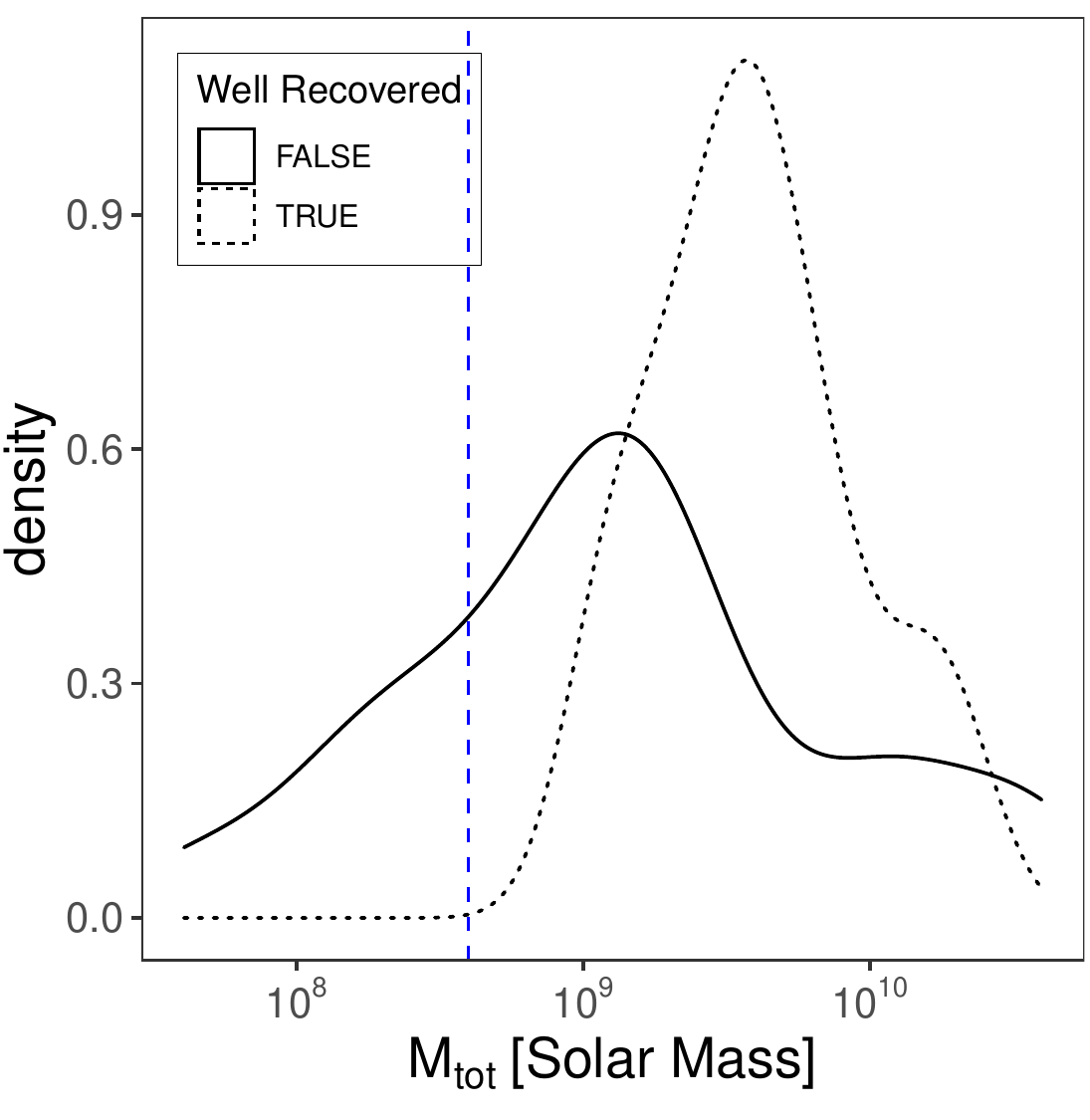}
\includegraphics[width=0.3\textwidth]{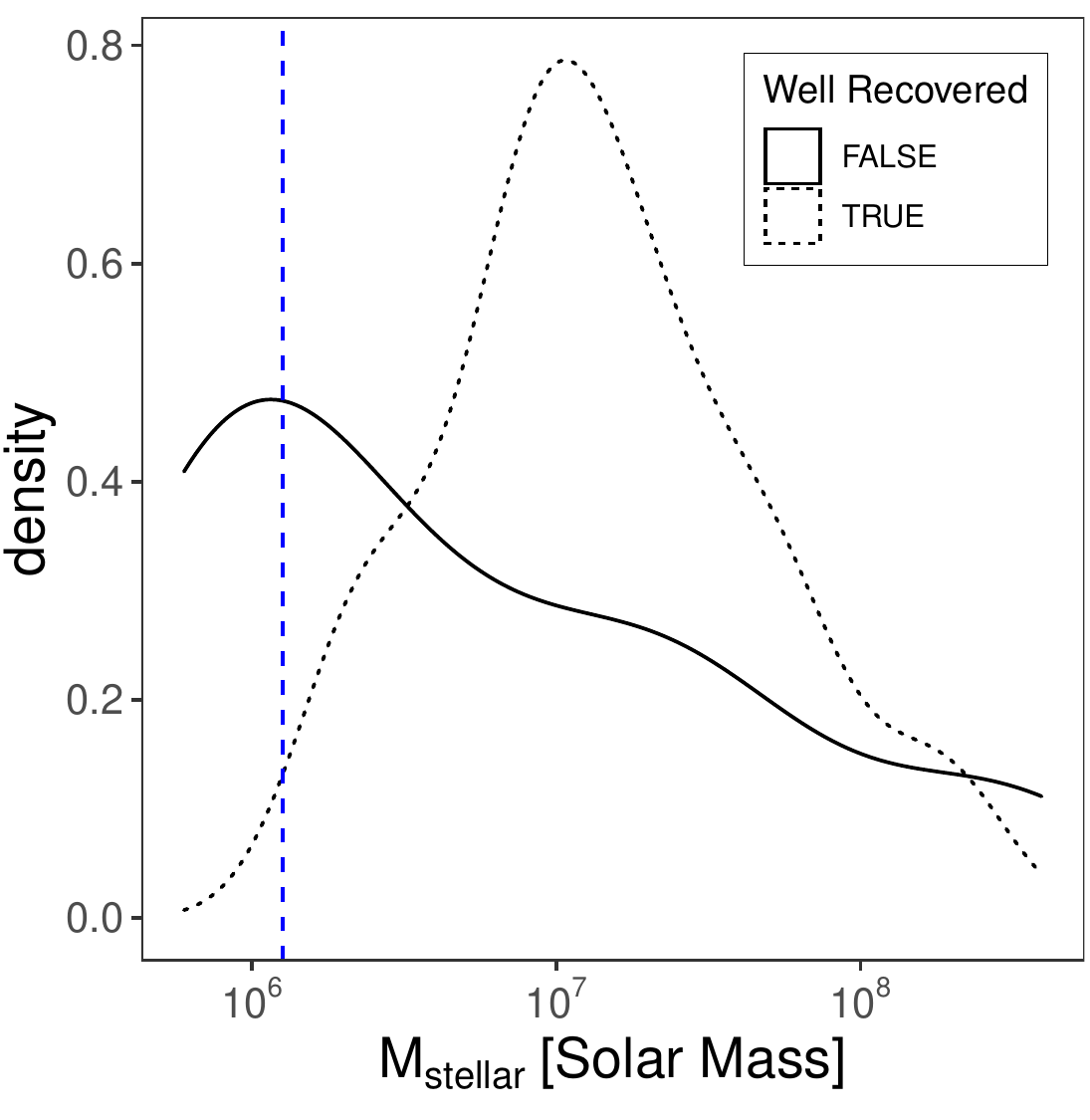}
\includegraphics[width=0.31\textwidth]{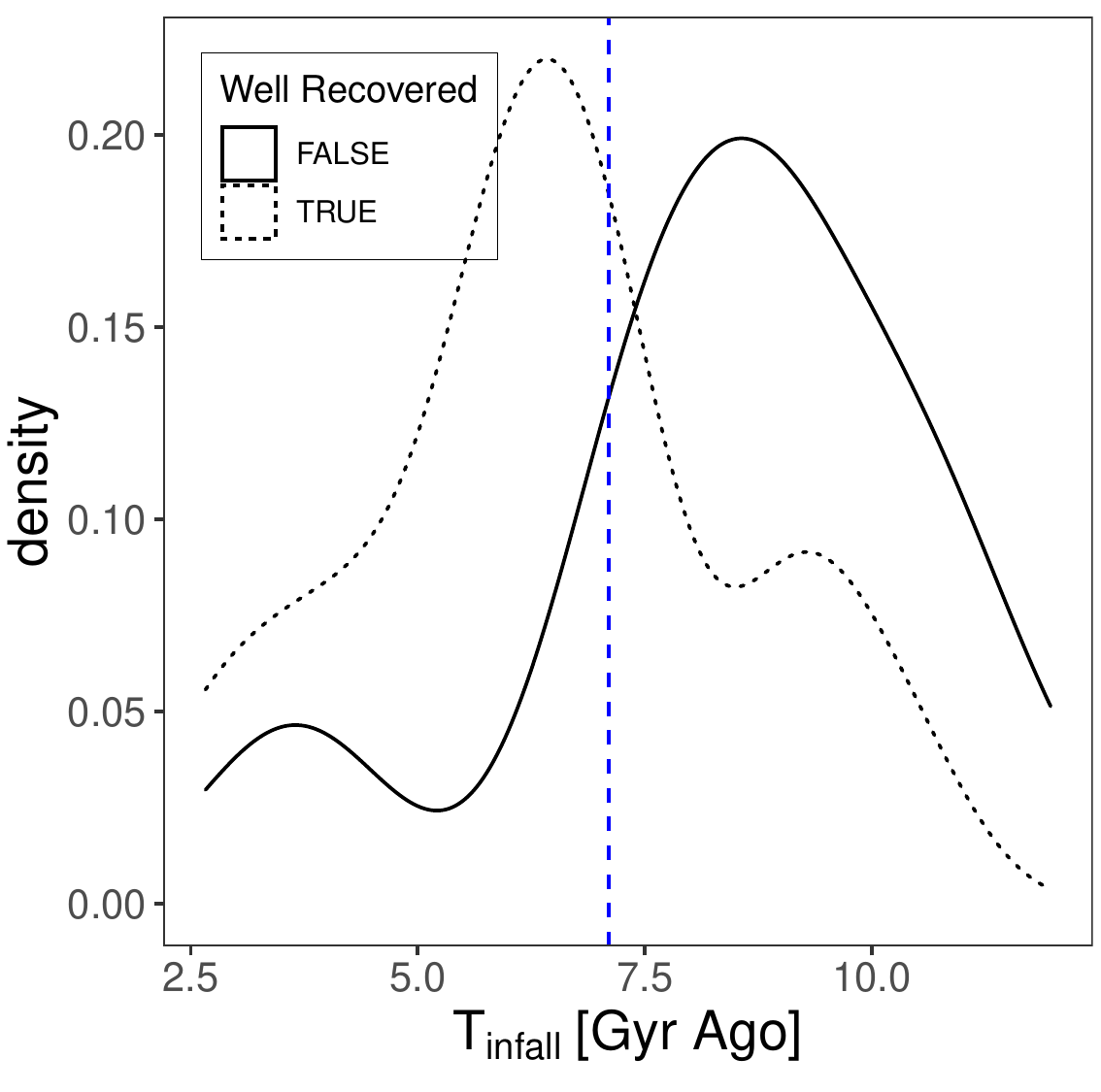}
\caption{Kernel density plots of 39 satellites from three galaxies in \mtot (left), $M_{\rm stellar}$ (middle) and \tinfall\, (right). The three blue vertical dashed lines label the boundaries in \figref{acc4} and \figref{acc5}: \mtot$=4.0\times 10^8 \msun$ (left), $M_{\rm stellar}=1.2\times 10^6 \msun$ (middle) and \tinfall$=7.1$ Gyr ago (right). In all three panels, the dotted density curve labels the well-recovered satellites, while the solid curve labels the poorly-recovered satellites. In the left panel, both the dotted and solid density curves peak above \mtot$=4.0\times 10^8 \msun$, with the dotted curve peaking at a higher \mtot, indicating that well-recovered satellites tend to be more massive. In the middle panel, the dotted curve peaks at a higher $M_{\rm stellar}$, indicating that well-recovered satellites tend to be have more stellar mass. In the right panel, the peaks of dotted and solid curves are on the two sides of \tinfall$=7.1$ Gyr ago}.
\label{fig:dens}
\end{figure*}
%%%%%% FIGURE %%%%%%
To appreciate how well the boundaries \mtot = $4.0\times 10^8 \msun$, $M_{\rm stellar}=1.2\times 10^6 \msun$ and \tinfall = 7.1 Gyr ago work in separating out well-recovered satellites, we show three kernel density plots of 39 satellites from three galaxies in \mtot (left), $M_{\rm stellar}$ (middle)  and \tinfall (right) in \figref{dens}. In all three panels, the dotted density curve represents well-recovered satellites, while the solid curve represents poorly-recovered satellites. The two curves are normalized separately. The vertical dashed lines in three panels show the boundaries in \figref{acc4} and \figref{acc5}: \mtot = $4.0\times 10^8 \msun$, $M_{\rm stellar}=1.2\times 10^6 \msun$ and \tinfall = 7.1 Gyr ago. In the left panel, the dotted curve peaks at a higher \mtot\, than the solid curve, indicating that well-recovered satellites tend to be more massive. However, the peaks of both curves are greater than $4.0\times 10^8 \msun$, indicating that the boundary in \mtot\, alone cannot distinguish between well-recovered and poorly-recovered satellites. In the middle panel, the solid curve peaks at a lower value than the dotted curve, indicating that well-recovered satellites tend to have more stellar mass than poorly-recovered ones. The peak of the solid curve is close to the $M_{\rm stellar}=1.2\times 10^6 \msun$ boundary, indicating that $M_{\rm stellar}$ alone cannot distinguish well-recovered satellites and poorly-recovered satellites either. In the right panel, the peaks of dotted and solid density curves are on the two sides of the \tinfall\, boundary, justifying the boundary \tinfall = 7.1 Gyr ago.

%%%%%% FIGURE %%%%%%
\begin{figure*}
\includegraphics[width=0.9\textwidth]{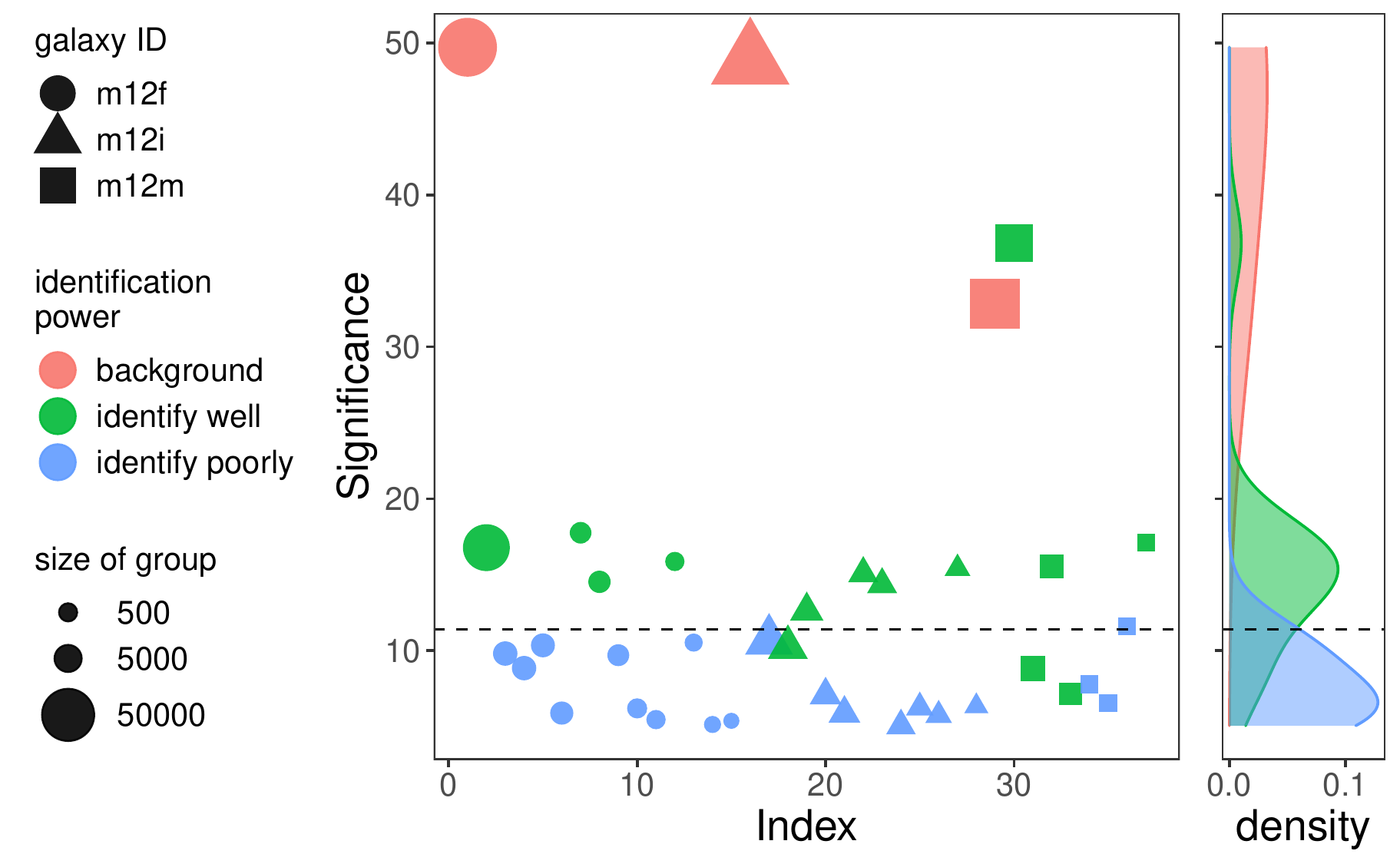}
\caption{The values of {\it significance} of 37 (34 non-background+3 background) groups identified by \Enlink\, in 3 MW-like galaxies (as shown in \figref{grp}) are plotted. ``Index" is a number between 1 and 37 to label each group. The size of a data point is proportional to the number of particles in the group. Groups from galaxies m12f, m12i and m12m are marked by circles, triangles and squares, respectively. The colour of a data point indicates whether the groups is a background group (salmon),  corresponds to a well-recovered satellite (green) or does not identify any satellite (blue). A kernel density plot is attached on the right hand side of the scatter plot. The black dashed line labeling the valley between green and blue curve peaks shows the cutoff in {\it significance} for groups corresponding to well-recovered satellites. The cutoff value in {\it significance} for groups corresponding to well-recovered satellites is 11.4, the 66th percentile of the {\it significance} of non-background groups.}
\label{fig:sig}
\end{figure*}
%%%%%% FIGURE %%%%%%
We have seen that massive satellites fell into MW-like galaxies relatively recently can be reliably recovered by cluster analysis in action space with \Enlink. In future applications to observational data it will be necessary to use a statistical metric provided by \Enlink\, to determine which groups are most likely to correspond to real satellites. The variable {\it significance}, assigned by \Enlink\, to each group and calculated by Equation \eqref{eq:sign}, is a good indicator of whether a group corresponds to a real satellite or not. In \figref{sig} we plot the values of {\it significance} of 34 non-background groups and 3 background groups. The size of a data point is proportional to the number of particles in the group, while the shape shows which galaxy this group is from, and the colour indicates whether the group is a background group (salmon), identifies a satellite well (green) or does not identify a single satellite (blue). A kernel density plot  is attached on the side of the scatter plot. The green curve showing the distribution of {\it significance} peaks at a higher {\it significance} than the blue curve. The black dashed line shows the valley between the green and blue peaks. This valley, located at {\it significance} = 11.4 or the 66th percentile of the values of {\it significance} of the non-background population, is the cutoff in {\it significance} for groups corresponding to well-recovered satellites. Note the cutoff in {\it significance} here should not be confused with $S_{Th}$ in Section \ref{ss:cluster}.  Among 12 non-background groups with {\it significance} above the cutoff, 11 of them are corresponding to ``well-recovered" satellites. This result implies that with the results of \Enlink\, cluster analysis  in action space alone (i.e. no other information on satellites, and ignoring the largest group identified by \Enlink\, assuming it to be ``background''), {\it groups with higher {\it significance} than 66\% of groups found by \Enlink\, are very likely to be corresponding to true satellites}.

\subsection{The Effect of {\it In-situ} Stars}
\label{subsec:insitu}

Besides accreted stars,  cosmological hydrodynamical simulations predict that \insitu  stars may also contribute to a significant part of the halo star population in a galactic halo. As mentioned earlier,  the fraction of \insitu halo stars is highly uncertain but estimates in the MW \citep{bell_etal_08,2020arXiv200608625N} suggest it could be as small as 5\%, while studies of resolved halo stellar populations in external galaxies find that the properties of these halo stars are consistent with being purely accreted \citep{harmsen_etal_17}. Recent work \citep{2020arXiv200608625N} shows that in the MW halo, the relative fraction of \insitu stars drops below 0.5 when $|Z_{gal}|>5 \mathrm{kpc}$, where $|Z_{\mathrm{gal}}|$ is the distance from the disk plane. We also note that over 95\% of the accreted star particles in the three MW-like galaxies have $|Z_{\mathrm{gal}}|>5 \mathrm{kpc}$.  Therefore to study the effects of \insitu stars on the robustness of satellite identification by cluster analysis in the action space, we build numerous mock data sets with \insitu {contamination ratios} equaling 0.1, 0.2, 0.3, 0.4 and 0.5  (where contamination ratio of 0.5 implies that 50\% of the halo star sample consists of \insitu stars) in three MW-like galaxies by randomly sampling the \insitu star particles with $|Z_{\mathrm{gal}}|>5 \mathrm{kpc}$. We then calculate the actions $\{J_r, J_z, J_{\phi}\}$ and do cluster analysis with \Enlink\, in the action space on these ``contaminated'' data sets. We pick out the well-recovered satellites in Section \ref{subsec:acc} and study their values of {\it merit} at different contamination ratios. To reduce the effect of the randomness in sampling the \insitu star particles, for every contamination ratio in each galaxy, we repeat the selection of the \insitu ``contaminant'' population 100 times, randomly picking the same fraction of \insitu stars each time. We then generate the value of {\it merit} for each well-recovered satellite from each data set. Each well-recovered satellite at a contamination ratio then has 100 values of {\it merit}.  We calculate the average and standard deviation of those 100 values. If the previously well-recovered satellite doesn't have a best fit group in one data set (i.e., the best {\it recovery} group doesn't match with the best {\it purity} group), then the {\it merit} of that satellite in that run is set to 0. 

%%%%%%%%%% FIGURE %%%%%%%%%%%%
\begin{figure*}
\includegraphics[width=0.32\textwidth]{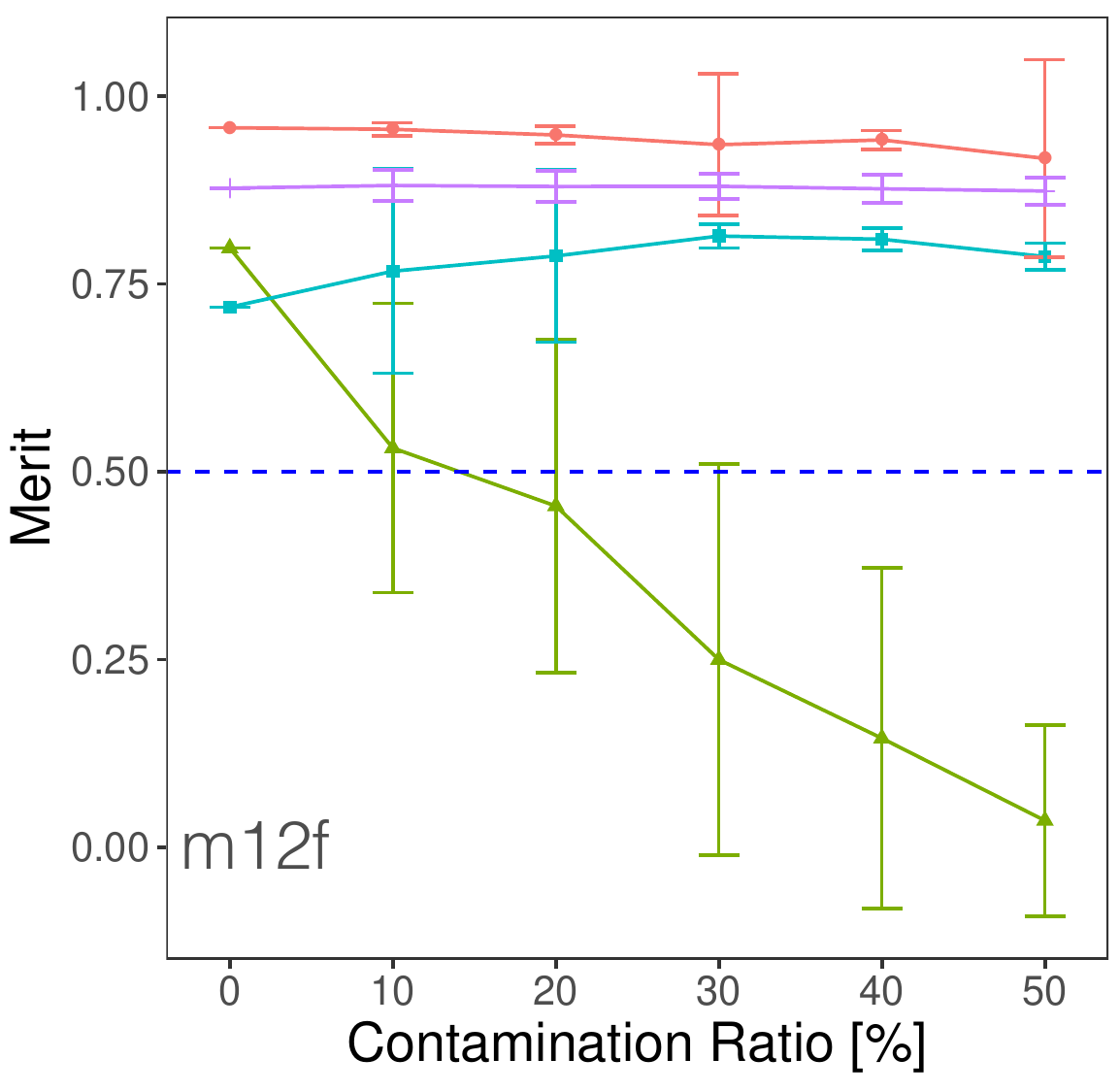}
\includegraphics[width=0.32\textwidth]{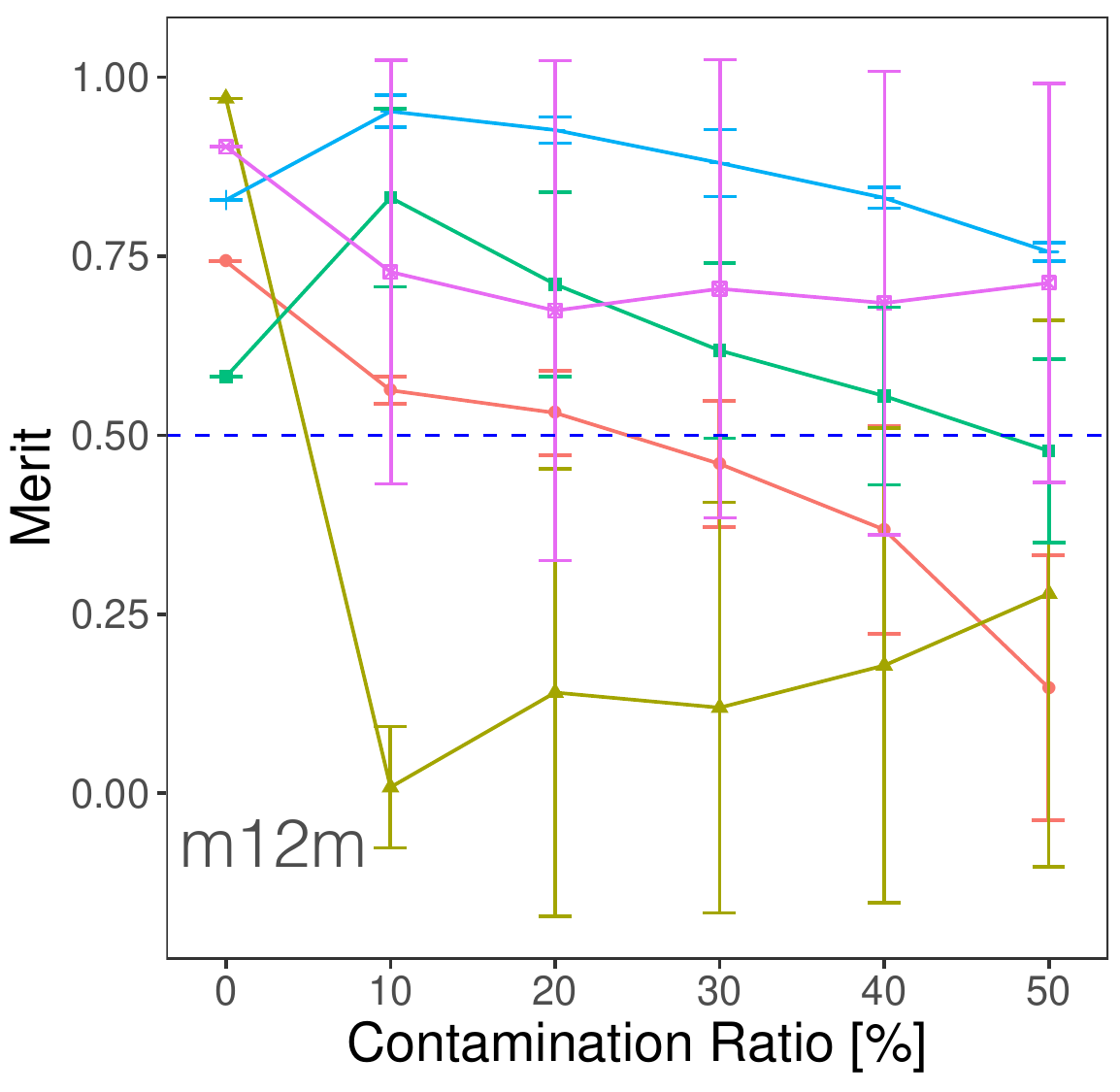}
\includegraphics[width=0.32\textwidth]{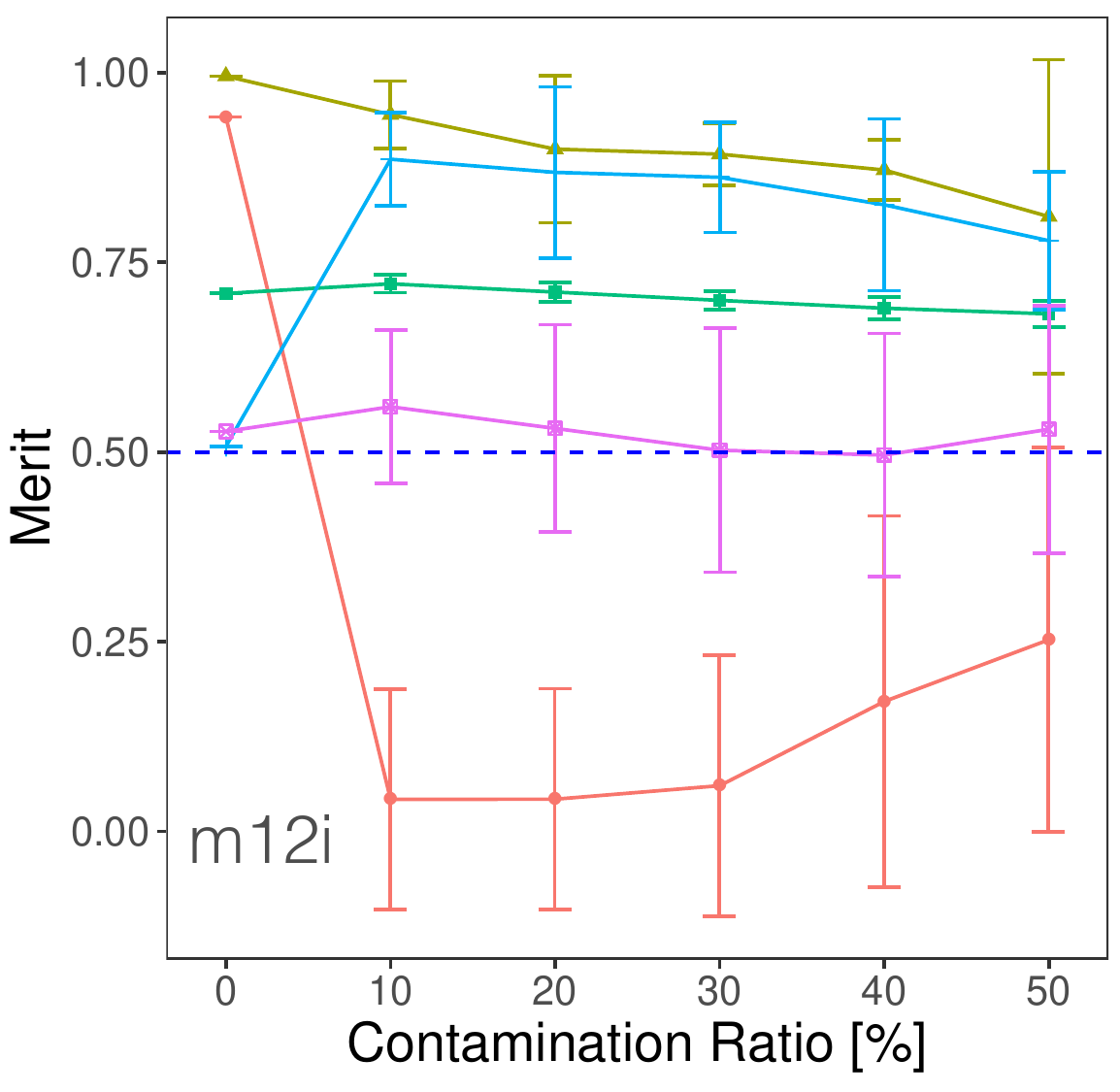}
\caption{Values of {\it merit} of well-recovered satellites in three galaxies (m12f, m12i, m12m from left to right) at different \insitu contamination ratios. Each differently colored line corresponds to one well-recovered satellite in each galaxy (the well-recovered satellites in three galaxies are summarized in \figref{acc4}). The values of merit come from the average of 100 runs and the error bars show the standard deviations. A blue dashed line in 3 panels shows {\it merit}=0.5, the threshold of ``well-recovered" objects. 64\% (9/14) of the well-recovered satellites have values of merit greater than 0.5 at all contamination ratios, being robust against the contamination from \insitu star particles. As the contamination ratio increases, the values of {\it merit} of some satellites drop below 0.5. These ``vulnerable'' satellites have higher standard deviations in merit than other robust satellites, showing that they are more affected by the randomness of picking \insitu star particles. Despite a few exceptions, there is a general trend that the merit decreases as the contamination ratio increases. The robustness of well-recovered satellites indicates that \Enlink\, can recover most of the well-recovered satellites when \insitu\, star particles are in the data set.}
\label{fig:cont}
\end{figure*}
%%%%%% FIGURE %%%%%%

\figref{cont} illustrates the change of values of {\it merit} of well-recovered satellites in three MW-like galaxies (as shown in Section \ref{subsec:acc}) as the contamination ratio increases from 0 to 0.5. Each data point and error bar in the figure shows the average and standard deviation of results of 100 runs, respectively. Lines of different colors correspond to different well-recovered satellites. 64\% (9/14) of the satellites are robust against the contamination from \insitu star particles, as their values of {\it merit} are above 0.5 at all contamination ratios. Other satellites fade away from the identification of \Enlink\, ({\it merit} $< 0.5$) as the contamination ratio increases. The standard deviation of the {\it merit} of these satellites are also higher than their robust companions, showing that they are more vulnerable under the random draw of the \insitu star particles. Besides a few satellites, as expected, most (79\%, 11/14) of the values of merit generally decrease as the contamination ratio increases.

\section{Discussion and Conclusions}
\label{sec:con}

\begin{figure*}
\centering
\includegraphics[width=0.9\textwidth]{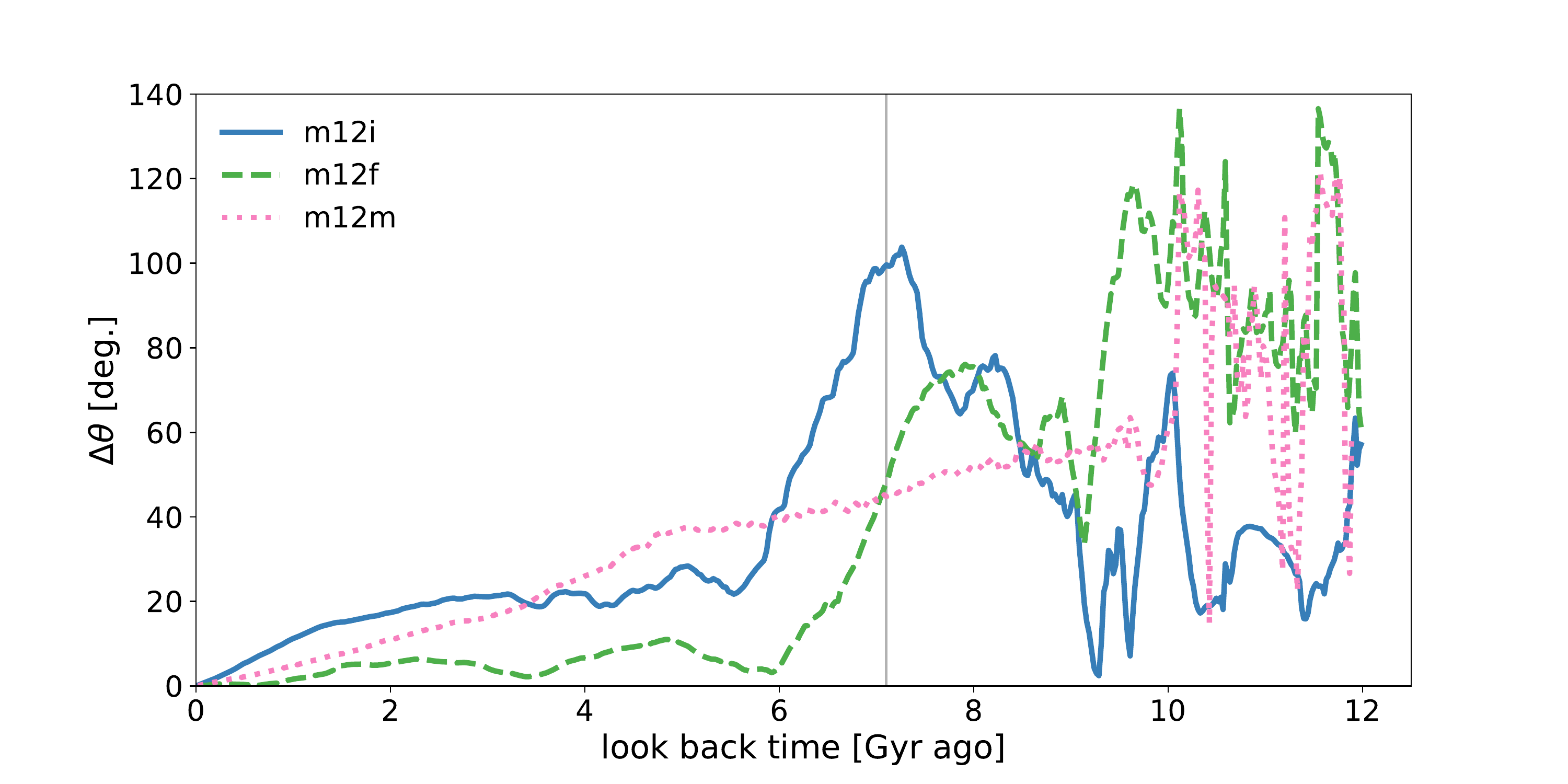}
\caption{The dynamical histories of the 3 MW-like simulated galaxies in FIRE-2. The angle $\Delta\theta$ of the normal vector of the galaxy disk relative to its present-day direction is plotted versus lookback time. If $\Delta\theta$ changes rapidly, then the evolution of the galaxy is chaotic; if $\Delta\theta$ changes slowly, then the galaxy evolves smoothly. The disk orientations of galaxies m12f and m12m changed rapidly at the beginning and evolved more smoothly after 8 Gyr ago. Galaxy m12i evolved more smoothly than the other two galaxies before 8 Gyr ago, allowing massive satellites that fell into m12i up to 10.3 Gyr ago to remain clustered in action space at $z=0$. A vertical line shows the time boundary for well-recovered satellites-7.1 Gyr ago.}
\label{fig:dyn}
\end{figure*}

Inspired by the principle of conservation of orbital actions, we use cluster analysis in action space to find accreted satellites with accreted star particles in three MW like galaxies in the FIRE-2 cosmological hydrodynamical simulations. We summarize the main findings of our work below:
\begin{enumerate}
\item{\textbf{Classification tree method finds the boundaries separating well-recovered and poorly-recovered satellites to be \tinfall = 7.1 Gyr ago and \mtot = $4.0\times 10^8 \msun$ (or $M_{\rm stellar}=1.2\times 10^6 \msun$) (see \figref{acc4} and \figref{acc5}).}} We note however, that the boundary in $M_{\rm stellar}$ coincides with the minimum mass of groups that \Enlink\, is set up to identify, determined by $N_{\rm min}$. This coincidence implies that the mass boundary could be an artifact of the choice of  $N_{\rm min}$.\\
The three galaxies have slightly different infall time and  mass boundaries for reliable detection. For example, galaxy m12i  has 3 well-recovered satellites which fell into the hosting halo more than 7.1 Gyr ago. This difference could be due to the distinct dynamical/evolutionary histories of the galaxies. \figref{dyn} shows the offset angle, $\Delta\theta$,  between the angular momentum vector of the disk at a given look-back time and the present time ($z=0$) as a function of look-back time for each of the three galaxies analyzed in this paper. The disk stars at each look-back time are defined as star particles that are within 30 kpc and also have formation distance within 30 kpc from the center of the host galaxy.
%\YW{We need a more detailed definition of angular momentum here, accounting stars in what radius range?} 
The graph shows that  the direction of the angular momentum vector  often changes suddenly and through large angles (presumably due to merger events) for look back time $>8$~Gyr ago.~\citet{2021arXiv210203369S} shows that many of the changes in this $\Delta\theta$ correlate with major gas-rich mergers. In particular, galaxies m12f and m12m experienced numerous chaotic changes in $\Delta\theta$ until 8 Gyr ago, after which  $\Delta\theta$ changed much more slowly and steadily. The slow change in $\Delta\theta$ implies the gravitational potentials of these two galaxies have changed adiabatically over the last 8 Gyrs, and therefore orbital actions of particles should be reasonably well conserved. The conservation of orbital actions implies that satellites accreted less than 8 Gyr ago should be found clustered in action space at $z=0$.  This ``8 Gyr ago'' time limit for finding satellites clustered in action space is in agreement with the time boundary for good recovery that we found in Section \ref{sec:res}. 
$\Delta\theta$ of galaxy m12i changes more smoothly compared with the other 2 galaxies, so it is expected that satellites accreted a long time ago would remain clustered in action space. 
This is in agreement with the fact that three satellites in this galaxy with \tinfall\, $>9$ Gyr ago are still well-recovered by \Enlink. In our Milky Way galaxy, Gaia Enceladus, which has \tinfall $\approx$ 10 Gyr ago and progenitor stellar mass $6\times10^8 \msun$~\citep{2018Natur.563...85H}, can still be recovered in phase space. The recovery of this massive substructure which fell into the Milky Way a long time ago indicates that the Milky Way might evolve smoothly from relatively early on, like m12i in the FIRE-2 simulations.
\item{\textbf{The value of {\it significance} of a group shows a high correlation with the identification power of this group (see \figref{sig}).} Of the groups with high {\it significance} ($>$11.4, 66th percentile of the {\it significance} of non-background groups in three galaxies),  most (92\%, 11/12) correspond to the well-recovered satellites. This implies that if  cluster analysis in action space is applied to observational data, this {\it significance} assigned by \Enlink\, can help us determine which groups are most likely to correspond to true accreted satellites.}
\item{\textbf{Most of the well-recovered satellites are robust against contamination of \insitu star particles (see \figref{cont}).} 64\% (9/14) of the well-recovered satellites in Section \ref{subsec:acc} stay well-recovered (with {\it merit} $> 0.5$) at 5 different contamination ratios (0.1, 0.2, 0.3, 0.4 and 0.5), where contamination ratio is the percentage of \insitu\, star particles in the data set of one galaxy. The satellites which fail to be identified by \Enlink\, in the presence of contamination have higher standard deviations in {\it merit}, indicating that they are more sensitive to the randomness in picking \insitu stars. This robustness against contamination from \insitu star particles indicates that it will be possible to apply cluster analysis in action space to observation data, even if there is significant contamination from \insitu stars. This is reassuring since the expected fraction of \insitu stars in the MW  (~5\%) is at the low end of the contamination fractions we have experimented with. }
\end{enumerate}
%Besides the infall time boundary, there is also a mass boundary for well-recovered satellites. Satellites with small masses (total mass$<$ $4.0\times 10^8 \msun$) are not robust enough to stay clustered in action space, because smaller halos are more easily stretched by the tidal force at the time of accretion and is less stable when the host halo potential undergoes changes.
%\YW{I cannot come up with a very good description of this process, but in general small halos are less stable}
%This could also be related to the fact that \nmin is set to be 300 and thus small satellites cannot be identified as distinct groups by \Enlink.
%%%%%% FIGURE %%%%%%
%%%%%% FIGURE %%%%%%
In this work we have demonstrated, using simulation data, that it is possible to find accreted satellites using cluster analysis in action space. We have deliberately focused on a fairly idealized set of circumstances: (a) we assumed that phase space coordinates of all star particles were known with no error; (b) we assumed nearly perfect knowledge of the gravitational potential arising from stars, dark matter and gas (although we approximated the potentials as axisymmetric even though they are triaxial); (c) we had perfect knowlegde of the merger history of the galaxies and therefore the true masses and infall times of the satellites. In future applications of this method to observational data, the following improvements should be explored. 
 \begin{enumerate}
 \item{As with all cosmological-hydrodynamical simulations, despite being until very recently the highest-resolution such simulations ever conducted of MW-mass halos \citep{2020arXiv200811207A}, resolution limits our results to accreted structures at the stellar mass of classical dwarf galaxies and higher, since each star particle in the FIRE-2 simulations is 7100 $\msun$. The total number of accreted star particles used in our analysis is also orders of magnitude smaller than the number of accreted stars with 6D phase space coordinates (from Gaia and future large ground based surveys such as 4MOST \citep{4MOST2012}, WEAVE \citep{WEAVE2014}, DESI \citep{DESI_2016a,DESI_2016b}, which are expected to yield 6 phase space coordinates for 3--4 million stars. The performance of our clustering technique when scaled up by 3--4 orders of magnitude in particle number has yet to be tested. On the other hand, if we restrict ourselves to, e.g., RR Lyrae variables, then the conversion rate between the real data set and the simulation data set is about one star per star particle, and the analysis in this work can be more easily done. In addition, real data have observational errors and for this study we have assumed that all phase space coordinates are known perfectly. Additional studies with more realistic mock datasets like the \texttt{ananke} data set \citep{sanderson_etal_20} are needed to assess how well \Enlink\ performs under realistic conditions.}
 \item{It is necessary to find an objective way to determine \nmin, the parameter used by \Enlink\, to determine a ``reasonable'' size for a cluster (Section \ref{ss:cluster}).  In this study \nmin\, was selected by trial and error in order to ensure that the number of groups that \Enlink\, produced was close to the number of satellites that were known to have been accreted  (as described in Section \ref{sec:data}). An objective way to determine \nmin\, and the other parameters that are used in the \Enlink\, cluster analysis algorithm, is needed to evaluate the quality of clusters found. We found that the conventional silhouette plots \citep{ROUSSEEUW198753} are not useful here, since many satellites are irregularly shaped in the action space and could be incorrectly classified by silhouette values. The choice of parameters for cluster analysis has been largely ignored in previous papers, and should be investigated in future.}
 \item{Stellar metallicities and the abundance of individual elements in a  star are also well conserved quantities that contain a substantial amount of information about the host satellite in which the star was born. However the stars in a satellite can have a range of [Fe/H] and $\alpha$-element abundance ratios as well as gradients in the abundances of other elements. \citet{sanderson_etal_17} show that including the abundance of certain  elements (e.g. $N$, $Ca$) can improve the recovery of the Galactic mass model built through cluster analysis. Incorporating metallicities of stars as features into cluster analysis might improve the performance of current results from cluster analysis in action space.}
 \item{The inclusion of metallicities of stars can also help to determine the masses of individual satellites (through the mass-metallicity relationship) and hence action-space clustering can be used to determine the number of accreted satellites per unit mass $N(M)$, a parameter that is a sensitive probe of the model of dark matter.}
 \item{We have assumed in this work that the overall gravitational potential of each galaxy is axisymmetric rather than triaxial. This restriction was driven by the fact that the most efficient action finders available (e.g. AGAMA) are restricted to oblate axisymmetric potentials. Triaxial action finders \citep{2015ascl.soft12020S,2015MNRAS.447.2479S} exist but are currently not fast enough to be useful for analysis of large datasets. Development of fast triaxial action finders would be extremely helpful.}
 \end{enumerate}

%\citep{Ostdiek:2019gnb} introduces a neural network that can distinguish between in situ stars and accreted stars in a Gaia DR2 like data set with 4D, 5D or 6D kinematics. In the test set, their algorithm can pick out 47\% of the accreted stars with a {\it purity} of 41\%, or 13\% of the accreted stars with a higher {\it purity} of 59\%. Since there is a well-performing algorithm to pick out accreted stars, it is reasonable that we focus on the accreted stars. In application to observational data, we first pick out the accreted stars, and then run the cluster analysis to identify groups. With our conclusions here, the significant non-background groups correspond to accreted satellites which are accreted relatively recently and massive enough. \YW{not sure whether to put this in conclusion or introduction}
\section*{ACKNOWLEDGEMENTS}
MV, YW and KF acknowledge support from a MICDE Catalyst grant at University of Michigan. MV acknowledges useful discussions with the ``Stellar Halos group'' of Department of Astronomy, UM and is supported by NASA-ATP awards NNX15AK79G and 80NSSC20K0509. KF is grateful for support as Jeff and Gail Kodosky Chair of Physics at the University of Texas, Austin. KF acknowledges support  from the $\rm{Vetenskapsr\mathring{a}de}$t (Swedish Research Council) through contract No. 638-2013-8993 and the Oskar Klein Centre for Cosmoparticle Physics. KF and YW further acknowledge support from DoE grant DE- SC007859 and the LCTP at the University of Michigan.
NP and RES acknowledge support from NASA-ATP award 80NSSC20K0513. RES acknowledges support from the Research Corporation for Science Advancement through the Scialog program on Time Domain Astrophysics, and from HST grant AR-15809 from STScI. AW receives support from NASA through ATP grants 80NSSC18K1097 and 80NSSC20K0513; HST grants GO-14734, AR-15057, AR-15809, and GO-15902 from STScI; a Scialog Award from the Heising-Simons Foundation; and a Hellman Fellowship.

The authors thank the Flatiron Institute Scientific Computing Core for providing computing resources that made this research possible, and especially for their hard work facilitating remote access during the pandemic. Some analysis for this paper was carried out on the Flatiron Institute's computing cluster \texttt{rusty}, which is supported by the Simons Foundation.

Simulations used in this work were run using XSEDE supported by NSF grant ACI-1548562, Blue Waters via allocation PRAC NSF.1713353 supported by the NSF, and NASA HEC Program through the NAS Division at Ames Research Center.

\bibliography{objectFinding.res}{}

\begin{thebibliography}{}
\makeatletter
\relax
\def\mn@urlcharsother{\let\do\@makeother \do\$\do\&\do\#\do\^\do\_\do\%\do\~}
\def\mn@doi{\begingroup\mn@urlcharsother \@ifnextchar [ {\mn@doi@}
  {\mn@doi@[]}}
\def\mn@doi@[#1]#2{\def\@tempa{#1}\ifx\@tempa\@empty \href
  {http://dx.doi.org/#2} {doi:#2}\else \href {http://dx.doi.org/#2} {#1}\fi
  \endgroup}
\def\mn@eprint#1#2{\mn@eprint@#1:#2::\@nil}
\def\mn@eprint@arXiv#1{\href {http://arxiv.org/abs/#1} {{\tt arXiv:#1}}}
\def\mn@eprint@dblp#1{\href {http://dblp.uni-trier.de/rec/bibtex/#1.xml}
  {dblp:#1}}
\def\mn@eprint@#1:#2:#3:#4\@nil{\def\@tempa {#1}\def\@tempb {#2}\def\@tempc
  {#3}\ifx \@tempc \@empty \let \@tempc \@tempb \let \@tempb \@tempa \fi \ifx
  \@tempb \@empty \def\@tempb {arXiv}\fi \@ifundefined
  {mn@eprint@\@tempb}{\@tempb:\@tempc}{\expandafter \expandafter \csname
  mn@eprint@\@tempb\endcsname \expandafter{\@tempc}}}

\bibitem[\protect\citeauthoryear{{Antoja} et~al.,}{{Antoja}
  et~al.}{2018}]{antoja_etal_18}
{Antoja} T.,  et~al., 2018, \mn@doi [\nat] {10.1038/s41586-018-0510-7}, \href
  {https://ui.adsabs.harvard.edu/abs/2018Natur.561..360A} {561, 360}

\bibitem[\protect\citeauthoryear{{Applebaum}, {Brooks}, {Christensen},
  {Munshi}, {Quinn}, {Shen}  \& {Tremmel}}{{Applebaum}
  et~al.}{2020}]{2020arXiv200811207A}
{Applebaum} E.,  {Brooks} A.~M.,  {Christensen} C.~R.,  {Munshi} F.,  {Quinn}
  T.~R.,  {Shen} S.,   {Tremmel} M.,  2020, arXiv e-prints, \href
  {https://ui.adsabs.harvard.edu/abs/2020arXiv200811207A} {p. arXiv:2008.11207}

\bibitem[\protect\citeauthoryear{{Behroozi}, {Wechsler}  \& {Wu}}{{Behroozi}
  et~al.}{2013a}]{2013ApJ...762..109B}
{Behroozi} P.~S.,  {Wechsler} R.~H.,   {Wu} H.-Y.,  2013a, \mn@doi [\apj]
  {10.1088/0004-637X/762/2/109}, \href
  {https://ui.adsabs.harvard.edu/abs/2013ApJ...762..109B} {762, 109}

\bibitem[\protect\citeauthoryear{{Behroozi}, {Wechsler}, {Wu}, {Busha},
  {Klypin}  \& {Primack}}{{Behroozi} et~al.}{2013b}]{2013ApJ...763...18B}
{Behroozi} P.~S.,  {Wechsler} R.~H.,  {Wu} H.-Y.,  {Busha} M.~T.,  {Klypin}
  A.~A.,   {Primack} J.~R.,  2013b, \mn@doi [\apj]
  {10.1088/0004-637X/763/1/18}, \href
  {https://ui.adsabs.harvard.edu/abs/2013ApJ...763...18B} {763, 18}

\bibitem[\protect\citeauthoryear{{Bell} et~al.,}{{Bell}
  et~al.}{2008}]{bell_etal_08}
{Bell} E.~F.,  et~al., 2008, \mn@doi [\apj] {10.1086/588032}, \href
  {https://ui.adsabs.harvard.edu/abs/2008ApJ...680..295B} {680, 295}

\bibitem[\protect\citeauthoryear{{Belokurov}, {Erkal}, {Evans}, {Koposov}  \&
  {Deason}}{{Belokurov} et~al.}{2018}]{2018MNRAS.478..611B}
{Belokurov} V.,  {Erkal} D.,  {Evans} N.~W.,  {Koposov} S.~E.,   {Deason}
  A.~J.,  2018, \mn@doi [\mnras] {10.1093/mnras/sty982}, \href
  {https://ui.adsabs.harvard.edu/abs/2018MNRAS.478..611B} {478, 611}

\bibitem[\protect\citeauthoryear{{Besla}, {Kallivayalil}, {Hernquist},
  {Robertson}, {Cox}, {van der Marel}  \& {Alcock}}{{Besla}
  et~al.}{2007}]{2007ApJ...668..949B}
{Besla} G.,  {Kallivayalil} N.,  {Hernquist} L.,  {Robertson} B.,  {Cox} T.~J.,
   {van der Marel} R.~P.,   {Alcock} C.,  2007, \mn@doi [\apj]
  {10.1086/521385}, \href
  {https://ui.adsabs.harvard.edu/abs/2007ApJ...668..949B} {668, 949}

\bibitem[\protect\citeauthoryear{{Besla}, {Kallivayalil}, {Hernquist}, {van der
  Marel}, {Cox}  \& {Kere{\v{s}}}}{{Besla} et~al.}{2010}]{2010ApJ...721L..97B}
{Besla} G.,  {Kallivayalil} N.,  {Hernquist} L.,  {van der Marel} R.~P.,  {Cox}
  T.~J.,   {Kere{\v{s}}} D.,  2010, \mn@doi [\apjl]
  {10.1088/2041-8205/721/2/L97}, \href
  {https://ui.adsabs.harvard.edu/abs/2010ApJ...721L..97B} {721, L97}

\bibitem[\protect\citeauthoryear{{Binney} \& {Tremaine}}{{Binney} \&
  {Tremaine}}{2008}]{BT08}
{Binney} J.,  {Tremaine} S.,  2008, {Galactic Dynamics: Second Edition}.
{Princeton University Press:Princeton, NJ, USA}

\bibitem[\protect\citeauthoryear{{Bonaca}, {Hogg}, {Price-Whelan}  \&
  {Conroy}}{{Bonaca} et~al.}{2019}]{Bonaca_spur_2018}
{Bonaca} A.,  {Hogg} D.~W.,  {Price-Whelan} A.~M.,   {Conroy} C.,  2019,
  \mn@doi [\apj] {10.3847/1538-4357/ab2873}, \href
  {https://ui.adsabs.harvard.edu/abs/2019ApJ...880...38B} {880, 38}

\bibitem[\protect\citeauthoryear{{Borsato}, {Martell}  \& {Simpson}}{{Borsato}
  et~al.}{2020}]{2020MNRAS.492.1370B}
{Borsato} N.~W.,  {Martell} S.~L.,   {Simpson} J.~D.,  2020, \mn@doi [\mnras]
  {10.1093/mnras/stz3479}, \href
  {https://ui.adsabs.harvard.edu/abs/2020MNRAS.492.1370B} {492, 1370}

\bibitem[\protect\citeauthoryear{Breiman, Friedman, Stone  \& Olshen}{Breiman
  et~al.}{1984}]{breiman1984classification}
Breiman L.,  Friedman J.,  Stone C.~J.,   Olshen R.~A.,  1984, Classification
  and regression trees.
CRC press

\bibitem[\protect\citeauthoryear{{Bullock} \& {Boylan-Kolchin}}{{Bullock} \&
  {Boylan-Kolchin}}{2017}]{bullock_boylan-kolchin_17}
{Bullock} J.~S.,  {Boylan-Kolchin} M.,  2017, \mn@doi [\araa]
  {10.1146/annurev-astro-091916-055313}, \href
  {https://ui.adsabs.harvard.edu/abs/2017ARA&A..55..343B} {55, 343}

\bibitem[\protect\citeauthoryear{{Cooper} et~al.,}{{Cooper}
  et~al.}{2010}]{cooper_etal_10}
{Cooper} A.~P.,  et~al., 2010, \mn@doi [\mnras]
  {10.1111/j.1365-2966.2010.16740.x}, \href
  {https://ui.adsabs.harvard.edu/abs/2010MNRAS.406..744C} {406, 744}

\bibitem[\protect\citeauthoryear{{DESI Collaboration} et~al.,}{{DESI
  Collaboration} et~al.}{2016a}]{DESI_2016a}
{DESI Collaboration} et~al., 2016a, ArXiv e-prints, arXiv:1611.00036, \href
  {http://adsabs.harvard.edu/abs/2016arXiv161100036D} {}

\bibitem[\protect\citeauthoryear{{DESI Collaboration} et~al.,}{{DESI
  Collaboration} et~al.}{2016b}]{DESI_2016b}
{DESI Collaboration} et~al., 2016b, ArXiv e-prints, arXiv:1611.00037, \href
  {http://adsabs.harvard.edu/abs/2016arXiv161100037D} {}

\bibitem[\protect\citeauthoryear{{Dalton} et~al.,}{{Dalton}
  et~al.}{2014}]{WEAVE2014}
{Dalton} G.,  et~al., 2014, {Project overview and update on WEAVE: the next
  generation wide-field spectroscopy facility for the William Herschel
  Telescope}.
SPIE, p. 91470L, \mn@doi{10.1117/12.2055132}

\bibitem[\protect\citeauthoryear{Erkal, Belokurov, Bovy  \& Sanders}{Erkal
  et~al.}{2016}]{Erkal2016_gaps}
Erkal D.,  Belokurov V.,  Bovy J.,   Sanders J.~L.,  2016, \mn@doi [Monthly
  Notices of the Royal Astronomical Society] {10.1093/mnras/stw1957}, 463, 102

\bibitem[\protect\citeauthoryear{{Evans} et~al.,}{{Evans}
  et~al.}{2018}]{2018A&A...616A...4E}
{Evans} D.~W.,  et~al., 2018, \mn@doi [\aap] {10.1051/0004-6361/201832756},
  \href {https://ui.adsabs.harvard.edu/abs/2018A&A...616A...4E} {616, A4}

\bibitem[\protect\citeauthoryear{{Fabricius} et~al.,}{{Fabricius}
  et~al.}{2016}]{2016A&A...595A...3F}
{Fabricius} C.,  et~al., 2016, \mn@doi [\aap] {10.1051/0004-6361/201628643},
  \href {https://ui.adsabs.harvard.edu/abs/2016A&A...595A...3F} {595, A3}

\bibitem[\protect\citeauthoryear{{Fakhouri} \& {Ma}}{{Fakhouri} \&
  {Ma}}{2008}]{fakhouri_08}
{Fakhouri} O.,  {Ma} C.-P.,  2008, \mn@doi [\mnras]
  {10.1111/j.1365-2966.2008.13075.x}, \href
  {https://ui.adsabs.harvard.edu/abs/2008MNRAS.386..577F} {386, 577}

\bibitem[\protect\citeauthoryear{{Feldmann} \& {Spolyar}}{{Feldmann} \&
  {Spolyar}}{2015}]{feldmann_spolyar_15}
{Feldmann} R.,  {Spolyar} D.,  2015, \mn@doi [\mnras] {10.1093/mnras/stu2147},
  \href {https://ui.adsabs.harvard.edu/abs/2015MNRAS.446.1000F} {446, 1000}

\bibitem[\protect\citeauthoryear{{Font}, {McCarthy}, {Crain}, {Theuns},
  {Schaye}, {Wiersma}  \& {Dalla Vecchia}}{{Font} et~al.}{2011}]{font_etal_11}
{Font} A.~S.,  {McCarthy} I.~G.,  {Crain} R.~A.,  {Theuns} T.,  {Schaye} J.,
  {Wiersma} R.~P.~C.,   {Dalla Vecchia} C.,  2011, \mn@doi [\mnras]
  {10.1111/j.1365-2966.2011.19227.x}, \href
  {https://ui.adsabs.harvard.edu/abs/2011MNRAS.416.2802F} {416, 2802}

\bibitem[\protect\citeauthoryear{{Garrison-Kimmel} et~al.,}{{Garrison-Kimmel}
  et~al.}{2018}]{2018MNRAS.481.4133G}
{Garrison-Kimmel} S.,  et~al., 2018, \mn@doi [\mnras] {10.1093/mnras/sty2513},
  \href {https://ui.adsabs.harvard.edu/abs/2018MNRAS.481.4133G} {481, 4133}

\bibitem[\protect\citeauthoryear{Garrison-Kimmel et~al.,}{Garrison-Kimmel
  et~al.}{2019}]{10.1093/mnras/stz1317}
Garrison-Kimmel S.,  et~al., 2019, \mn@doi [Monthly Notices of the Royal
  Astronomical Society] {10.1093/mnras/stz1317}, 487, 1380

\bibitem[\protect\citeauthoryear{{G{\'o}mez}, {Helmi}, {Brown}  \&
  {Li}}{{G{\'o}mez} et~al.}{2010}]{2010MNRAS.408..935G}
{G{\'o}mez} F.~A.,  {Helmi} A.,  {Brown} A. G.~A.,   {Li} Y.-S.,  2010, \mn@doi
  [\mnras] {10.1111/j.1365-2966.2010.17225.x}, \href
  {https://ui.adsabs.harvard.edu/abs/2010MNRAS.408..935G} {408, 935}

\bibitem[\protect\citeauthoryear{{Gudin} et~al.,}{{Gudin}
  et~al.}{2021}]{2021ApJ...908...79G}
{Gudin} D.,  et~al., 2021, \mn@doi [\apj] {10.3847/1538-4357/abd7ed}, \href
  {https://ui.adsabs.harvard.edu/abs/2021ApJ...908...79G} {908, 79}

\bibitem[\protect\citeauthoryear{{Harding}, {Morrison}, {Olszewski},
  {Arabadjis}, {Mateo}, {Dohm-Palmer}, {Freeman}  \& {Norris}}{{Harding}
  et~al.}{2001}]{harding_etal_01}
{Harding} P.,  {Morrison} H.~L.,  {Olszewski} E.~W.,  {Arabadjis} J.,  {Mateo}
  M.,  {Dohm-Palmer} R.~C.,  {Freeman} K.~C.,   {Norris} J.~E.,  2001, \mn@doi
  [\aj] {10.1086/322995}, \href
  {https://ui.adsabs.harvard.edu/abs/2001AJ....122.1397H} {122, 1397}

\bibitem[\protect\citeauthoryear{{Harmsen}, {Monachesi}, {Bell}, {de Jong},
  {Bailin}, {Radburn-Smith}  \& {Holwerda}}{{Harmsen}
  et~al.}{2017}]{harmsen_etal_17}
{Harmsen} B.,  {Monachesi} A.,  {Bell} E.~F.,  {de Jong} R.~S.,  {Bailin} J.,
  {Radburn-Smith} D.~J.,   {Holwerda} B.~W.,  2017, \mn@doi [\mnras]
  {10.1093/mnras/stw2992}, \href
  {https://ui.adsabs.harvard.edu/abs/2017MNRAS.466.1491H} {466, 1491}

\bibitem[\protect\citeauthoryear{Hastie, Tibshirani  \& Friedman}{Hastie
  et~al.}{2001}]{hastie01statisticallearning}
Hastie T.,  Tibshirani R.,   Friedman J.,  2001, The Elements of Statistical
  Learning.
Springer Series in Statistics, Springer New York Inc., New York, NY, USA

\bibitem[\protect\citeauthoryear{{Helmi} \& {de Zeeuw}}{{Helmi} \& {de
  Zeeuw}}{2000}]{helmi_dezeeuw_00}
{Helmi} A.,  {de Zeeuw} P.~T.,  2000, \mn@doi [\mnras]
  {10.1046/j.1365-8711.2000.03895.x}, \href
  {https://ui.adsabs.harvard.edu/abs/2000MNRAS.319..657H} {319, 657}

\bibitem[\protect\citeauthoryear{{Helmi}, {Babusiaux}, {Koppelman}, {Massari},
  {Veljanoski}  \& {Brown}}{{Helmi} et~al.}{2018}]{2018Natur.563...85H}
{Helmi} A.,  {Babusiaux} C.,  {Koppelman} H.~H.,  {Massari} D.,  {Veljanoski}
  J.,   {Brown} A. G.~A.,  2018, \mn@doi [\nat] {10.1038/s41586-018-0625-x},
  \href {https://ui.adsabs.harvard.edu/abs/2018Natur.563...85H} {563, 85}

\bibitem[\protect\citeauthoryear{{Hopkins}}{{Hopkins}}{2015}]{Hopkins2015}
{Hopkins} P.~F.,  2015, \mn@doi [\mnras] {10.1093/mnras/stv195}, \href
  {https://ui.adsabs.harvard.edu/abs/2015MNRAS.450...53H} {450, 53}

\bibitem[\protect\citeauthoryear{Hopkins et~al.,}{Hopkins
  et~al.}{2018}]{10.1093/mnras/sty1690}
Hopkins P.~F.,  et~al., 2018, \mn@doi [Monthly Notices of the Royal
  Astronomical Society] {10.1093/mnras/sty1690}, 480, 800

\bibitem[\protect\citeauthoryear{{Johnston}, {Hernquist}  \&
  {Bolte}}{{Johnston} et~al.}{1996}]{1996ApJ...465..278J}
{Johnston} K.~V.,  {Hernquist} L.,   {Bolte} M.,  1996, \mn@doi [\apj]
  {10.1086/177418}, \href
  {https://ui.adsabs.harvard.edu/abs/1996ApJ...465..278J} {465, 278}

\bibitem[\protect\citeauthoryear{{Lazar} et~al.,}{{Lazar}
  et~al.}{2020}]{lazar_etal_20}
{Lazar} A.,  et~al., 2020, \mn@doi [\mnras] {10.1093/mnras/staa2101}, \href
  {https://ui.adsabs.harvard.edu/abs/2020MNRAS.497.2393L} {497, 2393}

\bibitem[\protect\citeauthoryear{{Limberg} et~al.,}{{Limberg}
  et~al.}{2021}]{2021ApJ...907...10L}
{Limberg} G.,  et~al., 2021, \mn@doi [\apj] {10.3847/1538-4357/abcb87}, \href
  {https://ui.adsabs.harvard.edu/abs/2021ApJ...907...10L} {907, 10}

\bibitem[\protect\citeauthoryear{{Lindegren} et~al.,}{{Lindegren}
  et~al.}{2016}]{2016A&A...595A...4L}
{Lindegren} L.,  et~al., 2016, \mn@doi [\aap] {10.1051/0004-6361/201628714},
  \href {https://ui.adsabs.harvard.edu/abs/2016A&A...595A...4L} {595, A4}

\bibitem[\protect\citeauthoryear{{Lindegren} et~al.,}{{Lindegren}
  et~al.}{2018}]{2018A&A...616A...2L}
{Lindegren} L.,  et~al., 2018, \mn@doi [\aap] {10.1051/0004-6361/201832727},
  \href {https://ui.adsabs.harvard.edu/abs/2018A&A...616A...2L} {616, A2}

\bibitem[\protect\citeauthoryear{Lynden-Bell \& Lynden-Bell}{Lynden-Bell \&
  Lynden-Bell}{1995}]{LyndenBell:1995zz}
Lynden-Bell D.,  Lynden-Bell R.~M.,  1995, Mon. Not. Roy. Astron. Soc., 275,
  429

\bibitem[\protect\citeauthoryear{{Monachesi} et~al.,}{{Monachesi}
  et~al.}{2019}]{monachesi_etal_19}
{Monachesi} A.,  et~al., 2019, \mn@doi [\mnras] {10.1093/mnras/stz538}, \href
  {https://ui.adsabs.harvard.edu/abs/2019MNRAS.485.2589M} {485, 2589}

\bibitem[\protect\citeauthoryear{Myeong, Evans, Belokurov, Sanders  \&
  Koposov}{Myeong et~al.}{2018a}]{10.1093/mnras/sty1403}
Myeong G.~C.,  Evans N.~W.,  Belokurov V.,  Sanders J.~L.,   Koposov S.~E.,
  2018a, \mn@doi [Monthly Notices of the Royal Astronomical Society]
  {10.1093/mnras/sty1403}, 478, 5449

\bibitem[\protect\citeauthoryear{{Myeong}, {Evans}, {Belokurov}, {Sand ers}  \&
  {Koposov}}{{Myeong} et~al.}{2018b}]{2018ApJ...856L..26M}
{Myeong} G.~C.,  {Evans} N.~W.,  {Belokurov} V.,  {Sand ers} J.~L.,   {Koposov}
  S.~E.,  2018b, \mn@doi [\apjl] {10.3847/2041-8213/aab613}, \href
  {https://ui.adsabs.harvard.edu/abs/2018ApJ...856L..26M} {856, L26}

\bibitem[\protect\citeauthoryear{{Myeong}, {Vasiliev}, {Iorio}, {Evans}  \&
  {Belokurov}}{{Myeong} et~al.}{2019}]{2019MNRAS.488.1235M}
{Myeong} G.~C.,  {Vasiliev} E.,  {Iorio} G.,  {Evans} N.~W.,   {Belokurov} V.,
  2019, \mn@doi [\mnras] {10.1093/mnras/stz1770}, \href
  {https://ui.adsabs.harvard.edu/abs/2019MNRAS.488.1235M} {488, 1235}

\bibitem[\protect\citeauthoryear{{Naidu}, {Conroy}, {Bonaca}, {Johnson},
  {Ting}, {Caldwell}, {Zaritsky}  \& {Cargile}}{{Naidu}
  et~al.}{2020}]{2020arXiv200608625N}
{Naidu} R.~P.,  {Conroy} C.,  {Bonaca} A.,  {Johnson} B.~D.,  {Ting} Y.-S.,
  {Caldwell} N.,  {Zaritsky} D.,   {Cargile} P.~A.,  2020, arXiv e-prints,
  \href {https://ui.adsabs.harvard.edu/abs/2020arXiv200608625N} {p.
  arXiv:2006.08625}

\bibitem[\protect\citeauthoryear{{Necib}, {Ostdiek}, {Lisanti}, {Cohen},
  {Freytsis}  \& {Garrison-Kimmel}}{{Necib} et~al.}{2019}]{necib_etal_19}
{Necib} L.,  {Ostdiek} B.,  {Lisanti} M.,  {Cohen} T.,  {Freytsis} M.,
  {Garrison-Kimmel} S.,  2019, arXiv e-prints, \href
  {https://ui.adsabs.harvard.edu/abs/2019arXiv190707681N} {p. arXiv:1907.07681}

\bibitem[\protect\citeauthoryear{{Necib} et~al.,}{{Necib}
  et~al.}{2020}]{necib_etal_20_nyx}
{Necib} L.,  et~al., 2020, \mn@doi [Nature Astronomy]
  {10.1038/s41550-020-1131-2}, \href
  {https://ui.adsabs.harvard.edu/abs/2020NatAs.tmp..137N} {}

\bibitem[\protect\citeauthoryear{Ostdiek et~al.,}{Ostdiek
  et~al.}{2020}]{Ostdiek:2019gnb}
Ostdiek B.,  et~al., 2020, \mn@doi [Astron. Astrophys.]
  {10.1051/0004-6361/201936866}, 636, A75

\bibitem[\protect\citeauthoryear{{Panithanpaisal}, {Sanderson}, {Wetzel},
  {Cunningham}, {Bailin}  \& {Faucher-Gigu{\`e}re}}{{Panithanpaisal}
  et~al.}{2021}]{2021arXiv210409660P}
{Panithanpaisal} N.,  {Sanderson} R.~E.,  {Wetzel} A.,  {Cunningham} E.~C.,
  {Bailin} J.,   {Faucher-Gigu{\`e}re} C.-A.,  2021, arXiv e-prints, \href
  {https://ui.adsabs.harvard.edu/abs/2021arXiv210409660P} {p. arXiv:2104.09660}

\bibitem[\protect\citeauthoryear{{Perryman} et~al.,}{{Perryman}
  et~al.}{2001}]{2001A&A...369..339P}
{Perryman} M.~A.~C.,  et~al., 2001, \mn@doi [\aap]
  {10.1051/0004-6361:20010085}, \href
  {https://ui.adsabs.harvard.edu/abs/2001A&A...369..339P} {369, 339}

\bibitem[\protect\citeauthoryear{{Pillepich}, {Madau}  \& {Mayer}}{{Pillepich}
  et~al.}{2015}]{Pillepich15}
{Pillepich} A.,  {Madau} P.,   {Mayer} L.,  2015, \mn@doi [\apj]
  {10.1088/0004-637X/799/2/184}, \href
  {http://adsabs.harvard.edu/abs/2015ApJ...799..184P} {799, 184}

\bibitem[\protect\citeauthoryear{{Price-Whelan} \& {Bonaca}}{{Price-Whelan} \&
  {Bonaca}}{2018}]{WhelanBonacaGD12018}
{Price-Whelan} A.~M.,  {Bonaca} A.,  2018, \mn@doi [\apjl]
  {10.3847/2041-8213/aad7b5}, \href
  {http://adsabs.harvard.edu/abs/2018ApJ...863L..20P} {863, L20}

\bibitem[\protect\citeauthoryear{{Roederer}, {Hattori}  \&
  {Valluri}}{{Roederer} et~al.}{2018}]{2018AJ....156..179R}
{Roederer} I.~U.,  {Hattori} K.,   {Valluri} M.,  2018, \mn@doi [\aj]
  {10.3847/1538-3881/aadd9c}, \href
  {https://ui.adsabs.harvard.edu/abs/2018AJ....156..179R} {156, 179}

\bibitem[\protect\citeauthoryear{Rousseeuw}{Rousseeuw}{1987}]{ROUSSEEUW198753}
Rousseeuw P.~J.,  1987, \mn@doi [Journal of Computational and Applied
  Mathematics] {https://doi.org/10.1016/0377-0427(87)90125-7}, 20, 53

\bibitem[\protect\citeauthoryear{{Samuel}, {Wetzel}, {Chapman}, {Tollerud},
  {Hopkins}, {Boylan-Kolchin}, {Bailin}  \& {Faucher-Gigu{\`e}re}}{{Samuel}
  et~al.}{2020a}]{2020arXiv201008571S}
{Samuel} J.,  {Wetzel} A.,  {Chapman} S.,  {Tollerud} E.,  {Hopkins} P.~F.,
  {Boylan-Kolchin} M.,  {Bailin} J.,   {Faucher-Gigu{\`e}re} C.-A.,  2020a,
  arXiv e-prints, \href {https://ui.adsabs.harvard.edu/abs/2020arXiv201008571S}
  {p. arXiv:2010.08571}

\bibitem[\protect\citeauthoryear{{Samuel} et~al.,}{{Samuel}
  et~al.}{2020b}]{2020MNRAS.491.1471S}
{Samuel} J.,  et~al., 2020b, \mn@doi [\mnras] {10.1093/mnras/stz3054}, \href
  {https://ui.adsabs.harvard.edu/abs/2020MNRAS.491.1471S} {491, 1471}

\bibitem[\protect\citeauthoryear{{Sanders} \& {Binney}}{{Sanders} \&
  {Binney}}{2015a}]{2015ascl.soft12020S}
{Sanders} J.~L.,  {Binney} J.,  2015a, {TACT: The Action Computation Tool}
  (\mn@eprint {ascl} {1512.020})

\bibitem[\protect\citeauthoryear{{Sanders} \& {Binney}}{{Sanders} \&
  {Binney}}{2015b}]{2015MNRAS.447.2479S}
{Sanders} J.~L.,  {Binney} J.,  2015b, \mn@doi [\mnras]
  {10.1093/mnras/stu2598}, \href
  {https://ui.adsabs.harvard.edu/abs/2015MNRAS.447.2479S} {447, 2479}

\bibitem[\protect\citeauthoryear{{Sanderson}, {Wetzel}, {Sharma}  \&
  {Hopkins}}{{Sanderson} et~al.}{2017}]{sanderson_etal_17}
{Sanderson} R.,  {Wetzel} A.,  {Sharma} S.,   {Hopkins} P.,  2017, \mn@doi
  [Galaxies] {10.3390/galaxies5030043}, \href
  {https://ui.adsabs.harvard.edu/abs/2017Galax...5...43S} {5, 43}

\bibitem[\protect\citeauthoryear{{Sanderson} et~al.,}{{Sanderson}
  et~al.}{2020}]{sanderson_etal_20}
{Sanderson} R.~E.,  et~al., 2020, \mn@doi [\apjs] {10.3847/1538-4365/ab5b9d},
  \href {https://ui.adsabs.harvard.edu/abs/2020ApJS..246....6S} {246, 6}

\bibitem[\protect\citeauthoryear{{Santistevan}, {Wetzel}, {El-Badry},
  {Bland-Hawthorn}, {Boylan-Kolchin}, {Bailin}, {Faucher-Gigu{\`e}re}  \&
  {Benincasa}}{{Santistevan} et~al.}{2020}]{2020MNRAS.497..747S}
{Santistevan} I.~B.,  {Wetzel} A.,  {El-Badry} K.,  {Bland-Hawthorn} J.,
  {Boylan-Kolchin} M.,  {Bailin} J.,  {Faucher-Gigu{\`e}re} C.-A.,
  {Benincasa} S.,  2020, \mn@doi [\mnras] {10.1093/mnras/staa1923}, \href
  {https://ui.adsabs.harvard.edu/abs/2020MNRAS.497..747S} {497, 747}

\bibitem[\protect\citeauthoryear{{Santistevan}, {Wetzel}, {Sanderson},
  {El-Badry}, {Samuel}  \& {Faucher-Gigu{\`e}re}}{{Santistevan}
  et~al.}{2021}]{2021arXiv210203369S}
{Santistevan} I.~B.,  {Wetzel} A.,  {Sanderson} R.~E.,  {El-Badry} K.,
  {Samuel} J.,   {Faucher-Gigu{\`e}re} C.-A.,  2021, arXiv e-prints, \href
  {https://ui.adsabs.harvard.edu/abs/2021arXiv210203369S} {p. arXiv:2102.03369}

\bibitem[\protect\citeauthoryear{Sharma \& Johnston}{Sharma \&
  Johnston}{2009}]{Sharma_2009}
Sharma S.,  Johnston K.~V.,  2009, \mn@doi [The Astrophysical Journal]
  {10.1088/0004-637x/703/1/1061}, 703, 1061

\bibitem[\protect\citeauthoryear{{Tissera}, {Scannapieco}, {Beers}  \&
  {Carollo}}{{Tissera} et~al.}{2013}]{Tissera2013}
{Tissera} P.~B.,  {Scannapieco} C.,  {Beers} T.~C.,   {Carollo} D.,  2013,
  \mn@doi [\mnras] {10.1093/mnras/stt691}, \href
  {http://adsabs.harvard.edu/abs/2013MNRAS.432.3391T} {432, 3391}

\bibitem[\protect\citeauthoryear{{Tremaine}}{{Tremaine}}{1999}]{1999MNRAS.307..877T}
{Tremaine} S.,  1999, \mn@doi [\mnras] {10.1046/j.1365-8711.1999.02690.x},
  \href {https://ui.adsabs.harvard.edu/abs/1999MNRAS.307..877T} {307, 877}

\bibitem[\protect\citeauthoryear{Vasiliev}{Vasiliev}{2018}]{10.1093/mnras/sty2672}
Vasiliev E.,  2018, \mn@doi [Monthly Notices of the Royal Astronomical Society]
  {10.1093/mnras/sty2672}, 482, 1525

\bibitem[\protect\citeauthoryear{{Wetzel} \& {Garrison-Kimmel}}{{Wetzel} \&
  {Garrison-Kimmel}}{2020a}]{2020ascl.soft02014W}
{Wetzel} A.,  {Garrison-Kimmel} S.,  2020a, {HaloAnalysis: Read and analyze
  halo catalogs and merger trees} (\mn@eprint {ascl} {2002.014})

\bibitem[\protect\citeauthoryear{{Wetzel} \& {Garrison-Kimmel}}{{Wetzel} \&
  {Garrison-Kimmel}}{2020b}]{2020ascl.soft02015W}
{Wetzel} A.,  {Garrison-Kimmel} S.,  2020b, {GizmoAnalysis: Read and analyze
  Gizmo simulations} (\mn@eprint {ascl} {2002.015})

\bibitem[\protect\citeauthoryear{Wetzel, Hopkins, hoon Kim,
  Faucher-Gigu{\`{e}}re, Kere{\v{s}}  \& Quataert}{Wetzel
  et~al.}{2016}]{Wetzel_2016}
Wetzel A.~R.,  Hopkins P.~F.,  hoon Kim J.,  Faucher-Gigu{\`{e}}re C.-A.,
  Kere{\v{s}} D.,   Quataert E.,  2016, \mn@doi [The Astrophysical Journal]
  {10.3847/2041-8205/827/2/l23}, 827, L23

\bibitem[\protect\citeauthoryear{{Widrow}, {Gardner}, {Yanny}, {Dodelson}  \&
  {Chen}}{{Widrow} et~al.}{2012}]{widrow_etal_12}
{Widrow} L.~M.,  {Gardner} S.,  {Yanny} B.,  {Dodelson} S.,   {Chen} H.-Y.,
  2012, \mn@doi [\apjl] {10.1088/2041-8205/750/2/L41}, \href
  {https://ui.adsabs.harvard.edu/abs/2012ApJ...750L..41W} {750, L41}

\bibitem[\protect\citeauthoryear{{Yu} et~al.,}{{Yu} et~al.}{2020}]{yu_etal_20}
{Yu} S.,  et~al., 2020, \mn@doi [\mnras] {10.1093/mnras/staa522}, \href
  {https://ui.adsabs.harvard.edu/abs/2020MNRAS.494.1539Y} {494, 1539}

\bibitem[\protect\citeauthoryear{{Yuan} et~al.,}{{Yuan}
  et~al.}{2020}]{yuan_etal_20}
{Yuan} Z.,  et~al., 2020, \mn@doi [\apj] {10.3847/1538-4357/ab6ef7}, \href
  {https://ui.adsabs.harvard.edu/abs/2020ApJ...891...39Y} {891, 39}

\bibitem[\protect\citeauthoryear{{Zavala} \& {Frenk}}{{Zavala} \&
  {Frenk}}{2019}]{zavala_frenk_19}
{Zavala} J.,  {Frenk} C.~S.,  2019, \mn@doi [Galaxies]
  {10.3390/galaxies7040081}, \href
  {https://ui.adsabs.harvard.edu/abs/2019Galax...7...81Z} {7, 81}

\bibitem[\protect\citeauthoryear{{Zolotov}, {Willman}, {Brooks}, {Governato},
  {Brook}, {Hogg}, {Quinn}  \& {Stinson}}{{Zolotov} et~al.}{2009}]{Zolotov2009}
{Zolotov} A.,  {Willman} B.,  {Brooks} A.~M.,  {Governato} F.,  {Brook} C.~B.,
  {Hogg} D.~W.,  {Quinn} T.,   {Stinson} G.,  2009, \mn@doi [\apj]
  {10.1088/0004-637X/702/2/1058}, \href
  {http://adsabs.harvard.edu/abs/2009ApJ...702.1058Z} {702, 1058}

\bibitem[\protect\citeauthoryear{{Zolotov} et~al.,}{{Zolotov}
  et~al.}{2012}]{Zolotov2012}
{Zolotov} A.,  et~al., 2012, \mn@doi [\apj] {10.1088/0004-637X/761/1/71}, \href
  {http://adsabs.harvard.edu/abs/2012ApJ...761...71Z} {761, 71}

\bibitem[\protect\citeauthoryear{de Jong et~al.,}{de~Jong
  et~al.}{2012}]{4MOST2012}
de Jong R.~S.,  et~al., 2012, in McLean I.~S.,  Ramsay S.~K.,   Takami H.,
  eds,  Society of Photo-Optical Instrumentation Engineers (SPIE) Conference
  Series Vol. 8446, Ground-based and Airborne Instrumentation for Astronomy IV.
  SPIE, pp 252 -- 266, \mn@doi{10.1117/12.926239}, \url
  {https://doi.org/10.1117/12.926239}

\makeatother
\end{thebibliography}
\bibliographystyle{mnras}

\appendix
 \section{Classification Tree Method}
 \label{sec:dtm}
\begin{figure}
\centering
\includegraphics[width=0.4\textwidth]{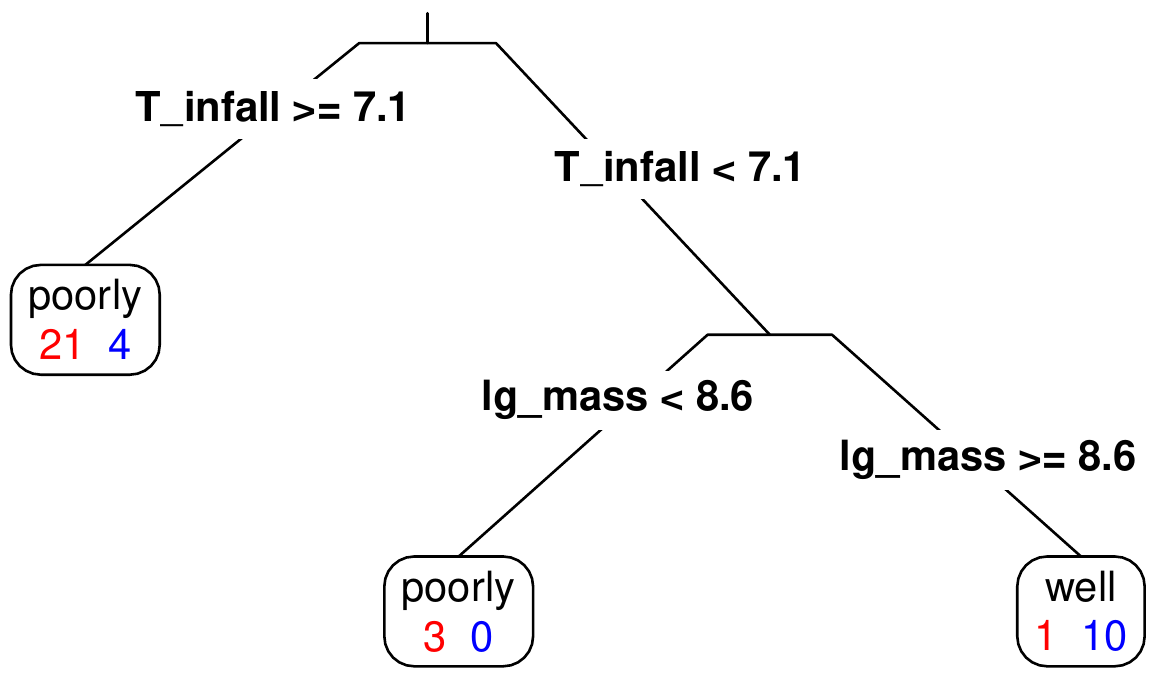}
\caption{A classification tree used in Section \ref{subsec:acc} to derive the boundaries between well-recovered and poorly-recovered satellites. ``T\_infall" indicates \tinfall\, in Gyr ago, ``lg\_mass" represents $\log_{10}(M_{\rm tot}/\msun)$. A ``well" or ``poorly" in each leaf node indicates the prediction of the label of data points in this node by majority vote, with ``well" corresponding to well-recovered and ``poorly" corresponding to poorly-recovered. The red and blue numbers in each leaf node indicate the numbers of poorly-recovered and well-recovered satellites in this leaf node, respectively. The tree with $\log_{10}(M_{\rm stellar}/\msun)$ and \tinfall\,in Gyr ago as input is similar, except that the split in $\log_{10}(M_{\rm stellar}/\msun)$ is located at 6.1.}
\label{fig:tree}
\end{figure}

 We use a binary classification tree \citep{breiman1984classification} to objectively determine the boundaries between well-recovered and poorly-recovered satellites in the \tinfall\, - \mtot\, space and \tinfall\, -$M_{\rm stellar}$ space. A classification tree tries to divide the multi-dimensional space covered by the input data into a series of regions, so that the data points inside each region are as pure (having the same label) as possible. To grow such a classification tree, we do several binary splits based on whether a feature is greater than or equal to ($\geq$) or smaller than ($<$) the split value. The rectangular region before a split is called a parent node, while the two sub-regions resulting from the binary split are called the children nodes. If a node has no children nodes, then it is called a leaf node. The impurity of a node indicates how diverse the labels of data points in a node are, and can be measured by cross-entropy, Gini-index or misclassification error. For details on these scales of impurity, see~\citet{hastie01statisticallearning}. For one parent node with $N_{\rm parent}$ data points and two children nodes with $N_{\rm left}$ and $N_{\rm right}$ data points resulting from a split at one feature $f_i = x$, calculate the quality of split $Q(f_i,x)$:
 \bea
 Q(f_i,x) = {\rm Impurity(parent~node)} \nonumber \\
 - \frac{N_{\rm left}}{N_{\rm parent}} {\rm Impurity(left~child)} \nonumber \\
        - \frac{N_{\rm right}}{N_{\rm parent}} {\rm Impurity(right~child)}
 \eea
 The feature $f_i$ and value $x$ that maximize $Q(f_i,x)$ are the split feature and split value of a binary split. A tree can be grown by adding binary splits in this way until each terminal node is pure and cannot be further split. We then prune the tree by giving a penalty proportional to the size of the tree, until a balance between accuracy and size of tree is reached. \figref{tree} shows the diagram of tree generated from $\log_{10}(M_{\rm tot}/\msun)$ and \tinfall\,in Gyr ago as input data. In each leaf node, a ``well" or ``poorly" indicates the prediction on the label of this node by majority vote. A red number in each leaf node is the number of satellites that are not well-recovered in this node, while the blue number corresponds to the number of well-recovered satellites in this node. The tree with $\log_{10}(M_{\rm stellar}/\msun)$ and \tinfall\,in Gyr ago as input is similar to the tree in \figref{tree}, except that the split in $\log_{10}(M_{\rm stellar}/\msun)$ locates at 6.1.

% Don't change these lines
\bsp	% typesetting comment
\label{lastpage}
\end{document}